\documentclass[number,3p,10pt]{elsarticle}

\usepackage{graphicx,epsfig}
\usepackage{rotate}
\usepackage{amsmath}
\usepackage{amssymb}
\usepackage{amsthm}

\journal{Physics Reports}

\newcommand{\bc}{\begin{center}}
\newcommand{\ec}{\end{center}}
\newcommand{\be}{\begin{equation}}
\newcommand{\ee}{\end{equation}}
\newcommand{\bea}{\begin{eqnarray}}
\newcommand{\eea}{\end{eqnarray}}
\newcommand{\la}{\langle}
\newcommand{\ra}{\rangle}

\begin{document}

\begin{frontmatter}

\title{Synchronization in complex networks}

\author{Alex Arenas}

\address{Departament d'Enginyeria Inform\`{a}tica i Matem\`{a}tiques,
  Universitat Rovira i Virgili, 43007 Tarragona, Spain}
\address{Institute for Biocomputation and Physics of Complex Systems (BIFI), University of Zaragoza, Zaragoza 50009, Spain}

\author{Albert D\'{\i}az-Guilera}
\address{Departament de F\'{\i}sica Fonamental, Universitat de
  Barcelona, 08028 Barcelona, Spain}
\address{Institute for Biocomputation and Physics of Complex Systems (BIFI), University of Zaragoza, Zaragoza 50009, Spain}

\author{Jurgen Kurths}
\address{Institute of Physics, University of Potsdam PF 601553, 14415 Potsdam,
Germany}

\author{Yamir Moreno}
\address{Institute for Biocomputation and Physics of Complex Systems (BIFI), University of Zaragoza, Zaragoza 50009, Spain}
\address{Department of Theoretical Physics, University of Zaragoza, Zaragoza 50009, Spain}

\author{Changsong Zhou}
\address{Department of Physics, Hong Kong Baptist University, Kowloon Tong, Hong Kong}

\date{\today}

\begin{abstract}
Synchronization processes in populations of locally interacting elements are in the focus of intense
research in physical, biological, chemical, technological and social systems. The many efforts devoted to understand synchronization phenomena in natural systems take now advantage of the recent theory of complex networks. In this review, we report the advances in the comprehension of synchronization phenomena when oscillating elements are constrained to interact in a complex network topology. We also overview the new emergent features coming out from the
interplay between the structure and the function of the underlying pattern of connections.
Extensive numerical work as well as analytical approaches to the problem are presented. Finally, we review several applications of synchronization in complex networks to different disciplines: biological systems and neuroscience, engineering and computer science, and
economy and social sciences.
 \end{abstract}

\begin{keyword}
synchronization \sep complex networks
\PACS 05.45.Xt \sep 89.75.Fb \sep 89.75.Hc
\end{keyword}

\end{frontmatter}

\tableofcontents

\section{Introduction}

Synchronization, as an emerging phenomenon of a population of dynamically interacting units, has fascinated humans from ancestral times. Synchronization processes are ubiquitous in nature and play a very important role in many different contexts as biology, ecology, climatology, sociology, technology, or even in arts \cite{prk01,okz07}. It is known that synchrony is rooted in human life from the metabolic processes in our cells to the highest cognitive tasks we perform as a group of individuals. For example, the effect of synchrony has been described in experiments of people communicating, or working together with a background of shared, non-directive conversation, song or rhythm, or of groups of children interacting to an unconscious beat. In all cases the purpose of the common wave length or rhythm is to strengthen the group bond. The lack of such synchrony can index unconscious tension, when goals cannot be identified nor worked towards because the members are "out of sync" \cite{hall83}.

Among the efforts
for the scientific description of synchronization phenomena, there are several capital works that represented a breakthrough in our understanding of these phenomena.
In 1665, the mathematician and physicist, inventor of the pendulum clock, C. Huygens, discovered an odd "kind of sympathy" in two pendulum clocks suspended side by side of each other. The pendulum clocks swung with exactly the same frequency and 180 degrees out of phase; when the pendula were disturbed, the antiphase state was restored within half an hour and persisted indefinitely. Huygens deduced that the crucial interaction for this effect came from "imperceptible movements" of the common frame supporting the two clocks. From that time on, the phenomenon got into the focus of scientists. Synchronization involves, at least, two elements in interaction, and the behavior of a few interacting oscillators has been intensively studied in the physics and mathematics literature.  However, the phenomenon of synchronization of large populations is a different challenge and requires different hypothesis to be solved. We will focus our attention on this last challenge.

In the obituary of Arthur T. Winfree, \citeauthor{obituary} \cite{obituary}
summarizes what can be considered the beginning of the modern quest to explain the synchronization of a population of interacting units: "\citeauthor{wiener48} \cite{wiener48} posed a problem in his book Cybernetics: How is it that thousands of neurons or fireflies or crickets can suddenly fall into step with one another, all firing or flashing or chirping at the same time, without any leader or signal from the environment? Wiener did not make significant mathematical progress on it, nor did anyone else until Winfree came along".  \citeauthor{winfree67} \cite{winfree67} studied the nonlinear dynamics of a large population of weakly coupled limit-cycle oscillators with intrinsic frequencies that were distributed about some mean value, according to some prescribed probability distribution. The milestone here was to consider biological oscillators as \emph{phase oscillators}, neglecting the amplitude. Working within the framework of a mean field model, Winfree discovered that such a population of
non-identical
oscillators can exhibit a remarkable cooperative phenomenon. When the variance of the frequencies distribution is large, the oscillators run incoherently,
each one near its natural frequency.
This behavior remains when reducing the variance
until a certain threshold is crossed.
However, below the threshold the oscillators begin to synchronize spontaneously (see \cite{winfree80}). Note that the original Winfree model was not solved analytically until recently \cite{as01}.

Although Winfree's approach proved to be successful in describing the emergence of spontaneous order in the system, it was based on the premise that every oscillator {\em feels} the same pattern of interactions. However, this all-to-all connectivity between elements of a large population is difficult to conceive in real world. When the number of elements is large enough, this pattern is incompatible with physical constraints as for example minimization of energy (or costs), and in general with the rare observation of long range interactions in systems formed by macroscopic elements. The particular local connectivity structure of the elements was missing (in fact, discarded) in these and subsequent approaches.

\begin{figure}[!t]
\begin{center}
\epsfig{file=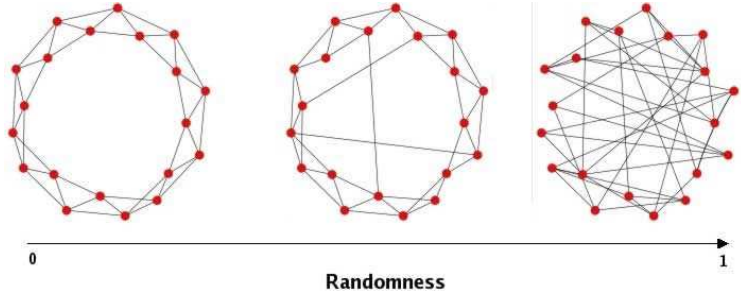,width=0.6\columnwidth,angle=0,clip=1}
\end{center}
\caption{Small-world network construction from a regular lattice by rewiring links with a certain probability
(randomness), as proposed by
\citeauthor{ws98} \cite{ws98}.
}
\label{sw}
\end{figure}

In 1998, Watts and Strogatz presented a simple model of network
structure, originally intended precisely to introduce the connectivity substrate in the problem of synchronization of cricket chirps, which show a high degree of coordination over long distances as though the insects were "invisibly" connected. Remarkably, this work did not end in a new contribution to synchronization theory but as the seed for the modern theory of complex
networks \cite{ws98}.
Starting with a regular lattice, they showed that adding a small number of random links reduces the distance between nodes drastically, see Fig.~\ref{sw}. This feature, known as \emph{small-world} (SW) effect, had been first reported in an experiment conducted by S. Milgram \cite{m67} examining the average path length for social networks of people in the United States. Nowadays, the phenomenon has been detected in many other natural and artificial networks. The inherent complexity
of the new model, from now on referred to as the Watts-Strogatz (WS) model,
was in its mixed nature in between regular lattices
and random graphs. Very soon, it turned out that the nature of many
interaction patterns observed in scenarios as diverse as the Internet, the World-Wide Web,
scientific collaboration networks, biological networks,
was even more "complex" than the WS model. Most of them showed a heavy tailed distribution
of connectivities with no characteristic scale. These networks have been since then called \emph{scale-free} (SF) networks
and the most connected nodes are called \emph{hubs}.
This novel structural complexity provoked
an explosion of works, mainly from the physicists community, since a completely new set of measures, models, and techniques, was needed to deal with these topological structures.

During one decade we have witnessed the evolution of the field of complex networks,
mainly from a static point of view, although some attempts to characterize
the dynamical properties of complex networks have also been made. One
of these dynamical implications,
addressed
since the very beginning of the subject, is the emergent phenomena of
synchronization of a population of units with an oscillating behavior. The analysis of synchronization processes has benefited from the advance in the understanding of the topology of
complex networks, but it has also contributed to the understanding of
general emergent properties of networked systems. The main goal of this review is precisely to revise the research undertaken so far in order to understand how synchronization phenomena are affected by the topological substrate of interactions, in particular when this substrate is a complex network.

The review is organized as follows.
We first introduce the basic mathematical descriptors of complex networks that will be used henceforth.
Next, we focus on the synchronization of populations of oscillators.
Section IV is devoted to the analysis of the conditions for the stability of the fully synchronized state using
the Master Stability Function (MSF) formalism.
Applications in different fields of science are presented afterwards and some perspectives provided.
Finally, the last section rounds off the review by giving our conclusions.

\section{Complex networks in a nutshell}

There exist excellent reviews devoted to the structural characterization and evolution of complex networks \cite{s01,ab02,dm02,n03a, blmch06,crtbv07}. Here we summarize the main features and standard measures used in complex networks. The goal is to provide the reader a brief overview of the subject as well as to introduce some notation that will be used throughout the review.

The mathematical abstraction of a complex network is a graph $\mathcal{G}$ comprising a set of $N$ nodes (or vertices) connected by a set of $M$ links (or edges), being $k_i$ the degree (number of links) of node $i$. This graph is represented by the adjacency matrix $A$, with entries $a_{ij} =1$ if a directed link from $j$ to $i$ exists, and 0 otherwise. In the more general case of a weighted network,
the graph is characterized by a matrix $W$, with entries $w_{ij}$, representing the strength (or weight) of the
link from $j$ to $i$. The investigation of the statistical properties of many natural and man-made complex networks revealed that, although representing very different systems, some categorization of them is possible.
The most representative of these properties refers to the {\em degree distribution} $P(k)$, that indicates the probability of a node to have a degree $k$. This fingerprint of complex networks has been taken for a long time as its most differentiating factor. However, several other measures help to precise the categorization.
Examples are the {\em average shortest path length} $\ell=\langle  d_{ij} \rangle$, where $d_{ij}$ is the length of the shortest path between node $i$ and node $j$, and the {\em clustering coefficient} $C$ that accounts for the fraction of actual triangles (three vertices forming a loop) over possible triangles in the graph.

The first classification of complex networks is related to the degree distribution $P(k)$. The differentiation between homogeneous and heterogeneous networks in degree is in general associated to the tail of the distribution.
If it decays exponentially fast with the degree
we refer to
as homogeneous networks, the most representative example being the  Erd\"{o}s-R\'{e}nyi (ER) \emph{random graph} \cite{er59}.
On the contrary, when the tail is heavy one can say that the network is heterogeneous.
In particular, SF networks are the class of networks whose distribution is a power-law,
$P(k)\sim k^{-\gamma}$, the Barab\'{a}si-Albert (BA) model \cite{ba99}
being the paradigmatic model of this type
of graph.
This network is grown by a mechanism in which all incoming nodes are linked preferentially to
the existing nodes.
Note that the limiting case of lattices, or regular networks, corresponds to a
situation where all nodes have the
same degree.

This categorization can be enriched by the behavior of $\ell$. For a lattice of dimension $d$ containing $N$ vertices, obviously, $\ell \sim N^{1/d}$.
For a random network, a  rough estimate for $\ell$ is also possible.
If the average number of nearest neighbors of a vertex is $\bar{k}$, then about $\bar{k} ^{\ell}$ vertices of the network are at a distance $\ell$ from the vertex or closer. Hence, $N\sim \bar{k}^ {\ell}$ and then $\ell \sim \ln(N)/\ln({\bar k)}$ , i.e., the average shortest-path length value is small even for very large networks. This smallness is usually referred to as the SW property. Associated to distances, there exist many measures that provide information about "centrality" of nodes. For instance, one can say that a node is central in terms of the relative distance to the rest of the network. One of the most frequently used centrality measures in the physics literature is the
\emph{betweenness} (load in some papers), that accounts
for the number of shortest paths between any pair of nodes in the network that go through a given node or link.

The clustering coefficient $C$ is also a discriminating property between different types of networks.
It is usually calculated as follows:
\begin{equation}
C=\frac{1}{N}\sum_{i=1}^{N} C_i = \frac{1}{N}\sum_{i=1}^{N} \frac{n_i}{k_i(k_{i}-1)/2},
\end{equation}
\noindent where $n_i$ is the number of connections between nearest neighbors of node $i$, and $k_i$ is its degree. A large clustering coefficient implies many transitive connections and consequently redundant paths in the network, while a low $C$ implies the opposite.

Finally, it is worth mentioning that
many networks have a \emph{community} structure, meaning that nodes are linked together in densely connected groups between which connections are sparser.
Finding the best partition of a network into communities is a very hard problem.
The most successful solutions, in terms of accuracy and
computational cost \cite{ddda05}, are those based on the optimization of a magnitude called {\em modularity},
proposed in \cite{ng04}, that precisely allows for the comparison of different partitionings of the network.
The modularity of a given partition is, up to a multiplicative constant, the number of links falling within groups minus its expected number in an equivalent network with links placed at random. Given a network partitioned into communities, the mathematical definition of modularity is expressed in terms of the adjacency matrix $a_{ij}$ and the total number of links $M=\frac{1}{2}\sum_i k_i$ as
\begin{equation}
Q=\frac{1}{2M}\sum_{ij} (a_{ij} -\frac{k_ik_j}{2M})\delta_{c_i, c_j}
\label{Q}
\end{equation}
where $c_i$ is the community to which node $i$ is assigned and the Kronecker delta function $\delta_{c_i,c_j}$ takes the value 1 if nodes $i$ and $j$ are in the same community, and 0 otherwise.
The larger the $Q$ the more modular the network is. This property promises to be specially adequate to unveil structure-function relationships in complex networks \cite{gn02,gddga03,gmta05,n06}.

\section{Coupled phase oscillator models on complex networks}

The need to understand synchronization, mainly in the context of biological neural networks, promoted the first studies of synchronization of coupled oscillators considering a network of interactions between them. In the late 80's,  \citeauthor{sm88} \cite{sm88} and later \citeauthor{nskk91} \cite{nskk91} studied the collective synchronization of phase non-linear oscillators with random intrinsic frequencies under a variety of coupling schemes in 2D lattices. Beyond the differences with the actual conception of a complex network, the topologies studied in \cite{nskk91} can be thought of as a first approach to reveal how the complexity of the connectivity affects synchronization.
The authors used a square lattice as a geometrical reference to construct three different connectivity schemes: four nearest neighbors, Gaussian connectivity truncated at $2\sigma$, and finally a random sparse connectivity.
These results showed that random long-range connections lead to a more rapid and robust phase locking between oscillators than nearest-neighbor coupling or locally dense connection schemes. This observation is at the root of the recent findings about synchronization in complex networks of oscillators. In the current section we review
the results obtained so far on three different kinds of oscillatory ensembles: limit cycle oscillators (Kuramoto), pulse-coupled models, and finally coupled map systems.
We reserve for Sect. \ref{sect_MSF} those works that use the MSF formalism.
Many other works whose major contribution is the understanding of synchronization phenomena in specific scenarios are discussed in the Applications section.

\subsection{Phase oscillators}
\label{sect_KM}
\subsubsection{The Kuramoto model}

The pioneering work by \citeauthor{winfree67} \cite{winfree67}  spurred the field of collective synchronization and called for mathematical approaches to tackle the problem.  One of these approaches, as already stated, considers a system made up of a huge population of weakly-coupled, nearly identical, interacting limit-cycle oscillators, where each oscillator exerts a phase dependent influence on the others and changes its rhythm according to a sensitivity function \cite{s00,abprs05}.

Even if these simplifications seem to be very crude, the phenomenology of the problem can be captured. Namely, the population of oscillators exhibits the dynamic analog
to an equilibrium phase transition. When the natural frequencies of the oscillators are too diverse compared to the
strength of the coupling, they are unable to synchronize and the system behaves incoherently. However, if the coupling is strong enough, all oscillators freeze into synchrony. The transition from one regime to the other takes place at a
certain threshold.  At this point some elements lock their relative phase and a cluster of synchronized nodes
develops. This constitutes the \emph{onset of synchronization}. Beyond this value, the
population of oscillators is split into a partially synchronized state
made up of oscillators locked in phase and a group of
nodes whose natural frequencies are too different as to be part of the
coherent cluster. Finally, after further increasing the coupling, more and more elements get entrained around the mean phase of the collective rhythm generated by the whole population and
the system settles in the completely synchronized state.

\citeauthor{Kuramoto75} \cite{Kuramoto75,Kuramoto84} worked out a mathematically tractable model to describe this phenomenology. He recognized that the most suitable case for analytical treatment should be the mean field approach.
He proposed an all-to-all purely sinusoidal coupling, and then the governing equations for each of the oscillators in the system are:
\be
\dot{\theta}_{i}=\omega_{i}+\frac{K}{N}\sum_{j=1}^{N}
\sin{(\theta_{j}-\theta_{i})}
\hspace{0.5cm} (i=1,...,N)\;,
\label{ekuramodel}
\ee
where the factor $1 / N$ is incorporated to ensure a good
behavior of the model in the thermodynamic limit,
$N\rightarrow\infty$, $\omega_i$ stands for the natural frequency of
oscillator $i$, and $K$ is the coupling constant. The frequencies $\omega_i$ are distributed according to some function $g(\omega)$, that is usually assumed to be unimodal and symmetric about its mean frequency $\Omega$. Admittedly, due to the rotational symmetry in the model, we can use a rotating frame and redefine $\omega_i\rightarrow\omega_i+\Omega$ for all $i$ and set $\Omega=0$, so that the $\omega_i$'s denote deviations from the mean frequency.

The collective dynamics of the whole population is measured by the \emph{macroscopic} complex order parameter,
\be
r(t)\text{e}^{\text{i}\phi(t)}=\frac{1}{N}\sum_{j=1}^{N}\text{e}^{\text{i}\theta_{j}(t)}\;,
\label{ekuraorderparam}
\ee
where the modulus $0\le r(t) \le 1$ measures the phase coherence of
the population and $\phi(t)$ is the average phase. The values $r\simeq 1$ and $r\simeq 0$ (where $\simeq$ stands for fluctuations of size $O(N^{-1/2})$) describe the limits in which all oscillators are either phase locked or move incoherently, respectively. Multiplying both parts of Eq.\ (\ref{ekuraorderparam}) by $\text{e}^{-\text{i}\theta_i}$ and equating imaginary parts gives
\be
r \sin(\phi-\theta_i)=\frac{1}{N}\sum_{j=1}^{N}\sin(\theta_{j}-\theta_i)\;,
\ee
yielding
\be
\dot{\theta}_{i}=\omega_{i}+Kr\sin{(\phi-\theta_{i})}
\hspace{0.5cm} (i=1,...,N)\;.
\label{ekuramf}
\ee
Equation\ (\ref{ekuramf}) states that each oscillator interacts with all the others only through the mean field quantities $r$ and $\phi$. The first quantity provides a positive feedback loop to the system's collective rhythm: as $r$ increases because the population becomes more coherent, the coupling between the oscillators is further strengthened and more of them can be recruited to take part in the coherent pack. Moreover, Eq.\ (\ref{ekuramf}) allows to calculate the critical coupling $K_c$ and to characterize the order parameter
$
\lim_{t\rightarrow\infty}r_t(K)=r(K).
$
Looking for steady solutions, one assumes that $r(t)$  and $\phi(t)$ are constant. Next, without loss of generality, we can set $\phi=0$, which leads to the equations of motion \cite{Kuramoto75,Kuramoto84}
\be
\dot{\theta}_{i}=\omega_{i}-Kr\sin{(\theta_{i})}
\hspace{0.5cm} (i=1,...,N)\;.
\label{ekurasteady}
\ee
The solutions of Eq.\ (\ref{ekurasteady}) reveal two different types of long-term behavior when the coupling is larger than the critical value, $K_c$. On the one hand, a group of oscillators for which $|\omega_i|\leq Kr$ are phase-locked at frequency $\Omega$ in the original frame according to the equation $\omega_{i}=Kr\sin{(\theta_{i})}$. On the other hand, the rest of the oscillators for which $|\omega_i| > Kr$ holds, are drifting around the circle, sometimes accelerating and sometimes rotating at lower frequencies. Demanding some conditions for the stationary distribution of drifting oscillators with frequency $\omega_i$ and phases $\theta_i$ \cite{s00}, a self-consistent equation for $r$ can be derived as
\be
r=Kr\int_{-\frac{\pi}{2}}^{\frac{\pi}{2}}\left( \cos^2\theta \right) g(\omega)d\theta,
\nonumber
\ee
where $\omega=Kr\sin{(\theta)}$. This equation admits a non-trivial solution,
\be
K_c=\frac{2}{\pi g(0)}.
\label{ekurakc}
\ee
beyond which $r>0$.
Equation \ (\ref{ekurakc}) is the Kuramoto mean field expression for the critical coupling at the onset of synchronization. Moreover, near the onset, the order parameter, $r$, obeys the usual square-root scaling law for mean field models, namely,
\be
r\sim(K-K_c)^{\beta}
\label{ekuraexponent}
\ee
with $\beta=1/2$.
Numerical simulations of the model verified these results.
The Kuramoto model (KM, from now on) approach to synchronization was a breakthrough for the
understanding of synchronization in large populations
of oscillators.

Even in the simplest case of a mean field interaction, there are still
unsolved problems that have resisted any analytical attempt. This is the case, e.g., for finite populations of oscillators and some questions regarding global stability results \cite{abprs05}. In what follows, we focus on another aspect of the model's assumptions, namely that of the connection topology of real systems \cite{n03a,blmch06}, which usually do not show the all-to-all pattern of interconnections underneath the mean field approach.

\subsubsection{Kuramoto model on complex networks}
\label{kmnets}

To deal with the KM on complex topologies, it is necessary to reformulate Eq. (\ref{ekuramodel}) to include the connectivity

\be
\dot{\theta}_i=\omega_i + \sum_{j}
\sigma_{ij}a_{ij}\sin(\theta_j-\theta_i) \hspace{0.5cm} (i=1,...,N)\;,
\label{ekurageneral}
\ee
where $\sigma_{ij}$ is the coupling strength between pairs of
connected oscillators and $a_{ij}$ are the elements of the connectivity matrix. The original Kuramoto model is recovered by letting $a_{ij}=1, \forall i\neq
j$ (all-to-all) and $\sigma_{ij}=K/N, \forall i,j$.

The first problem when defining the KM in complex networks is how to state
the interaction dynamics properly. In contrast with the mean field model, there are several ways to define how the connection topology enters in the governing equations of the dynamics. A good theory for Kuramoto oscillators in complex networks should be phenomenologically relevant and provide formulas amenable to rigorous mathematical treatment. Therefore, such a theory should at least  preserve the essential fact of treating the heterogeneity of the network independently of the interaction dynamics, and at
the same time, should remain calculable in the thermodynamic limit.

For the original model, Eq. (\ref{ekuramodel}), the coupling term on
the right hand side of Eq. (\ref{ekurageneral}) is an intensive magnitude because the dependence on the size of the system cancels out. This
independence on the number of oscillators $N$ is achieved by choosing
$\sigma_{ij}=K/N$. This prescription turns out to be
essential for the analysis of the system in the thermodynamic limit
$N\rightarrow \infty$ in the all-to-all case. However, choosing $\sigma_{ij}=K/N$
for the governing equations of the KM in a complex network makes them to become dependent on
$N$. Therefore, in the thermodynamic limit, the coupling term tends to
zero except for those nodes with a degree that scales with $N$.  Note that the existence of such nodes
is only possible in networks with power-law
degree distributions  \cite{n03a,blmch06}, but this happens with a very small probability as
$k^{-\gamma}$, with $\gamma>2$. In these cases, mean field solutions
independent of $N$ are recovered, with slight differences in the onset
of synchronization of all-to-all and SF networks
\cite{roh05a} .

A second prescription consists in taking $\sigma_{ij}=K/k_i$ (where
$k_i$ is the degree of node $i$) so that $\sigma_{ij}$ is a weighted
interaction factor that also makes the right hand side of Eq.
(\ref{ekurageneral}) intensive. This form has been used to solve the {\em paradox of
heterogeneity} \cite{mzk05b} that states that the heterogeneity in the degree distribution,
which often reduces the average distance between nodes, may suppress
synchronization in networks of oscillators coupled symmetrically with uniform
coupling strength. This result refers to the stability of the \emph {fully synchronized
state}, but not to the dependence of the order parameter on the coupling strength (where partially synchronized
and unsynchronized states exist).
Besides, the inclusion of weights in the interaction strongly
affects the original KM dynamics in complex networks because it can impose a
dynamic homogeneity that masks the real topological heterogeneity of the network.

The prescription $\sigma_{ij}=K/\text{const}$, which may seem more
appropriate, also causes some conceptual problems because the sum in the
right hand side of Eq.\ (\ref{ekurageneral}) could eventually diverge in the
thermodynamic limit.
The constant in the denominator could in principle be any quantity related to the topology, such as the average connectivity of the graph,$\langle  k \rangle$, or the maximum degree $k_{\max}$. Its physical meaning is a re-scaling of the temporal scales involved in the dynamics. However, except for the case of $\sigma_{ij}=K/k_{\max}$, the other possible settings do not avoid the problems when $N\rightarrow \infty$. On the other hand, for a proper comparison of the results obtained for different complex topologies (e.g. SF or uniformly random), the global and local measures of coherence should be represented according to their respective time
scales. Therefore, given two complex networks A and B with $k_{\max}=k_A$ and
$k_{\max}=k_B$ respectively, it follows that to make meaningful comparisons between observables, the equations of motion Eq.\ (\ref{ekurageneral}) should refer to the same time scales, i.e., $\sigma_{ij}=K_A/k_A=K_B/k_B=\sigma$. With this formulation in mind,
Eq.\ (\ref{ekurageneral}) reduces to
\begin{equation}
\dot{\theta}_i=\omega_i + \sigma \sum_{j}
a_{ij}\sin(\theta_j-\theta_i) \hspace{0.5cm} (i=1,...,N)\;,
\label{ekurafinal}
\end{equation}
independently of the specific topology of the network. This allows us to study the dynamics of Eq. (\ref{ekurafinal}) on different topologies, compare the results, and properly inspect the interplay
between topology and dynamics in what concerns synchronization.

As we shall see, there are also several ways to define the order parameter that characterizes the global dynamics of the system, some of which were introduced to allow for analytical treatments at the onset of synchronization. We advance, however, that the same order parameter, Eq. (\ref{ekuraorderparam}), is often used to describe the coherence of the synchronized state.

\subsubsection{Onset of synchronization in complex networks}

Studies on synchronization in complex topologies where each node is considered to be a Kuramoto oscillator, were first reported for WS
networks \cite{w99,hck02} and BA graphs \cite{mp04,mvp04}. These works are mainly numerical explorations of the onset of synchronization, their main goal being the characterization of the critical coupling beyond which groups of nodes beating coherently first appear.  In \cite{hck02}, the authors considered oscillators with intrinsic frequencies distributed according to a Gaussian distribution with unit variance arranged in a WS network with varying rewiring probability, $p$, and explored how the order parameter, Eq. (\ref{ekuraorderparam}), changes upon addition of long-range links. Moreover, they assumed a normalized coupling strength $\sigma_{ij}=K/\langle  k \rangle$  , where $\langle  k \rangle$   is the average degree of the graph. Numerical integration of the equations of motion (\ref{ekurageneral}) under variation of $p$ shows that collective synchronization emerges even for very small values of the rewiring probability.

The results confirm that networks obtained from a regular ring by
just rewiring a tiny fraction of links ($p\gtrsim 0$) can be synchronized with a finite $K$.
Moreover, in contrast with the arguments provided in \cite{hck02}, we notice that their results had been obtained for a fixed average degree and thus the Kuramoto's critical coupling can not be recovered by simply taking $p\rightarrow 1$, which produces a random ER graph with a fixed minimum connectivity. This limit is recovered by letting $\langle  k \rangle$   increase. Actually, numerical simulations of the same model in \cite{w99} showed that the Kuramoto limit is approached when the average connectivity grows.

\begin{figure}[t]
\begin{center}
\epsfig{file=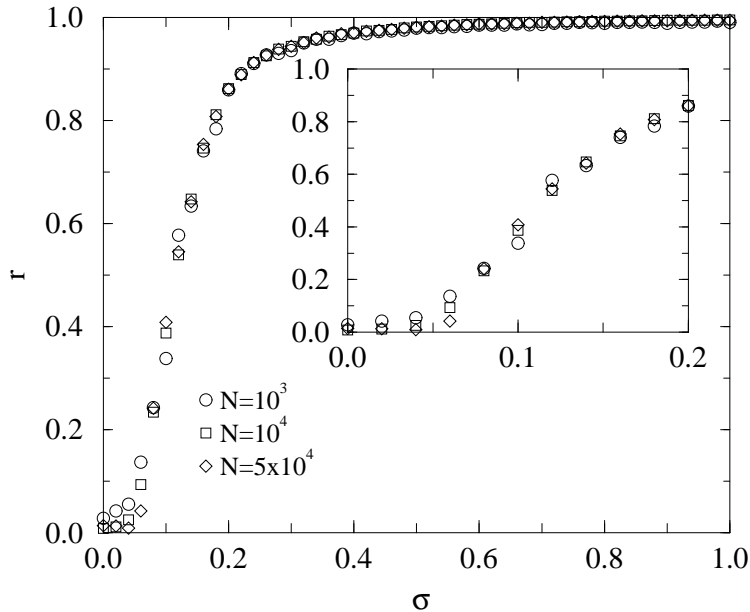,width=3.3in,angle=-90,clip=1}
\end{center}
\caption{Order parameter $r$ (Eq.\ (\ref{ekuraorderparam})) as a function of $\sigma$ for several BA networks of different
sizes. Finite size scaling analysis shows that the onset of synchronization takes place at a critical value
$\sigma_c=0.05(1)$. The inset is a zoom around $\sigma_c$. From \cite{mp04}.}
\label{figkmsf}
\end{figure}

In \cite{mp04} the same problem in BA networks it is considered.
The natural frequencies and the
initial values of $\theta_i$ were randomly drawn from a uniform
distribution in the interval $(-1/2,1/2)$ and $(-\pi,\pi)$,
respectively. The global dynamics of the system,
Eq. (\ref{ekurafinal}), turns out to be qualitatively the same as for the original KM as shown in Fig.\ \ref{figkmsf}, where the dependence of the order parameter  Eq.\ (\ref{ekuraorderparam}) with $\sigma$ is shown for several system sizes.

The existence of a critical point for the KM on SF networks came as a surprise. Admittedly, this is one of the few cases in which a dynamical process shows a critical behavior when the substrate is described by a power-law connectivity distribution with an exponent $\gamma\leq 3$ \cite{n03a,blmch06,dgm07}.
In principle it could be a finite size effect, but it turned out from numerical simulations that this was not the case. To determine the exact value of $\sigma_c$, one can make use of standard finite-size scaling analysis. At least two complementary strategies have been reported. The first one allows bounding the critical point and is computationally more expensive. Consider a network of size $N$, for which
no synchronization is attained below $\sigma_c$, where $r(t)$ decays to a small residual value of size $O(1/\sqrt{N})$. Then, the critical point may be found by examining the $N$-dependence of $r(\sigma,N)$. In the sub-critical regime ($\sigma < \sigma_c$), the stationary value of $r$ falls off as $N^{-1/2}$, while for $\sigma >\sigma_c$, the order parameter reaches a stationary value as $N\rightarrow\infty$ (though still with $O(1/\sqrt{N})$ fluctuations). Therefore, plots of $r$ versus $N$ allow us to locate the critical point $\sigma_c$. Alternatively, a more accurate approach can be adopted. Assume the scaling form for the order parameter \cite{md99}:

\be
r=N^{-\alpha}f(N^{\nu}(\sigma-\sigma_c))\;,
\label{ekfss}
\ee
where $f(x)$ is a universal scaling function bounded as $x \rightarrow \pm \infty$ and $\alpha$ and $\nu$ are two critical exponents to be determined. Since at $\sigma=\sigma_c$, the value of the function $f$ is independent of $N$, the estimation of $\sigma_c$ can be done by plotting $N^{\alpha}r$ as a function of $\sigma$ for various sizes $N$ and then finding the value of $\alpha$ that gives a well-defined crossing point, the critical coupling $\sigma_c$. As a by-product, the method also allows us to calculate the two scaling exponents $\alpha$ and $\nu$, the latter can be obtained from the equality
\be
\ln[(dr/d\sigma)|_{\sigma_c}]=(\nu-\alpha)\ln N+\text{const},
\ee
once $\alpha$ is computed.

Following these scaling procedures,
it was estimated a value for the
critical coupling strength $\sigma_c=0.05(1)$ \cite{mp04,gma07a,gma07b}. Moreover, $r\sim(\sigma-\sigma_c)^{\beta}$ when approaching the critical point from above with
$\beta=0.46(2)$ indicating that the square-root behavior typical of
the mean field version of the model ($\beta=1/2$) seems to hold as well for BA networks.

Before turning our attention to some theoretical attempts to tackle the onset of synchronization, it is worth to briefly summarize other numerical results that have explored how the critical coupling depends on other topological features of the underlying SF graph. Recent results have shed light on the influence of the topology of the local
interactions on the route to and the onset of synchronization. In particular, the authors in  \cite{mm05,mm07,gm07} explored the Kuramoto dynamics on networks in which the degree distribution is kept fixed, while the
clustering coefficient ($C$) and the average path length ($\ell$) of the graph change. The results suggest that the onset of
synchronization is mainly determined by $C$, namely, networks with a high clustering coefficient promote synchronization at lower values of the coupling strength. On the other hand, when the coupling is increased beyond
the critical point, the effect of $\ell$ dominates over $C$ and
the phase diagram is smoothed out (a sort of stretching), delaying the
appearance of the fully synchronized state as the average shortest path length increases.

In a series of recent papers \cite{roh05a,roh05b,roh06a,i04,i05,l05}, the onset of synchronization in large networks of coupled oscillators has been analyzed from a theoretical point of view under certain (sometimes strong) assumptions. Despite these efforts no exact analytical results for the KM on general complex networks are available up to date. Moreover, the analytical approaches predict that for uncorrelated SF networks with an exponent $\gamma\leq 3$, the critical coupling vanishes as $N\rightarrow\infty$, in contrast to numerical simulations on BA model networks. It appears that the strong heterogeneity of real networks and the finite average connectivity strongly hampers analytical solutions of the model.

Following \cite{roh05a}, consider the system in Eq. (\ref{ekurafinal}),
with a symmetric\footnote{The reader can find the extension of the forthcoming formalism to directed networks in \cite{roh05b}.}
adjacency matrix $a_{ij}=a_{ji}$.
Defining a local order parameter $r_i$ as
\be
r_i\text{e}^{\text{i}\phi_i}=\sum_{j=1}^{N}a_{ij}\la\text{e}^{\text{i}\theta_{j}}\ra_t\;,
\label{ekrlocal}
\ee
where $\la\cdot\cdot\cdot\ra_t$ stands for a time average, a new global order parameter to measure the macroscopic coherence is readily introduced as
\be
r=\frac{\sum_{i=1}^{N}r_i}{\sum_{i=1}^{N}k_i}\;.
\label{ekrglobal}
\ee
Now, rewriting Eq.\ (\ref{ekurafinal}) as a function of $r_i$, yields,
\be
\dot{\theta}_i=\omega_i-\sigma r_i \sin(\theta_i-\phi_i)-\sigma h_i(t)\;.
\label{erestrepo1}
\ee
In Eq.\ (\ref{erestrepo1}), $h_i(t)=\text{Im}\{\text{e}^{-\text{i}\theta_i}\sum_{j=1}^{N}a_{ij}
(\la\text{e}^{\text{i}\theta_{j}}\ra_t-\text{e}^{\text{i}\theta_j})\}$ depends on time and contains time fluctuations. Assuming the terms in the previous sum to be statistically independent, $h_i(t)$ is expected to be proportional to $\sqrt{k_i}$ above the transition, where $r_i\sim k_i$. Therefore, except very close to the critical point, and assuming that the number of connections of each node is large enough\footnote{This obviously restricts the range of real networks to which the approximation can be applied.} ($k_i\gg 1$ as to be able to neglect the time fluctuations entering $h_i$, i.e., $h_i\ll r_i$), the equation describing the dynamics of node $i$ can be reduced to
\be
\dot{\theta}_i=\omega_i-\sigma r_i \sin(\theta_i-\phi_i)\;.
\label{erestrepo2}
\ee
Next, we look for stationary solutions of Eq.\ (\ref{erestrepo2}), i.e. $\sin(\theta_i-\phi_i)={\omega_i}/{\sigma r_i}$. In particular, oscillators whose intrinsic frequency satisfies $|\omega_i|\leq \sigma r_i$ become locked. Then, as in the Kuramoto mean field model, there are two contributions (though in this case to the local order parameter), one from locked and the other from drifting oscillators such that
\bea
r_i& = &\sum_{j=1}^{N}a_{ij}\la\text{e}^{\text{i}(\theta_j-\phi_i)}\ra_{t} =\label{erestrepo3}
 \\
& =& \sum_{|\omega_j|\leq \sigma r_j}a_{ij}\text{e}^{{\text{i}(\theta_j-\phi_i)}}+\sum_{|\omega_j| > \sigma r_j}a_{ij}\la\text{e}^{\text{i}(\theta_j-\phi_i)}\ra_{t}\;.\nonumber
\eea
To move one step further, some assumptions are needed. Consider a graph such that the average degree of nearest neighbors is high (i.e., if the neighbors of node $i$ are well-connected). Then it is reasonable to assume that these nodes are not affected by the intrinsic frequency of $i$. This is equivalent to assume solutions $(r_i,\phi_i)$ that are, in a statistical sense, independent of the natural frequency $\omega_i$. With this assumption, the second summand in Eq.\ (\ref{erestrepo3}) can be neglected. Taking into account that the distribution $g(\omega)$ is symmetric and centered at $\Omega=0$, after some algebra one is left with \cite{roh05a}
\be
r_i=\sum_{|\omega_j|\leq \sigma r_j}a_{ij}\cos(\phi_j-\phi_i)\sqrt{1-\left(\frac{\omega_j}{\sigma r_j}\right)^2}\;.
\label{erestrepo4}
\ee
The critical coupling $\sigma_c$ is given by the solution of Eq.\ (\ref{erestrepo4}) that yields the smallest $\sigma$. It can be argued that it is obtained when $\cos(\phi_j-\phi_i)=1$ in Eq.\ (\ref{erestrepo4}), thus
\be
r_i=\sum_{|\omega_j|\leq \sigma r_j}a_{ij}\sqrt{1-\left(\frac{\omega_j}{\sigma r_j}\right)^2}\;,
\label{erestrepo5}
\ee
which is the main equation of the \emph{time average} approximation (recall that time fluctuations have been neglected). Note, however, that to obtain the critical coupling, one has to know the adjacency matrix as well as the particular values of $\omega_i$ for all $i$ and then solve Eq.\ (\ref{erestrepo5}) numerically for the $\{ r_i\}$. Finally, the global order parameter defined in Eq.\ (\ref{ekrglobal}) can be computed from $r_i$.

Even if the underlying graph satisfies the other aforementioned topological constraints, it seems unrealistic to require knowledge of the $\{ \omega_i \}$'s. A further approach, referred to as the \emph{frequency distribution} approximation can be adopted. According to the assumption that $k_i \gg 1$ for all $i$, or equivalently, that the number of connections per node is large (a dense graph), one can also consider that the natural frequencies of the neighbors of node $i$ follows the distribution $g(\omega)$. Then, Eq.\ (\ref{erestrepo5}) can be rewritten avoiding the dependence on the particular realization of $\{ \omega_i \}$ to yield,
\bea
r_i&=&\sum_{j}a_{ij}\int_{-\sigma r_j}^{\sigma r_j}g(\omega)\sqrt{1-\left(\frac{\omega}{\sigma r_j}\right)^2}d\omega\nonumber\\
&=&\sigma\sum_{j}a_{ij}r_j\int_{-1}^{1}g(x\sigma r_j)\sqrt{1-x^2}dx\;,
\label{erestrepo6}
\eea
with $x=\omega/(\sigma r_j)$. This equation allows us to readily determine the order parameter $r$ as a function of the network topology ($a_{ij}$), the frequency distribution ($g(\omega)$) and the control parameter ($\sigma$). On the other hand, Eq.\ (\ref{erestrepo5}) still does not provide explicit expressions for the order parameter and the critical coupling strength. To this end, one introduces a first-order approximation $g(x\sigma r_j)\approx g(0)$ which is valid for small, but nonzero, values of $r$. Namely, when $r_j\rightarrow 0^+$
\bea
r_i^0=\frac{\sigma}{K_c}\sum_{j}a_{ij}r_j^0\;,
\nonumber
\eea
where $K_c={2}/(\pi g(0))$ is Kuramoto's critical coupling. Moreover, as the smallest value of $\sigma$ corresponds to $\sigma_c$, it follows that the critical coupling is related to both $K_c$ and the largest eigenvalue $\lambda_{\max}$ of the adjacency matrix, yielding
\be
\sigma_c=\frac{K_c}{\lambda_{\max}}\;.
\label{erestrepo7}
\ee
Equation\ (\ref{erestrepo7}) states that in complex networks, synchronization is first attained at a value of the coupling strength that inversely depends on
$g(0)$
and on the largest eigenvalue $\lambda_{\max}$ of the adjacency matrix.
Note that this equation also recovers Kuramoto's result when $a_{ij}=1$, $\forall i\neq j$, since $\lambda_{\max}=N-1$. It is worth stressing that although this method allows us to calculate  $\sigma_c$ analytically, it fails to explain why for uncorrelated random SF networks with $\gamma\leq 3$ and in the thermodynamic limit $N\rightarrow\infty$, the critical value remains finite. This disagreement comes from the fact that in these SF networks, $\lambda_{\max}$ is proportional to the cutoff of the degree distribution, $k_{\max}$ which in turn scales with the system size. Putting the two dependencies together, one obtains $\lambda_{\max}\sim k_{\max}^{\frac{1}{2}}\sim N^{\frac{1}{2(\gamma-1)}}\rightarrow\infty$ as $N\rightarrow\infty$, thus predicting $\sigma_c=0$ in the thermodynamic limit, in contrast to finite size scaling analysis for the critical coupling via numerical solution of the equations of motion.
Note, however, that the difference may be due to the use of distinct order parameters.
Moreover, even in the case of SF networks with $\gamma>3$, $\lambda_{\max}$ still diverges when we take the thermodynamic limit, so that $\sigma_c\rightarrow 0$ as well. As we shall see soon, this is not the case when other approaches are adopted, at least for $\gamma>3$.

It is possible to go beyond with the latter approximation and to determine the behavior of $r$ near the critical point. In \cite{roh05a} a perturbative approach to higher orders of Eq.\ (\ref{erestrepo6}) is developed, which is valid for relatively homogeneous degree distributions ($\gamma>5$).\footnote{The approach holds if the fourth moment of the degree distribution, $\langle  k^4\ra=\int_1^{\infty}P(k)k^4dk$ remains finite when $N\rightarrow\infty$.} They showed that for $(\sigma/\sigma_c)-1\sim 0^+$
\be
r^2=\left(\frac{\eta}{\eta_1K_{c}^{2}}\right)\left(\frac{\sigma}{\sigma_c}-1\right)
\left(\frac{\sigma}{\sigma_c}\right)^3\;,
\ee
where $\eta_1=-\pi g''(0)K_c/16$ and
\be
\eta=\frac{\langle  u\ra^2\lambda_{\max}^2}{N\langle  k \ra^2\langle  u^4 \ra} \;,
\ee
 where $u$ is the normalized eigenvector of the adjacency matrix corresponding to $\lambda_{\max}$ and $\langle  u^4\ra=\sum_j^N u_j^4/N$.

The analytical insights discussed so far can also be reformulated in terms of a mean field approximation \cite{i04,l05,roh05a,i05} for complex networks. This approach (valid for large enough $\langle  k \rangle$  ) considers that every oscillator is influenced by the local field created in its neighborhood, so that  $r_i$ is proportional to the degree of the nodes $k_i$, i.e., $r_i\sim k_i$. Assuming this is the case and introducing the order parameter $r$ through
\be
r=\frac{r_i}{k_i}=\frac{1}{k_i}\left |\sum_{j=1}^{N}a_{ij}\langle  \text{e}^{\text{i}\theta_j}\ra_t\right |\;,
\label{elee1}
\ee
after summing over $i$ and substituting $r_i=rk_i$ in Eq.\ (\ref{erestrepo6}) we obtain \cite{roh05a}
\be
\sum_{j}^{N}k_j=\sigma\sum_{j}^{N}k_j^2\int_{-1}^{1}g(x\sigma r k_j)\sqrt{1-x^2}dx\;.
\label{elee2}
\ee
The above relation, Eq.\ (\ref{elee2}), was independently derived in \cite{i04}, who first studied analytically the problem of synchronization in complex networks, though using a different approach. Taking the continuum limit, Eq. (\ref{elee2}) becomes
\be
\int kP(k)dk=\sigma\int k^2P(k)dk\int_{-1}^{1}g(x\sigma r k)\sqrt{1-x^2}dx\;,
\label{elee3}
\ee
which for $r\rightarrow 0^{+}$ verifies
\bea
\int kP(k)dk&=&\sigma\int k^2P(k)dk\int_{-1}^{1}g(0)\sqrt{1-x^2}dx\nonumber\\
&=&\frac{\sigma g(0)\pi}{2}\int k^2P(k)dk\;,
\label{elee4}
\eea
which leads to the condition for the onset of synchronization ($r>0$) as
\be
\frac{\sigma g(0)\pi}{2}\int k^2P(k)dk>\int kP(k)dk\nonumber\;,
\ee
that is,
\be
\sigma_c=\frac{2}{\pi g(0)}\frac{\langle  k\ra}{\langle  k^2\ra}=K_c\frac{\langle  k\ra}{\langle  k^2\ra}\;.
\label{elee5}
\ee
The mean field result, Eq.\ (\ref{elee5}), gives as a surprising result that the critical coupling $\sigma_c$ in complex networks is nothing else but the one corresponding to the all-to-all topology $K_c$ re-scaled by the ratio between the first two moments of the degree distribution, regardless of the many differences between the patterns of interconnections. Precisely, it states that the critical coupling strongly depends on the topology of the underlying graph.
In particular, $\sigma_c\rightarrow 0$ when the second moment of the distribution $\langle  k^2\rangle$   diverges, which is the case for SF networks with $\gamma\le 3$. Note, that in contrast with the result in Eq.\ (\ref{erestrepo7}), for $\gamma >3$, the coupling strength does not vanish in the thermodynamic limit. On the other hand, if the mean degree is kept fixed and the heterogeneity of the graph is increased by decreasing $\gamma$, the onset of synchronization occurs at smaller values of $\sigma_c$. Interestingly enough, the dependence gathered in Eq.\ (\ref{elee5}) has the same functional form for the critical points of other dynamical processes such as percolation and epidemic spreading processes \cite{n03a,blmch06, dgm07}. While this result is still under numerical scrutiny, it would imply that the critical properties of many dynamical processes on complex networks are essentially determined by the topology of the graph, no matter whether the dynamics is nonlinear or not. The corroboration of this last claim will be of extreme importance in physics, probably changing many preconceived ideas about the nature of dynamical phenomena.

Within the mean field theory, it is also possible to obtain the behavior of the order parameter $r$ near the transition to synchronization. Equation\ (\ref{elee5}) was also independently derived in \cite{l05} starting from the differential equation Eq.\ (\ref{ekurafinal}). Using the weighted order parameter
\be
\bar{r}(t)\text{e}^{\text{i}\bar{\phi}(t)}=\frac{\sum_{i=1}^{N}\sum_{j=1}^{N}a_{ij}\text{e}^{\text{i}\theta_j}}
{\sum_{i=1}^{N}k_i}\;,
\nonumber
\ee
and assuming the same magnitude of the effective field of each pair of coupled oscillators one obtains
\be
\dot{\theta}_i=\omega_i-\frac{\sigma}{\langle  k \ra} k_i \bar{r} \sin(\theta_i)\;,
\ee
where we have set $\bar{\phi}=0$. Now, it is considered again that in the stationary state the system divides into two groups of oscillators, which are either locked or rotating in a nonuniform manner. Following the same procedure employed in all the previous derivations, the only contribution to $r$ comes from the former set of oscillators. After some algebra \cite{l05}, it is shown that the critical coupling $\sigma_c$ is given by Eq.\ (\ref{elee5}) and that near criticality
\be
r\sim(\sigma-\sigma_c)^{\beta}\;,
\label{beta_meanfield}
\ee
for $\gamma>3$, with a critical exponent $\beta=\frac{1}{2}$ if $\gamma\ge 5$, and $\beta=\frac{1}{\gamma-3}$ when $3<\gamma\le 5$. For the most common cases in real networks of $2<\gamma <3$, the critical coupling tends to zero in the thermodynamic limit so that $r$ should be nonzero as soon as $\sigma\neq 0$. In this case, one gets $r\sim \sigma^{1/(3-\gamma)}$. Notably, the latter equation is exactly the same found for the absence of critical behavior in the region $2<\gamma <3$ for a model of epidemic spreading \cite{mpv02}.

One recent theoretical study in \cite{olkk07} is worth mentioning here. They  have extended the mean field approach to the case in which the coupling is asymmetric and depends on the degree. In particular, they studied a system of oscillators arranged in a complex topology whose dynamics is given by
\be
\dot{\theta}_i=\omega_i+\frac{\sigma}{k_i^{1-\eta}} \sum_{j=1}^{N} \sin(\theta_j-\theta_i).
\ee
$\eta=1$ corresponds to the symmetric, non-degree dependent, case. Extending the mean field formalism to the cases $\eta\neq1$, they investigated the nature of the synchronization transition as a function of the magnitude and sign of the parameter $\eta$. By exploring the whole parameter space $(\eta,\sigma)$, they found that for $\eta=0$ and SF networks with $2 <\gamma<3$, a finite critical coupling $\sigma_c$ is recovered in sharp contrast to the non-weighted coupling case for which we already know that $\sigma_c=0$. This result seems phenomenologically meaningful, since setting $\eta=0$ implies that the coupling in Eq.\ (\ref{ekurageneral}) is $\sigma_{ij}=\sigma/k_i$, which, as discussed before \cite{mzk05b}, might have the effect of partially destroying the heterogeneity inherent to the underlying graph by normalizing all the contributions $\sum_{j=1}^{N} a_{ij} \sin(\theta_j-\theta_i)$ by $k_i=\sum_{j=1}^{N} a_{ij}$.

\subsubsection{Path towards synchronization in complex networks}
\label{sect_km_path}

Up to now, we have discussed both numerically and theoretically the onset of synchronization. In the next section, we shall also discuss how the structural properties of the networks influence the stability of the fully synchronized state. But, what happens in the region where we are neither close to the onset of synchronization nor at complete synchronization? How is the latter state attained when different topologies are considered?

As we have seen, the influence of the topology is not only given by the degree distribution, but also by how the oscillators interact locally.  To reduce the number of degrees of freedom to a minimum, let us focus on the influence of
heterogeneity and study the evolution of synchronization for a family of complex networks which  are comparable in their clustering, average
distance and correlations so that the only difference is due to the
degree distribution.\footnote{This isolation of individual features of complex networks is essential to understand the interplay between topology and dynamics. As we will discuss along the review, many times this aspect has not been properly controlled raising results that are confusing, contradictory or even incorrect.}
For these networks, the previous theoretical approaches argued that the critical coupling $\sigma_c$ is proportional to $\langle  k \ra/\langle  k^2\rangle$  , so that different topologies should give rise to distinct critical points. In particular, in \cite{gma07a,gma07b} it was studied numerically the path towards synchronization in ER and SF networks. They also studied several networks whose degree of heterogeneity can be tuned between the two limiting cases \cite{gm06}. These authors put forward the question: How do SF networks compare with ER
ones and what are the roots of the different behaviors observed?

\begin{table}
\caption{\label{table1} Topological properties of the networks and critical coupling strengths derived from a
   finite size scaling analyses, Eq.\ (\ref{ekfss}). Different values of $\chi$ corresponds to grown networks whose degree of heterogeneity varies smoothly
between the two limiting cases of ER and SF graphs. From \cite{gma07b}.}
\begin{tabular}{llll}
$\chi$ & $ \langle  k^2 \rangle$   & $k_{\max}$ &
  $\sigma_c $\\
\hline
0.0$\;$ (SF) &	115.5 &	326.3 &	0.051\\
0.2 &	56.7 &	111.6 &	0.066\\	
0.4 &	44.9 &	47.7 &	0.088\\
0.6 &	41.1 &	25.6 &	0.103\\
0.8 &	39.6 &	16.8 &	0.108\\
1.0$\;$ (ER) &	39.0 &	14.8 &	0.122
\end{tabular}
\end{table}

Numerical simulations  \cite{gma07a,gma07b} confirm qualitatively the theoretical predictions for the onset of synchronization, as summarized in Table\ \ref{table1}. In fact, the onset of synchronization first occurs for SF networks. As the network substrate
becomes more homogeneous, the critical point $\sigma_c$ shifts to
larger values and the system seems to be less synchronizable. On the
other hand,  they also showed that the route to complete
synchronization, $r=1$, is sharper for homogeneous networks. No critical exponents for the behavior of $r$ near the transition points have been reported yet for the ER network, so that comparison with the mean field value $\beta=1/2$ for a SF network with $\gamma=3$ is not possible.\footnote{The numerical value of $\beta$ contradicts the prediction of the mean-field approach (see the discussion after Eq. (\ref{beta_meanfield}))
The reason of such discrepancy is not clear yet.}
Numerically, a detailed finite size scaling analysis in SF and ER
topologies shows that the critical coupling strength corresponds in SF networks to
$\sigma_c^{\mbox{\scriptsize{SF}}}= 0.051$, and in random ER networks to $\sigma_c^{\mbox{\scriptsize{ER}}} = 0.122$, a fairly significant numerical difference.

\begin{figure}[!t]
\begin{center}
\epsfig{file=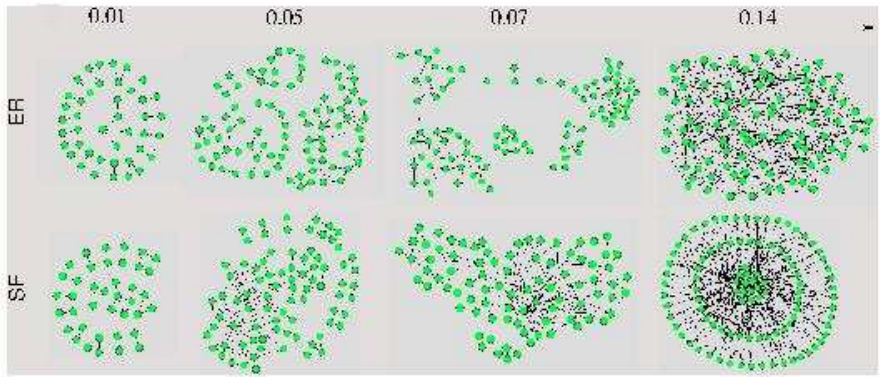,width=0.8\columnwidth,angle=0,clip=1}
\end{center}
  \caption{(color online) Synchronized components for several values of
  $\sigma$ for the two limiting cases of ER and SF networks. The figure clearly illustrates the differences in forming synchronization patterns for both types of networks: in the SF case links and nodes are incorporated
  together to the largest of the synchronized clusters, while for the ER network, what is added are
  links between nodes already belonging to such cluster. From \cite{gma07a}.}
\label{nets}
\end{figure}

The mechanisms behind the differences in the emergence of collective behavior for ER and SF topologies can be explored numerically by defining a local order parameter that captures and quantifies the way in which clusters of locked oscillators emerge. The main difference with respect to $r$  is that one measures the degree of synchronization of nodes ($r$) with
respect to the average phase $\phi$ and the other ($r_{\mbox{\scriptsize{link}}}$) to the degree of
synchronization between every pair of connected nodes. Thus, $r_{\mbox{\scriptsize{link}}}$ gives the fraction of all possible links that are
synchronized in the network as
\be
r_{\mbox{\scriptsize{link}}}=\frac{1}{2N_l}\sum_{i,j=1}^{N}a_{ij}\left |\lim_{\Delta
  t\rightarrow\infty}\frac{1}{\Delta t}\int_{t_r}^{t_r+\Delta
  t}e^{i\left[\theta_i(t) -\theta_j(t)\right]}dt\right |\;,
 \label{r_link}
\ee
being  $t_r$ the time the system needs to settle into the stationary state, and $\Delta t$ a large averaging time. In \cite{gma07a,gma07b} the degree of synchronization
of  pairs of connected oscillators was measured in terms of the symmetric matrix
\be
{\cal D}_{ij} =a_{ij}\left |\lim_{\Delta t\rightarrow\infty}\frac{1}{\Delta
  t}\int_{t_r}^{t_r+\Delta t}e^{i\left[\theta_i(t)-\theta_j(t)\right]}dt\right
  |\;,
\label{ematrix}
\ee
which, once filtered using a threshold $T$
such that the fraction of synchronized pairs equals $r_{\mbox{\scriptsize{link}}}$, allows us to identify the
synchronized links and reconstruct the clusters of synchrony for any value of $\sigma$, as illustrated in Fig.\ref{nets}. From a microscopic analysis, it turns out that for homogeneous topologies, many small
clusters of synchronized pairs of oscillators are spread over the graph and merge together to form a giant synchronized cluster when the
effective coupling is increased. On the contrary, in heterogeneous graphs, a central core containing the hubs first comes up driving the evolution of synchronization patterns by absorbing small clusters.
Moreover, the evolution of $r_{\mbox{\scriptsize{link}}}$ as $\sigma$ grows explains why the transition is sharper for ER networks: nodes are added first to the giant synchronized cluster and later on the links among these nodes
that were missing in the original clusters of synchrony. In SF graphs, oscillators are added to
the largest synchronized component together with most of their links, resulting in a much slower growth of $r_{\mbox{\scriptsize{link}}}$. Finally,  it is also computed the probability that a node with degree $k$ belongs to
the largest synchronized cluster and reported that this probability is an increasing function of $k$ for
every $\sigma$, namely, the more connected a node is, the more likely
it takes part in the cluster of synchronized links \cite{gma07a,gma07b}. It is interesting to mention here that a similar dependence is obtained if one analyzes the stability of the synchronized state under perturbations of nodes of degree $k$.  In \cite{mp04} it was found that the average time $\langle  \tau \rangle$   a node needs to get back into the fully synchronized state is inversely proportional to its degree, i.e., $\langle  \tau \ra\sim k^{-1}$.

Very recently \cite{ad07}, the path towards synchronization was also studied looking for the relation between the time needed for complete synchronization and the spectral properties of the Laplacian matrix of the graph,
\be
L_{ij}=k_i\delta_{ij}-a_{ij}.
\label{eq_lapl}
\ee
The Laplacian matrix is symmetric with zero row-sum and hence all the eigenvalues are real and non-negative. Considering the case of identical Kuramoto oscillators, whose dynamics has only one attractor, the fully synchronized state, they found that the synchronization time scales with the inverse of the smallest nonzero eigenvalue of the Laplacian matrix. Surprisingly, this relation qualitatively holds for very different networks where synchronization is achieved, indicating that this eigenvalue alone might be a relevant topological property for synchronization phenomena.
 The authors in \cite{dnm06} remark the role of this eigenvalue not only for synchronization purposes but also for the flow of random walkers on the network.

\subsubsection{Kuramoto model on structured or modular networks}
\label{kmsmn}

In this section, we discuss a context in which synchronization has turned out to be a relevant phenomenon to explore the relation between dynamical and topological properties of complex networks. Many complex networks in nature are modular, i.e. composed of certain
subgraphs with differentiated internal and external connectivity that
form communities \cite{ng04,ddda05,blmch06}. This is a limiting situation in which the local structure
may greatly affect the dynamics, irrespective of whether or not we deal
with homogeneous or heterogeneous networks.

Synchronization processes on modular networks have
been recently studied as a mechanism for community detection \cite{orhk05,adp06a,adp06b,lbl06}. The situation in which a set of identical Kuramoto oscillators (i.e., $\omega_i=\omega,\; \forall i$) with random
initial conditions evolves after a transient to the
synchronized state was addressed In \cite{adp06a,adp06b}.\footnote{It is worth stressing here that for this purpose the assumption of $\omega_i=\omega$ can be adopted without loss of generality as it makes the analysis
easier. Synchronization of non-identical oscillators also reveals the existence of community structures. See \cite{gma07b}.}
Note that in this case full synchronization is always achieved as this state is the only attractor of the dynamics so that the coupling strength sets the time scale to attain full synchronization: the smaller $\sigma$ is, the longer the time scale. The authors in \cite{adp06a} guessed that if high densely  interconnected motifs synchronize more easily than those with sparse connections \cite{mvp04}, then the synchronization of complex networks with community structure should behave differently at different time and spatial scales. In synthetic modular networks, starting from random initial phases, the highly connected units forming local clusters synchronize first and later on, in a sequential process, larger and larger topological structures do the same up to the point in which complete synchronization is achieved and the whole population of oscillators beat at the same pace. This process occurs at different time scales and the dynamical route towards the global attractor reveals the topological structures that represent communities, from the microscale at very early states up to the macroscale at the end of the time evolution.

\begin{figure}[t]
\begin{center}
\epsfig{file=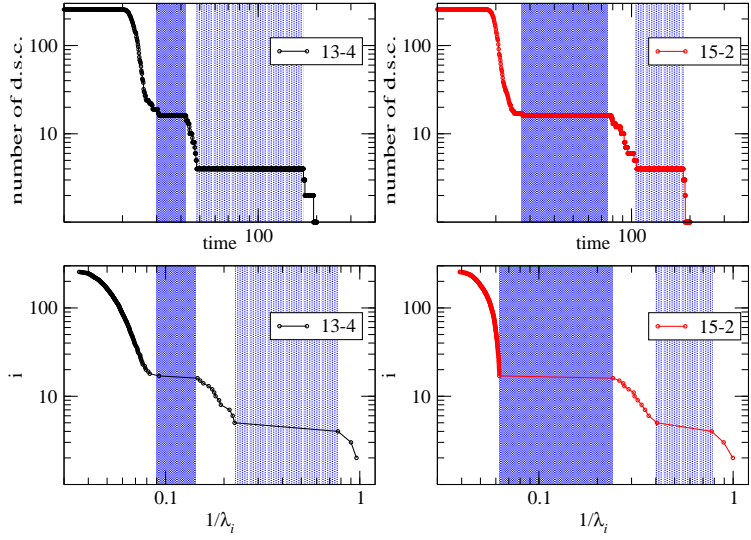,width=3.6in,angle=-90,clip=1}
\end{center}
\caption{(color online) Top: Number of disconnected synchronized components (d.s.c.) as a function of time.
Bottom: Rank index versus the corresponding eigenvalue of the Laplacian matrix. Each column corresponds to a network with two hierarchical levels of communities.
The difference lies in the relative weight of the two modular levels. From \cite{adp06a}.}
\label{figrank}
\end{figure}

The authors studied the time evolution of pairs of oscillators defining the local order parameter
\be
\rho_{ij}(t)=\langle  \cos[\theta_i(t)-\theta_j(t)]\rangle \;,
\ee
averaged over different initial conditions, which measures the correlation between pairs of oscillators. To identify the emergence of compact clusters reflecting communities, a binary dynamic connectivity matrix is introduced such that
\be
{\cal D}_t(T)_{ij}=\left\{
\begin{array}{l}
1\quad \text{if}\quad \rho_{ij}(t)>T\\
0\quad \text{if}\quad \rho_{ij}(t)<T,
\end{array}
\right.
\ee
for a given threshold $T$.
Changing the threshold $T$ at fixed times reveals the correlations between the dynamics and the underlying structure, namely, for large enough $T$, one is left with a set of disconnected clusters or communities that are the innermost ones, while for smaller values of $T$ inter-community connections show up. In other words, the inner community levels are the
first to become synchronized, subsequently the second level groups, and finally the whole system shows global
synchronization.
Note that
since the function $\rho_{ij}(t)$ is continuous and monotonic, we can define a new matrix ${\cal D}_T(t)$,
that takes into account the time evolution for a fixed threshold.
The evolution of this matrix unravels the topological structure of the underlying network at different time scales.
In the top panels of Fig. \ref{figrank} we plot the number of connected components corresponding to the binary
connectivity matrix with a fixed threshold as a function of time for networks with two hierarchical levels
of communities. There we can notice how this procedure shows the existence of two clear time scales corresponding
to the two topological scales.

It is also possible to go one step further and show that the evolution of the system to the global attractor is intimately linked to the whole spectrum of the Laplacian matrix (\ref{eq_lapl}).  The bottom panels of Fig. \ref{figrank}
show the ranked index of the eigenvalues of $L_{ij}$ versus their inverse.
As can be seen, both representations (top and bottom) are qualitatively equivalent, revealing the topological structure of the networks.
The only difference is that one comes from a dynamical matrix and the other from the spectrum of the Laplacian, that fully characterizes the topology. Thus, synchronization can be used to unveil topological scales when the architecture of the network is unknown.

The relationship between the eigenvalue spectrum of $L_{ij}$ and the dynamical structures of Fig.\ \ref{figrank} can be understood from the linearized dynamics of the Kuramoto model, which reads \cite{adp06a,adp06b}
\be
\dot{\theta}_i=-{\sigma}\sum_{j=1}^{N}L_{ij}\theta_j\qquad i=1,..., N\;,
\label{alex1}
\ee
that is a good approximation after a fast transient starting from random initial phases in the range $[0,2\pi]$.

The solution of Eq. (\ref{alex1}), in terms of the normal modes $\psi_i(t)$, is
\be
\theta_i(t)=\sum_{j=1}^{N}B_{ij}\psi_j(t)=\sum_{j=1}^{N}B_{ij}\psi_j(0)\text{e}^{-\sigma\lambda_jt}\;,
\label{alex2}
\ee
where $B_{ij}$ is the normalized eigenvector matrix and $\lambda_i$ the eigenvalues of the Laplacian matrix.
Going back to the original coordinates, the phase difference between any pair of oscillators is
\begin{equation}
\left|\theta_i(t)-\theta_m(t)\right| \le  \sum_{j,k=1}^N \left|B_{ij}-B_{mj}\right| e^{-\sigma\lambda_j t}\left|B^{\dagger}_{jk}\right| \left|\varphi_k(0)\right|.
\end{equation}

Assuming a global bound for the initial conditions $\left|\theta_k(0)\right|\le \Theta$, $\forall k$ and taking into account the normalization of the eigenvector matrix, $\left|B_{ij}\right|\le 1$, we can sum over the index $k$ to get
\begin{equation}
\left|\theta_i(t)-\theta_m(t)\right| \le N\Theta \sum_{j=2}^N \left| B_{ij}-B_{mj}\right| e^{-\sigma\lambda_j t}.
\end{equation}
The sum starts at $j=2$ because all the components of the first eigenvector (the one corresponding to the zero eigenvalue) are identical, which is the warranty of the final synchronization of the system. Here we can see the clear relation between topology (represented by the eigenvectors and eigenvalues of the Laplacian matrix) and dynamics.
For long times all exponentials go to zero and the oscillators get synchronized. At short times the main contribution
comes from small eigenvalues; then those nodes with similar projections on the eigenvectors of small indices will
get synchronized even at short times. In networks which are hierarchically organized this happens at all
topological and dynamical scales.

An additional significance of the importance of this relationship between spectral and dynamical properties of
identical oscillators comes from the work in \cite{gd08}. The authors propose a method of network reduction
based on the similarity of eigenvector projections. From the original network, nodes are merged according to
the similarity of their components in the different eigenvectors, producing a reduced network; this merging procedure basically preserves the original
eigenvalues of the Laplacian matrix in the new coarse grained Laplacian matrix.
The authors determine the best clustering of the nodes and show that the evolution of identical Kuramoto oscillators in the original network
and according to the original Laplacian is equivalent to the evolution of the reduced network in terms of the reduced Laplacian matrix.

The above results refer to situations in which networks have clearly defined community structure. The approach we have shown
enables one to deal with different time and topological scales. In the current literature about community detection
\cite{ng04,ddda05}, the main goal is to maximize the modularity, see Eq. (\ref{Q}).
In this case the different algorithms
try to find the best partition of a network. Using a dynamical procedure, however, we are able to
devise all partitions at different scales. In \cite{ad06} it is found that the partition with the largest
modularity turns out to be the one for which the system is more stable, if the networks are homogeneous in
degree.\footnote{Here stability (relative) of a given structure is understood to be the ratio between the final and initial times a partition remains synchronized. In terms of the number of connected components in Fig. \ref{figrank}
it corresponds to the length of the plateaus.}
If the networks have hubs, these more connected nodes need more time to synchronize with their neighbors
and tend to form communities by themselves. This is in contradiction with the optimization of the modularity
that punishes single node communities. From this result we can conclude that the modularity is a good measure
for community partitioning. But when dealing with dynamical evolution in complex networks other related
functions different from modularity are needed.

For real networks, it has been shown that the same phenomenology applies \cite{orhk05}. These authors studied a system of Kuramoto oscillators, Eq.\ (\ref{ekurageneral}) with $\sigma_{ij}=\sigma/k_i$, arranged on the nodes of two real networks with community structures, the yeast protein interaction network and the Autonomous System  representation of the Internet map. Both networks have a modular structure, but differ in the way communities are assembled together. In the former one, the modules are connected diversely (as for the synthetic networks analyzed before), while in the latter one different communities are interwoven mainly through a single module. The authors found that the transition to synchrony depends on the type of intermodular connections such that communities can mutually or independently synchronize.

Modular networks are found in nature and they are commonly the result of a growth process.
Nevertheless, these structural properties can also emerge as an adaptive mechanism
generated by dynamical processes taking place in the existing network,
and synchronization could be one of them.
In particular, in \cite{gz06} the authors studied the evolution of a network of Kuramoto oscillators.
For a coupling strength below its critical value,
the network is rewired by replacing links between neighbors with a large frequency difference
with links between units with a small frequency difference.
In this case, the network dynamically evolves to configurations that increase the order parameter.
Along this evolution they noticed the appearance of synchronized groups (communities) that make
the structure of the network to be more complex than the random starting one.

Very recently \cite{bilpr07},
a slightly different model was considered, where the dynamics of each node is governed by
\be
\dot{x}_i=\omega_i+\frac{\sigma}{\sum_{j\in \Gamma_i} b_{ij}^{\alpha (t)}}\sum_{j\in \Gamma_i} b_{ij}^{\alpha (t)}
\sin (x_j-x_i) \beta e^{\beta |x_j-x_i|}
\nonumber
\ee
being $b_{ij}$ the betweenness centrality of the link, $\Gamma_i$ the set of nodes that are connected to $i$, and $\alpha$ a time-dependent exponent. The authors use this dynamical evolution to identify communities. The betweenness is used as a measure of community coparticipation, since links between nodes that are in the same community have low betweenness \cite{gn02}. Starting from a synchronized state, $\alpha$ is decreased from zero and then the corresponding interaction strength on those links is increasingly enhanced. An additional mechanism that adjusts frequencies between neighboring nodes causes the final state
to show partial synchronization among nodes that are in the same community.

\subsubsection{Synchronization by pacemakers}

It is worth mentioning the existence of other different approaches to the
synchronization of populations of Kuramoto oscillators.
So far we have referred to populations where the oscillators are nearly identical in the sense that
they can have slightly different frequencies.
Whenever there is a subset of units that play a special role,
in the sense that they have substantially different frequencies than the rest in the population or
they affect some units but are not affected by any of them, one usually refers to them as \emph{pacemakers}.
The effect of pacemakers has been studied in regular networks, as for instance in one-dimensional rings,
two-dimensional tori and Cayley trees \cite{rm06}.
So far, the only approach in a complex topology has been performed in \cite{km04}.
There, the authors considered a system of identical units (same frequency)
and a singular pacemaker.
For an ER network they found that for a large coupling the pacemaker entrains the whole system
(all units with the same effective frequency, that of the pacemaker),
but the
phase distribution is hierarchically organized. Units at the same downward distance from
the pacemaker form shells of common phases. As the coupling strength is decreased the entrainment breaks down
at a value that depends exponentially on the depth of the network.
This result also holds for complex networks, as for instance in WS or SF networks, although the
analytical explanation is only valid for ER networks.

\subsection{Pulse-coupled models}

In parallel to the studies  described so far, some other approaches to synchronization in networks have invoked  models where the interaction between units takes the form of a pulse.
In particular, much attention has been devoted to models  akin to reproduce the dynamics of neurons, e.g. integrate-and-fire oscillators (IFOs). The basics of an IFO system is as follows. The phase dynamics of any oscillator $i$ is linear in time $d \phi_i(t)/dt =1$ in absence of external perturbations. However, when the oscillator $i$ reaches the threshold $\phi_i(t)=1$ it sends a signal (or pulse) to the rest of the oscillators to which it is connected, and relaxes to $\phi_i(t)=0$. The pulse can be considered to propagate instantaneously or with a certain time delay $\tau$, and when it reaches other oscillators induces a phase jump $\phi_j \rightarrow \phi_j + \Delta(\phi_j) $. The effects of the topology on the synchronization phenomena emerging in a network of IFOs
are at least as rich as those presented in Sect. \ref{sect_KM}, although far more difficult to be revealed
analytically. The main problem here is that the dynamics presents discontinuities in the variable states that are difficult to deal with. Nevertheless, many insights are recovered from direct simulations and clever mappings of the system. From direct simulations the first insights pointed directly to certain scaling relations between the synchronization time and topological parameters of networks. In ER networks, the scaling relation between the time to needed to achieve complete synchronization $T$, the number of nodes $N$, and the number of links $M$, was found to be
\begin{equation}
\frac{T}{N^{2\alpha-\beta}}\sim\left ( \frac{M}{N^2} \right )^{\alpha},
\end{equation}
with $\alpha=1.30(5)$ and $\beta=1.50(5)$. Comparing this synchronization process with the same system on a regular square lattice, one realizes that the time needed to synchronize a random network is larger, specially in sparse networks \cite{gdlp00}. In between of these two extremal topologies, some WS networks with a rewiring probability $p$ were studied and were found to expand the synchronization time more than the original regular lattice. However, it was first pointed out that an appropriate normalization of the pulses received by each neuron, rescales the time to very short values. This phenomenon of normalization of the total input signal received by each oscillator has been repeatedly used to homogenize the dynamics in heterogeneous substrates.

IFOs in SW networks were revisited later in \cite{rrs04} to study the possibility of self-sustained activity induced by the topology itself. Considering a unidirectional ring of IFOs with density $p$ of random long-range directed connections, the authors showed that periodical patterns persist at low values of $p$, while long-transients of disordered activity patterns are observed for high values of $p$. Responsible for this behavior is a tradeoff between the average path length and the speed of activity propagation. For low $p$ configurations, the distances in the networks decrease logarithmically with size, while the superposition of activities is almost the same than in the regular configuration, i.e.  the same activity occurs but in a "smaller" network able to self-sustain its excitation. However for large $p$,
the superposition of activity between excited domains plays also an important role, and then both effects make
the synchronized self-sustained activity collapse, leading to disordered patterns.

\begin{figure}[!t]
\begin{center}
\epsfig{file=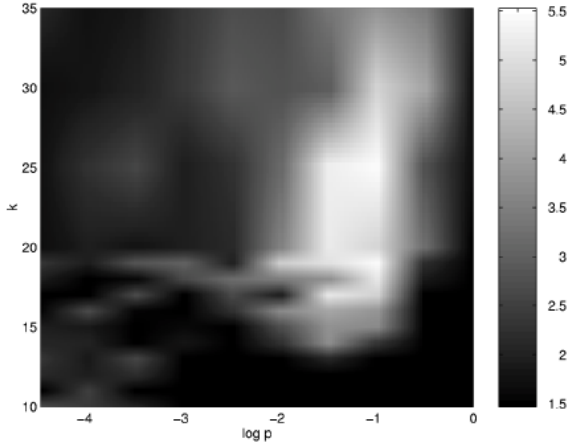,width=3.3in,angle=0,clip=1}
\end{center}
\caption{Phase diagram which shows the regions of oscillatory
(clear) and nonoscillatory (dark) activity of the
network in the $(k,p)$ plane. The island that appears on the right
side indicates that the SW (for some range of values of $k$) is the
only regime capable to produce fast coherent oscillations in the
average activity after the presentation of the stimulus. From \cite{lhcs00}.}
\label{EPC_1}
\end{figure}

In  \cite{lhcs00}, networks of nonidentical Hodgkin-Huxley \cite{hh52} elements coupled by excitatory synapses in random, regular, and SW topologies, were investigated for the first time. The parameters of the model neurons were kept to stay below the bifurcation point, until the input arrives and forces the system to undergo a saddle-node bifurcation on a limit cycle. The dynamics of the system ends up into a coherent oscillation or in the activation of asynchronous states. In absence of a detailed analysis of the mechanism that generates coherence, the simulations showed several effects of the topology on the dynamics, the most interesting of which is that  achieving synchronization in regular networks takes longer compared to SW, where the existence of short-cuts favors faster synchronization. The results obtained in all cases show that the randomness of the topology has strong effects on the dynamics of these models, in particular the average connectivity is a control parameter for the transition between asynchronous and synchronous states. In Fig.~\ref{EPC_1} we present a phase diagram for the Hodgkin-Huxley model in SW networks with varying average degree $\langle  k \rangle$   and rewiring probability $p$.
A detailed analysis of sparse random networks of general IFOs was exposed in \cite{gh00}. Their analytic results are in agreement with the previous observations. Very recently \cite{lsas06},  a SW network
of non-identical Hodgkin-Huxley units in which some of the couplings could be negative was analyzed;
they surprisingly found that a small fraction of such phase-repulsive links can enhance synchronization.

In a slightly different scenario \cite{vld06} a system of pulse-coupled Bonhoeffer-van der Pol-FitzHugh-Nagumo oscillators \cite{c00} in WS networks was studied numerically. This study reports a major influence of the average path length of the network on the degree of synchronization, whereas local properties characterized by clustering and loop coefficients seem to play a minor role.
In any case, the authors warn that the results are far from being conclusive, since single characteristics
of the network are not easily isolated. We will come back to this issue in the next section, when dealing with the
stability of the synchronized state.

The works reviewed so far in this subsection are based on the assumption that the coupling is fixed, and that the only source of topological complexity is embedded in the connectivity matrix.
The authors in \cite{dtdwg04} showed that for networks of pulse-coupled oscillators with complex connectivity,
coupling heterogeneity induces  periodic firing patterns, which replace the state of global synchrony. The  coupling heterogeneity has a critical value from which the periodic firing patterns become asynchronous aperiodic states. These results are in agreement with the observations described in previous works and allow us to state that a certain degree of complexity in the interaction between pulse coupled oscillators is needed to observe regular (or ordered) patterns. However, once a critical level of complexity is surpassed, asynchronous aperiodic states dominate the dynamic phenomena.

\subsection{Coupled maps}

Maps represent simple realizations of dynamical systems exhibiting chaotic behavior. At a first sight they can represent
discrete versions of continuous oscillators. Coupled populations of such rather simple dynamical systems have been
one of the paradigmatic models to explain the emergence and self-organization in complex systems due to the rich
variety of global qualitative behavior they give rise to.
From a more practical point of view coupled map systems have found a widespread range of applications, ranging from
fluid dynamics and turbulence to stock markets or ecological systems \cite{prk01}.
Since these systems are nowadays known to have complex topologies,
populations of maps coupled
through a complex pattern of interactions are natural candidates to study the onset of synchronization as an overall
characteristic of the population.

Coupled maps have been widely analyzed in regular lattices, trees and also in global connectivity schemes.
The first attempt to consider connectivities in between these extreme cases is \cite{g96}.
He proposed a system formed by units, whose individual dynamics are given by the logistic map, that are connected to a fixed number $k$
of other units randomly chosen (multiple and self-links are permitted). The evolution rule for the units is
\begin{equation}
    x_i(t+1)=\frac{1}{k}\sum_ja_{ij}f\left(x_j(t)\right).
    \label{gade}
\end{equation}
A linear stability analysis of this system is performed in
terms of the eigenvalues of the matrix $A$. For the logistic map it is shown that for $k>4$ the maps synchronize. The time the system needs to synchronize decreases with the connectivity $k$ and also with the system size,
although in the latter case the time saturates for large values of the system size.
When the connectivity pattern is changed to a modified WS model (by adding long-range short-cuts but not rewiring),
the authors in \cite{gh00b} showed that just a nonzero value of the addition probability is enough
to guarantee synchronization in the thermodynamic limit.

In another early attempt to include non-regular topologies in chaotic dynamics \cite{bpvl03},
a SW network was analyzed, in which
\begin{equation}\label{bpvl}
    \theta_{i}(t+1)= (1-\sigma) f\left(\theta_{i}(t)\right) +\frac{\sigma}{4+\kappa}\sum_{j=1}^N a_{ij}f\left(\theta_{j}(t)\right),
\end{equation}
where $\kappa$ is the number of shortcuts in the network, and $\sigma$ the coupling constant. Each unit evolves according to a sine-circle map \cite{o93}
\begin{equation}\label{sine-circle}
    f(\theta)=\theta +\Omega - \frac{K}{2\pi} \sin (2\pi \theta) \pmod 1,
\end{equation}
which provides a simple example for describing the dynamics of a phase oscillator perturbed by a time-periodic force.
Here $K$ is a constant related to the external force amplitude and $0\le\Omega<1$ is the ratio between the natural oscillator frequency and the forcing frequency. It is observed that synchronization, in terms of a parameter related to the winding number dispersion, is induced by long-range coupling in a system that, in the absence of the shortcuts,
does not synchronize.

A slightly different approach was conducted in \cite{jj02}, who considered a population of units
evolving according to
\begin{equation}\label{jj}
    x_i(t+1)= (1-\sigma ) f\left(x_i(t)\right) + \frac{\sigma}{{k}_i}\sum_{j\in {\Gamma}_i} f\left(x_j(t)\right).
\end{equation}
They obtain the stability condition of the synchronized state
in terms of the eigenvalues of the normalized Laplacian matrix
($\delta_{ij}-a_{ij}/k_i$)
and the Lyapunov exponent of the map $f(x)$.
Furthermore, they also find a sufficient condition for the system to synchronize independently of the initial conditions, namely
\begin{equation}\label{jj2}
    (1-\sigma \lambda_2) \sup |f'|<1,
\end{equation}
where $\lambda_2$ is the smallest non-zero eigenvalue of the normalized Laplacian matrix.
They demonstrate their results for regular connectivity patterns as global coupling and one-dimensional rings
with a varying number of nearest neighbors, since in these cases the eigenvalues can be computed analytically.
Complexity in the connectivity pattern is introduced in different ways. In these cases one needs numerical estimates
of the eigenvalues to compare the synchronization condition (\ref{jj2}) with the simulation of the model (\ref{jj}).
By using a quadratic map $f(x)=1-ax^2$ \cite{o93} and choosing the free parameter $a$  in a range where different regimes are realized, they find that in a random network
the system synchronizes for an arbitrary large number of units, whenever the number of neighbors is larger than some threshold determined by the maximal Lyapunov exponent. This implies a remarkable difference to the
one-dimensional case where synchronization is not possible when the number of units is large enough.
For a WS model their main finding is that a quite high value of the rewiring probability
($p>0.8$) is needed to achieve complete synchronization. Finally for BA networks the behavior is comparable to the random ER case.

Following a similar line, in \cite{lgh04} it is studied the behavior of a model where the interaction
between the units can be strengthened according to the degree. In this case
\begin{equation}\label{lgh}
    x_{t+1,i}= (1-\sigma ) f(x_{t,i}) + \frac{\sigma}{{\cal N}_i}\sum_{j\in {\Gamma}_i}k_j^\alpha f(x_{t,j}),
\end{equation}
where ${{\cal N}_i}=\sum_{j\in {\Gamma}_i} k_j^\alpha$ is the appropriate normalizing constant.
Here
the function $f(x)$ is also a quadratic map.
The authors study first BA networks. When $\alpha=0$ the model
is equivalent to that discussed previously; in this case they find the existence of a first-order transition
between the coherent and the noncoherent phases that depends on both the mean connectivity and the coupling
$\sigma$. As varying $a$, the parameter of the quadratic map, they find that these two critical values are related by the power law $\sigma_c \propto k_c^{-\mu}$. The effect of $\alpha$ being larger than zero is only quantitative,
since in this case the transition appears at smaller values of the interaction as compared with the usual case.
Additionally, the authors studied two types of deterministic
SF networks: a pseudofractal SF network introduced in \cite{dgm02} and the Apollonian network
introduced in \cite{aha04}. In both cases, there is no coherence when $\alpha=0$ and $a=2$. This fact leads the authors to conclude that some degree of randomness that shortens the mean distance between units is needed for achieving a synchronized state, since in these networks the SF nature is not related to a short mean distance.
Nevertheless, this situation is avoided if the contributions from the hubs are strengthened, by making $\alpha>0$.

Another set of papers deals with units that are coupled with some transmission delay \cite{ajw04,mm05b,mpm06}.
For instance, in \cite{ajw04} the authors
propose a model in which all units have the same time delay (in discrete units) with respect to the unit considered:
\begin{equation}
    x_i(t+1)=f\left(x_i(t)\right)+\frac{\sigma}{k_i}\sum_{j\in\Gamma_i} \left[ f(x_j\left(t-\tau_{ij})\right) - f\left(x_i(t)\right)\right].
\end{equation}
For a uniform delay $\tau_{ij}=\tau \; \forall i,j$, they show analytically,  and numerically, that the delay
facilitates synchronization
for general topologies. In any case, this fact confirms the results obtained in \cite{jj02} that ER and SF networks are easier to synchronize than regular or SW ones. Furthermore, one of the implications of connection delays is the possibility of the emergence of new collective phenomena.
In \cite{mm05b,mpm06} the authors considered uniform distributions of (discrete) time delays. Their main result is that in the presence of random delays the synchronization depends mainly on the average number of links per node, whereas for fixed delays there is also a dependence on other topological characteristics.

In a more general framework \cite{ja03,aj03b,jah05}, the following problem is considered
\begin{equation}
    x_i(t+1)=(1-\sigma) f\left(x_i(t)\right)+\sigma \frac{1}{k_i}\sum_{j\in \Gamma_i} g\left(x_j(t)\right).
\end{equation}

For a logistic map $g(x)=\mu x (1-x)$ (although analyses on other maps as the circle map and the tent map
have also been performed) the authors show a phase diagram in which the different stationary configurations
are obtained as a function of the coupling strength $\sigma$ and the parameter of the map $\mu$.
The stationary configurations are classified in the following way: turbulence (all units behave chaotically),
partially ordered states (few synchronized clusters with some isolated nodes),
ordered states (two or more synchronized clusters with no isolated nodes),
coherent states (nodes form a single synchronized cluster),
and variable states (nodes form different states depending strongly on initial conditions).
The critical value of the coupling above where phase synchronized clusters are observed depends on the
type of network and the coupling function.
As a remarkable point, it is found that two different mechanisms of cluster formation (partial synchronization)
can be distinguished: self-organized and driven clusters. In the first case, the nodes of a cluster
get synchronized because of intracluster coupling. In the latter case, however, synchronization is due to intercluster coupling; now the nodes of one cluster are driven by those of the others.
For a linear coupling function $g(x)=x$, self-organization of clusters dominates at weak coupling;
when increasing
the coupling strength, a transition to driven-type clusters, almost independent of the type of network, appears.
However, for a nonlinear coupling function the driven type dominates for weak coupling, and only
networks with a tree-like structure show some cluster formation for strong coupling.

\begin{figure}[!t]
\begin{center}
\epsfig{file=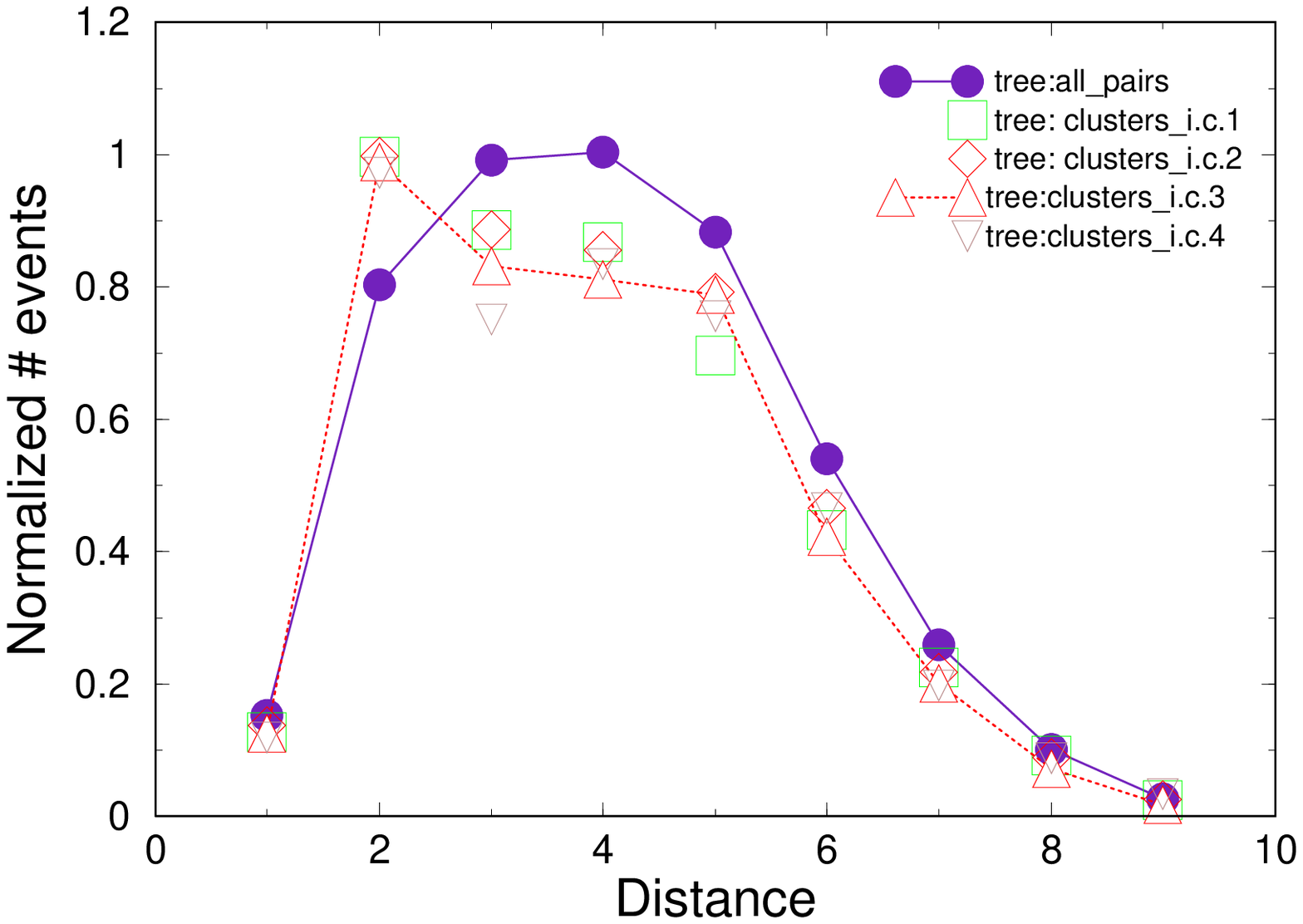,width=0.4\textwidth,angle=0,clip=1}
\epsfig{file=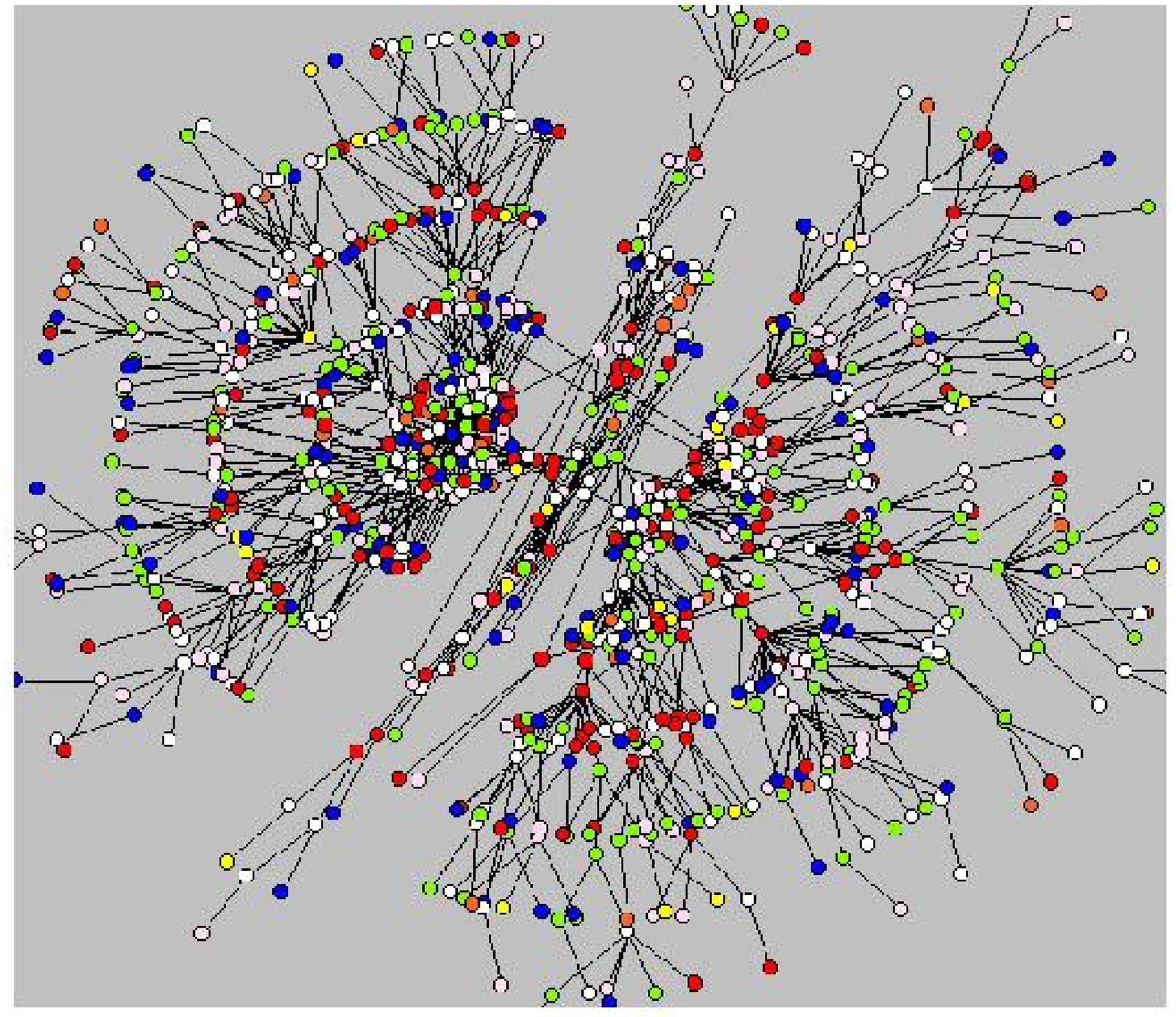,width=0.4\textwidth,angle=0,clip=1}
\end{center}
\caption{(color online) Left: Distribution of the topological distances for a tree with 1000 nodes (bullets)
and the distance inside the synchronized clusters (other symbols) for $\sigma = 0.017$ and
different initial conditions. Right: visualization of the tree with five interconnected clusters of synchronized nodes
marked by different colors. From \cite{lt07}.}
\label{fig_tadic}
\end{figure}

Finally, it is worth mentioning the very Ref.  \cite{lt07}, in which the authors consider a SF tree (preferential-attachment growing network with one link per node) of two-dimensional standard maps:
\begin{equation}
\begin{array}{lll}\label{tlt}
  x'&=& x + y + \mu \sin (2\pi x) \;\; \pmod 1 \\
  y' &=&  y + \mu \sin (2\pi x).
\end{array}
\end{equation}
The nodes are coupled through  the angle coordinate ($x$)
so that the complete time-step of the node $i$ is
\begin{equation}
\begin{array}{lll}
  x_{i}(t+1)&=& (1-\varepsilon ) x'_{i}(t) + \frac{\varepsilon}{k_i} \sum_{j\in\Gamma_i} \left( x_{j}(t)-x'_{i}(t)\right)\\
  y_{i}(t+1) &=& (1-\varepsilon ) y'_{i}(t).
\end{array}
\end{equation}
Here, ($'$) denotes the next iteration of the (uncoupled) standard map (\ref{tlt}) and
$t$ denotes the global discrete time.
The update of each node is
the sum of a contribution given by the updates of the nodes,
the $'$ part,
plus a coupling contribution given by the sum of differences, taking into account a delay in the coupling
from the neighbors.
By keeping
$\mu=0.9$
such that the individual dynamics is in the strongly chaotic regime,
the authors analyze the dependence on the interaction strength $\sigma$. For small values of the coupling the motion of the individual units is still chaotic, but the trajectories are contained in a bounded region. With further increments of the coupling, the units follow periodic motions which are highly synchronized.
In this case, however, synchronization takes place in clusters, each cluster having a common value of the band center
around which the periodic motion occurs, and
center values appear in a discrete set of possible values.
These clusters form patterns of dynamical
regularity affecting mainly nodes at distances 2, 3, and 4,
as shown in Fig. \ref{fig_tadic}(left).
In fact, the histograms of distances between nodes along the tree and between nodes belonging to the same synchronized cluster have different statistical weights only for these values of the distances.

All previously discussed chaotic models are discrete time maps, which are appropriate discrete versions
of chaotic oscillators. Nevertheless, we also notice a couple of works dealing with time continuous maps. In \cite{lc04} the authors analyze a WS network of R\"{o}ssler oscillators, where the parameters
are chosen  to ensure that the system generates chaotic dynamics. The basic observation is that the network synchronizes when the coupling strength is increased, as expected. Another interesting result is that the mean phase difference among the chaotic oscillators decreases with the increasing of the probability of adding long-range random short-cuts. Along the same line, in \cite{ym06} it is considered a system of R\"{o}ssler
oscillators on BA networks. The only tuning parameter of the BA networks is $m$, the number of links  that a newly added node has.
For $m=1$ (SF trees) there is no synchronized state for a large number of oscillators.
Increasing $m$ synchronization is favored.
The topological effect of increasing $m$ is to create loops, but it is shown that this is not the only fact
that improves synchronization.

Finally, a different and interesting proposal was made in \cite{bbh04b} where a fixed 1-d connectivity pattern
is complemented by a set of switching long-range connections. In this case, it is proven that interactions
between nodes that are only sporadic and of short duration are very efficient for achieving synchronization.

As a summary, we can say that
most of the works deal with particular models of coupled maps (logistic maps, sine-circle maps, quadratic maps, ...).
Thus, it is possible, to obtain in some cases not only the conditions of local stability of the completely synchronized state
but the conditions for the synchronization independently of the initial conditions.
In general, the addition of short-cuts to regular lattices improves synchronization.
There are even some cases, for which synchronization is only attainable when a small fraction of randomness is add to the
system. On the contrary, in the next section we will discuss the linear stability of the synchronized state for general dynamical systems.

\section{Stability of the synchronized state in complex networks}
\label{sect_MSF}

In the previous section we have reviewed the synchronization of various types of oscillators on complex networks.
Another line of research on synchronization in complex networks, developed
in parallel to the studies of  synchronization in networks of
phase oscillators, is the investigation of the stability of the completely
synchronized state of populations of \emph{identical} oscillators.
The seminal work by \citeauthor{bp02} \cite{bp02} initiated this research line by analyzing the stability of synchronization in SW networks
using the Master Stability Function (MSF). The framework of MSF was developed earlier
for the study of synchronization of identical oscillators on regular or other simple  network  configurations~\cite{pc98,fjcmp00}.
The extension of the framework to complex topologies is  natural and important, because it relates the
stability of the fully synchronized state to the spectral properties of the  underlying
structure. It provides with an objective criterion to characterize the stability of the global synchronization state, from now on called
{\em synchronizability of  networks} independently of the particularities of the oscillators.
Relevant insights about the structure-dynamics relationship
has been obtained using this technique.

In this section, we review the MSF  formalism and the main results obtained so far.
Note that the MSF approach assesses the \emph{linear stability} of the completely synchronized state, which is a necessary, but not
a sufficient condition for synchronization.

\subsection{Master Stability Function formalism}

To introduce the MSF formalism, we start with an arbitrary connected network of coupled oscillators.
The assumption here for the
stability analysis of synchronization is that all the oscillators are identical, represented by the
state vector  ${\bf x}$ in an $m$-dimensional space. The
equation of motion is described by the  general form
\begin{equation}
\dot {\bf x}={\bf F}({\bf x}) \label{MSF_eq_0}.
\end{equation}
For simplicity, we consider time-continuous systems. However, the formalism applies also to time-discrete maps.
We will also assume an identical output function
${\bf H}({\bf x})$ for all the oscillators, which generates the signal from the state ${\bf x}$
and sends it  to other oscillators in the networks. In this representation, ${\bf H}$ is a vector function of dimension $m$.
For example, for the 3-dimensional system ${\bf x}=(x, y, z)$,
we can take ${\bf H}({\bf x})=(x, 0, 0)$, which means that the  oscillators are coupled only through the
component $x$.  ${\bf H}({\bf x})$ can be any linear or nonlinear mapping  of the state vector ${\bf x}$.
The $N$  oscillators, $i=1,\ldots ,N$,  are coupled in a network specified by the adjacency matrix $A=(a_{ij})$. We have
\begin{eqnarray}
\dot{\bf x}_i&=&{\bf F}({\bf x}_i)+
\sigma \sum_{i=1}^{N}a_{ij}w_{ij}[{\bf H}({\bf x}_j)-{\bf H}({\bf x}_i)]\label{MSF_eq_1}\\
         &=&{\bf F}({\bf x}_i)-\sigma\sum_{j=1}^{N}G_{ij} {\bf H}({\bf x}_j), \label{MSF_eq_2}
\end{eqnarray}
being $w_{ij}\ge 0$ the connection weights,
i.e., the network is, in general,  \emph{weighted}.
The coupling matrix $G$ is
 $G_{ij}=-a_{ij}w_{ij} $ if $i \neq j$ and $G_{ii}=\sum_{j=1}^{N}a_{ij}w_{ij}$.
 When the coupling strength is uniform for all the connections ($w_{ij}=1$), the network is \emph{unweighted}, and
 the  coupling matrix $G$ is just the usual Laplacian matrix $L$.
 By definition,  the coupling  matrix $G$ has  zero row-sum. Thus there  exists a \emph{completely synchronized state}  in this network of identical oscillators, i.e.,
\begin{equation}
{\bf x}_1(t)={\bf x}_2(t)=\ldots={\bf x}_N(t)= {\bf s}(t),
\end{equation}
which   is a solution of Eq.~(\ref{MSF_eq_2}).
In this synchronized state, ${\bf s}(t)$  also approaches the solution of  Eq.~(\ref{MSF_eq_0}), i.e.,
$\dot{\bf s}={\bf F}({\bf s})$. This subspace in the state space of
Eq.~(\ref{MSF_eq_2}), where all the oscillators  evolve synchronously on the same solution of the isolated
oscillator ${\bf F}$, is called the \emph{synchronization manifold}.

\subsubsection{ Linear Stability and Master Stability Function}

When all the oscillators are initially set at the synchronization manifold, they will remain synchronized.
Now the crucial question is whether the synchronization manifold is stable in the presence of small perturbations
$\delta{\bf x}_i$.  To assess the stability, we need to  know whether the perturbations grow or decay in time.
The linear  evolution of small  $\delta{\bf x}_i$ can be obtained by setting ${\bf x}_i(t)={\bf s}(t)+ \delta{\bf x}_i(t)$
in Eq.~(\ref{MSF_eq_2}), and expanding the functions ${\bf F}$ and ${\bf H}$ to first order in a Taylor series, i.e.,
${\bf F}({\bf x}_i)={\bf F} ({\bf s})+ D{\bf F}({\bf s})  \delta {\bf x}_i$ and
 ${\bf H}({\bf x}_i)={\bf H} ({\bf s})+ D{\bf H}({\bf s})  \delta {\bf x}_i$.
Here $D{\bf F}({\bf s})$ and $D{\bf H}({\bf s})$ are the Jacobian matrices of ${\bf F}$ and ${\bf H}$ on ${\bf s}$, respectively.
This expansion   results in the following linear variational equations for $\delta {\bf x}_i$,
\begin{equation}
\delta\dot{\bf x}_i=D{\bf F}({\bf s})  \delta {\bf x}_i-\sigma  D{\bf H}({\bf s})
\sum_{j=1}^{N}G_{ij}  \delta {\bf x}_j.
\label{MSF_vari_eq}
\end{equation}
The variational equations display a block form,  each block $(ij)$ having   $m$ components.
The  main idea here  is to project $\delta {\bf x}$
into the eigenspace spanned by the eigenvectors ${\bf v}_i$ of the coupling matrix $G$.
This projection can operate in block form  without affecting the structure inside the blocks.
By doing so, Eqs.~(\ref{MSF_vari_eq}) can be diagonalized  into $N$ decoupled
eigenmodes in the block form
\begin{equation}
\dot{\bf \xi}_l=\left[ D{\bf F}({\bf s}) - \sigma\lambda_l D{\bf H}({\bf s})\right]  {\bf \xi}_l,
\;\;\; l=1,\cdots,N,
\label{MSF_mode_i}
\end{equation}
where $\xi_l$ is the eigenmode associated with the eigenvalue
$\lambda_l$ of $G$.
Since $G$ has the property of zero row-sum, the minimal eigenvalue is always zero, i.e.,  $\lambda_1=0$, with the corresponding
eigenvector ${\bf v}_1=(1,1,\ldots, 1)$.
So the first  eigenmode
$\dot{\bf \xi}_1= D{\bf F}({\bf s})  {\bf \xi}_1$
corresponds to  the  perturbation parallel to  the synchronization manifold.  The  other $N-1$  eigenmodes are transverse to the
synchronization manifold and should be damped out to have a stable synchronization manifold.

In general, the eigenvalues spectrum is complex when  the coupling matrix is not symmetrical, e.g.,
when the network is directed ($a_{ij}\neq a_{ji}$) or when the coupling weights are asymmetrical ($w_{ij}\neq w_{ji}$).
In the literature, the characterization of network topology and the analysis of synchronization and other dynamical processes
have mainly focused on undirected and unweighted networks. This case corresponds to a symmetric $G$ and therefore all its eigenvalues
are real, which makes the analysis simpler.
In the following we also assume that  all the eigenvalues are real, which is always the case for
symmetric $G$ but it can also be true for non-symmetric cases, as will be discussed later.

When the eigenvalues spectrum is real, it  has the following properties: (i) $\lambda_1=0$ due to zero row sum of $G$;
(ii) $\lambda_l\ge0$ since $G$ is positive semidefinite and (iii) there is only one zero eigenvalue if the network is connected.
Accordingly, the eigenvalues can be ordered as

\begin{equation}
0=\lambda_1< \lambda_2\cdots\le\lambda_N.
\label{eigenvalue_order}
\end{equation}

\noindent In a connected network of $N$ oscillators,  $\lambda_2$ and $\lambda_N$ are then, the minimal and the maximal non-zero eigenvalues, respectively.

Importantly, we observe that all the individual variational equations in the system of equations Eq.~(\ref{MSF_mode_i}) have the same form
\begin{equation}
\dot{\bf \xi}=\left[ D{\bf F}({\bf s}) - {\bf \alpha}  D{\bf H}({\bf s})\right] {\bf \xi}.
\label{MSF_norm_mode}
\end{equation}
They only differ by the parameter $\alpha_l=\sigma\lambda_l$.
Now if we know the stability of the solution ${\bf \xi}= {\bf 0}$ for any reasonable  value of $\alpha$, then we
can infer  the stability for  any eigenmode  with $\alpha_l=\sigma\lambda_l$. To assess the  stability of this master
variational equation (\ref{MSF_norm_mode}),  we  calculate  its largest Lyapunov exponent $\lambda_{\max}$ as  a function
of ${\alpha}$, the resulting function is the \emph{master stability function}. The evolution of small ${\bf \xi}$ is then
described on average as $||{\bf \xi}(t)|| \sim   \exp[\lambda_{\max}({\alpha}) t]$, and the mode is stable with
$||\xi|| \to 0$  if $\lambda_{\max}({\alpha})<0$.

So far, we have assumed that  the eigenvalues $\lambda_l$  are real, and the problem
is to  compute the MSF $\lambda_{\max}(\alpha)$  for real values of  $\alpha$.
For the general case of complex  eigenvalues $\lambda_l$
one needs to evaluate the MSF in the complex plane  $\alpha=(\alpha_R, \alpha_I$). This scenario is mathematically more intricate
and less results are available. For many oscillators types $\lambda_{\max}(\alpha)<0$ in a bounded region of this plane \cite{pc98,fjcmp00}. However, semi-bounded (generally speaking unbounded) regions where $\lambda_{\max}(\alpha)<0$ can also occur for some specific
oscillators. We will see later how this difference between region bounds affects the measurement of synchronizability.

\begin{figure}
\begin{center}
\epsfig{figure=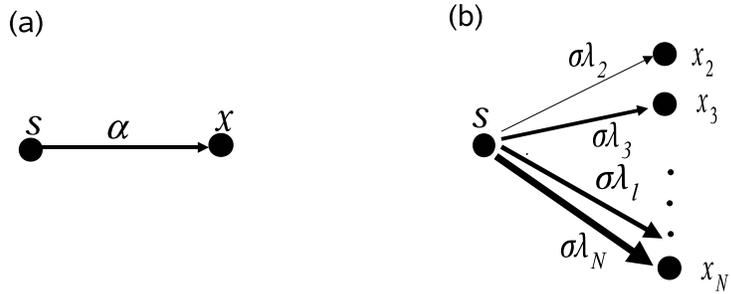,width=0.6\textwidth}
\end{center}
\caption{Schematic plot of the eigenmode decomposition. (a) For 2 unidirectionally coupled nodes; (b)
for a network of $N$ coupled nodes.}
\label{MSF_fig_cs1}
\end{figure}

The reader may find the framework of master stability function abstract. Here
 we present it in a physically intuitive manner with the help of a schematic  diagram.
For this, we restrict ourself  to the case where $\alpha$ is real.  Let us consider first two coupled
oscillators. The first one is autonomous
and  evolving along the trajectory ${\bf s}(t)$, and the  second one is driven
by  the first one with a coupling strength $\alpha$, as depicted in Fig.~\ref{MSF_fig_cs1}(a). The dynamical equations read,
\begin{eqnarray}
\dot{\bf s}&=&{\bf F}({\bf s}), \\
\dot{\bf x}&=&{\bf F}({\bf x})+{\bf \alpha} [{\bf H}({\bf s})-{\bf H}({\bf x})].
\end{eqnarray}
Then immediately, we obtain the  linear variational equation for the synchronization difference
${\bf \xi} (t)={\bf x}(t)-{\bf s}(t)$ as in Eq.~(\ref{MSF_norm_mode}). This means that the MSF
describes the stability of the state in which the  two oscillators are synchronized.

In a similar spirit, the equations for the $N-1$  transverse eigenmodes in Eq.~(\ref{MSF_mode_i}) can be
graphically represented as
$N-1$ oscillators driven by a common autonomous oscillator $\dot{\bf s}={\bf F}({\bf s})$, and the corresponding
coupling strengths are $\sigma\lambda_l$, as depicted in Fig.~\ref{MSF_fig_cs1}(b).  In this representation, the mode decomposition decouples
the complex network connections into many pairs of oscillators and the stability thus can be understood from that of the two coupled oscillators,
the MSF. The complete synchronization state ${\bf s}$  is stable when all the $N-1$
oscillators in  Fig.~\ref{MSF_fig_cs1}(b) are synchronized by the common forcing signal ${\bf s}$.
This picture will be useful later on in this section.

\subsubsection{Measures of synchronizability}

\begin{figure}
\begin{center}
\epsfig{figure=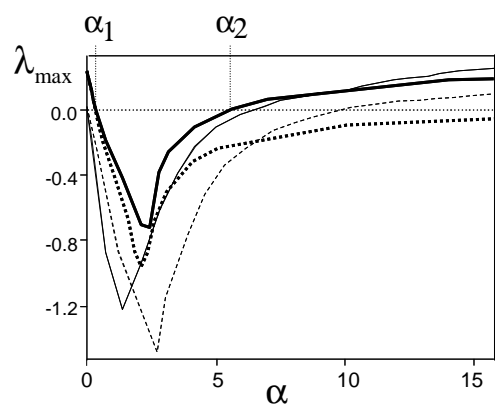,width=0.4\textwidth}
\end{center}
\caption{
Four master stability functions for coupled
R\"ossler oscillators: chaotic (bold) and periodic (regular lines);
with y coupling (dashed) and x coupling (solid lines). (The curves are scaled
for clearer visualization). From \cite{bp02}.
}
\label{bp02_fig1}
\end{figure}

In the case of real eigenvalue spectrum of $G$, the MSF only presents real values of $\alpha$. Here we distinguish two cases that will lead to different
criteria for synchronization in networks. The first case refers to \emph{bounded} MSFs, where  $\lambda_{\max}(\alpha)<0$ within a finite interval
$\alpha_1 < \alpha < \alpha_2$ (Fig.~\ref{bp02_fig1}).
A physical picture of this case for two coupled oscillators, as those in Fig.~\ref{MSF_fig_cs1}(a), is that the driven oscillator
will be synchronized, ${\bf x}(t)={\bf s}(t)$, if the coupling  strength $\alpha \in (\alpha_1, \alpha_2$).
Bounded MSFs are quite common for many oscillators $F$ and many coupling functions $H$.

For discrete time chaotic maps, the MSFs are always bounded, the reason is the following:
in discrete time systems,  Eq.~(\ref{MSF_norm_mode}) reads as
$${\bf \xi}_{n+1}=\left[ D{\bf F}({\bf s}(n)) - {\bf \alpha}  D{\bf H}({\bf s}(n))\right] {\bf \xi}_{n}=J(\alpha,{\bf s}(n)){\bf \xi}_{n}=J^{n}\xi_0,$$
where  $n$ is the iteration step and   $J^{n}=\Pi_{l=0}^{n}J(\alpha,{\bf s}(l))$.
The largest Lyapunov exponent is \cite{pc89}
$$\lambda_{\max}=\lim_{T\to \infty}\frac{1}{T} \ln \frac{||{\bf \xi}_{T}||}{||{\bf \xi}_0||}=\lim_{T\to \infty}\frac{1}{T} \ln ||J^T||.$$
where $T$ is the number of iterations.
Now suppose $\alpha>0$ is a very large value, then at each step $n$, $J(\alpha,{\bf s}(n)) \approx -\alpha  D{\bf H}({\bf s}(n))$, and
$J^{T}\approx (-\alpha)^T \Pi_{n=0}^{T-1} D{\bf H}({\bf s}(n))$.
As a result,  $\lambda_{\max} \approx \ln \alpha + \mu$. Here
$\mu=  \lim_{T\to \infty}\frac{1}{T} \ln ||J^T_H||$ is a finite  measure on the
chaotic attractor  similar to the largest  Lyapunov exponent $\lambda_1^F$ of the isolated chaotic map ${\bf F}$,
$\lambda_1^F=  \lim_{T\to \infty}\frac{1}{T} \ln ||J^T_F||$, where $J^T_H= \Pi_{n=0}^{T-1} D{\bf H}({\bf s}(n))$ and $J^T_F= \Pi_{n=0}^{T-1} D{\bf F}({\bf s}(n))$,
respectively. Consequently, $\lambda_{\max}>0$ when $\alpha$ is large enough,
i.e. the MSF always becomes positive when the coupling strength  is large enough, thus it is bounded.

There are also situations where the MSFs are \emph{unbounded}, and $\lambda_{\max}<0$ for $\alpha>\alpha_1$ without the upper limit $\alpha_2$.
This happens, for example, in time-continuous systems with ${\bf H} ({\bf x})={\bf x}$ where
the oscillators are linearly coupled through all the $m$ corresponding components.
In this case, the driven oscillator in  Fig.~\ref{MSF_fig_cs1}(a)
will be synchronized if the coupling  strength $\alpha > \alpha_1$.

To synchronize a network of oscillators,
all the $N-1$  oscillators in Fig.~\ref{MSF_fig_cs1}(b) must be synchronized by the common forcing signal ${\bf s}(t)$.
In the case of bounded MSFs, this requires that
$\sigma\lambda_l \in (\alpha_1, \alpha_2)$ for $2 \le l \le N$. Explicitly, the following condition is necessary for the stability of
the synchronization state of the network,
\begin{equation}
\alpha_1< \sigma\lambda_2 \le  \sigma\lambda_3 \le \cdots\le \sigma \lambda_N<\alpha_2.
\end{equation}
This  condition can be only fulfilled,  for some values of
$\sigma$, when the eigenratio $R$ satisfies the following relation
\begin{equation}
R\equiv \frac{\lambda_N}{\lambda_2}<\frac{\alpha_2}{\alpha_1}.
\label{MSF_R}
\end{equation}
Therefore, we conclude that it is impossible to synchronize the network if $R> \alpha_2/\alpha_1$, since
there is  no  $\sigma$ for which the fully synchronized state is linearly stable.
On the contrary, if $R<\alpha_2/\alpha_1$ the synchronous state is stable for $\sigma_{\min} < \sigma < \sigma_{\max}$
where $\sigma_{\min}= \alpha_1/\lambda_2$ and $\sigma_{\max}=\alpha_2/\lambda_N$, respectively.

If the MSF is unbounded, then the synchronization manifold is stable if it exists $\sigma$ satisfying
\begin{equation}
\alpha_1< \sigma\lambda_2 \le  \sigma\lambda_3 \le \cdots\le \sigma \lambda_N,
\label{unbound_seq}
\end{equation}
which will be true for any $\sigma$ larger than a certain synchronization threshold $\sigma_{\min}$.
Thus, the larger  $\lambda_2$ the smaller synchronization threshold $\sigma_{\min}$ \cite{mzk05a,mzk05b,nm06a}.
Moreover, as previously discussed in Sect. \ref{sect_km_path}, $\lambda_2$ also plays a special role
in the time needed to achieve complete synchronization~\cite{ad07}.

Note that the eigenratio $R$ and the nonzero minimal  eigenvalue $\lambda_2$ depend only on the network structure, as defined by $G$.
For bounded MSF,  if $R$ is small,  the condition  in Eq.~(\ref{MSF_R})
is, in general, easier  to satisfy.
From this, it follows that the smaller the eigenratio $R$ the more
synchronizable the network and vice versa \cite{bp02}.
In this sense, we can characterize the synchronizability
of the networks with $R$ and $\lambda_2$, without referring to specific oscillators.
As will be discussed soon, these two types of synchronizability on a specific network can depend on different parameters and then can be correlated to different network descriptors. In this review, synchronization based on the eigenratio $R$ is called \emph{Type I} and synchronizability based solely on $\lambda_2$ is called \emph{Type II} synchronizability.

For general directed networks, the spectrum of the coupling matrix $G$
is complex, and it is still unclear  how to develop simple measures, without referring to the MSF
of specific oscillators and coupling functions, to evaluate  the synchronizability of
different (directed) networks. This is probably one of the reasons why the vast majority of works using the MSF have focused
on undirected and unweighted networks whose spectra are real. In a recent work \cite{hcab05}, both the ratios of
the real and imaginary parts of the eigenvalues are used.  However, it should be noted that  networks with the same
ratios (or even the same minimal and maximal real and complex parts), but different boundaries
of eigenvalues in the complex plane, will have different synchronization thresholds for the same MSF. This is contrast to the cases
of real spectra, where the synchronization thresholds for the same MSF will be identical for  networks
with the same  $\lambda_2$ and $\lambda_N$, irrespective to the other spectral properties.

The key advantage of the MSF framework  is that it provides an objective criteria ( $\lambda_2$ and $\lambda_N$)
to assess the synchronizability of complex networks without referring to specific oscillators. The drawback is that
it only informs about the dynamics towards synchronization from small perturbations of the synchronization manifold (linear stability).
The study of synchronizability is then converted into the investigation of the eigenvalues of the
coupling matrix $G$ of the networks. In the current scenario of the MSF, natural questions are: What is the synchronizability
of different types of complex networks?  Which structural properties are related to or control the synchronizability?

Note that although the more appropriate approach seems to be that of following the general framework of graph theory to investigate the spectral properties of networks, many authors have tried to relate a single statistical property of networks with synchronizability, sometimes misinterpreting and generalizing the results without a proper estimation of their constraints. In the following section, we will  summarize results on the synchronizability of typical network models and then describe the relationship between graph theoretical measures and the synchronizability of complex networks.
Further insights have been obtained in the scope of graph theory by providing with bounds that constrain the eigenvalues of the Laplacian of networks.

\subsubsection{Synchronizability of typical network models}
\label{typical_net}

\begin{figure}
\begin{center}
\epsfig{figure=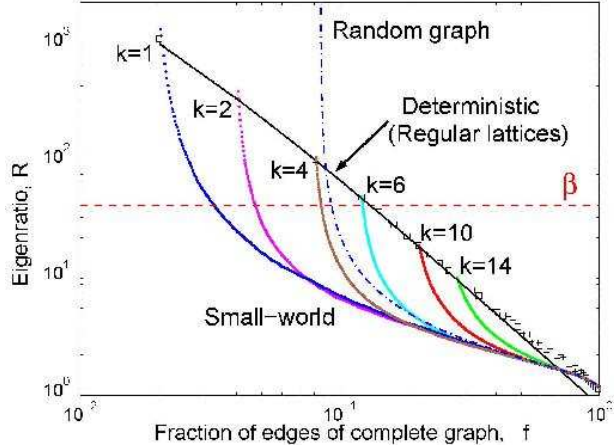,width=0.5\textwidth}
\end{center}
\caption{
(color online) Decay of the eigenratio  $R$
in a $N=100$ lattice as a fraction $f$ of the $N(N-1)/2$ possible  edges  are added following
purely deterministic, semirandom, and purely random schemes. Networks become
synchronizable below the dashed line  ($\beta=\alpha_2/\alpha_1$, where $\alpha_1$ and $\alpha_2$
are from Fig.~\ref{bp02_fig1}). The squares
(numerical) and the solid line [analytic Eq.~(\ref{bp02_eq1})] show the
eigenratio decay of regular networks through the deterministic
addition of short-range connections. The dot-dashed line corresponds
to purely random graphs [Eq.~(\ref{bp02_eq3})], which become almost
surely disconnected and unsynchronizable at $f \simeq 0.0843$. The
semirandom approach (dots, shown for ranges $k=1, 2, 4, 6, 10, 14$)
is more efficient in producing synchronization when $k<\ln N$. From \cite{bp02}.
}
\label{bp02_fig2}
\end{figure}

We will present the main findings related to synchronizability in the most common classes of complex networks found in the literature: Regular, SW, ER, and SF networks.

\paragraph*{Regular networks}

For a long time, synchronization of chaotic oscillators has been studied on regular networks~\cite{pc98,fjcmp00}.
A typical example of a regular network  is a cycle (or ring) of $N$ nodes each coupled
 to its $2k$ nearest neighbors with a total of $Nk$ links.
 The eigenvalues of the coupling matrix are~\cite{m99,bp02}
\begin{equation}
\lambda_l=2k-2\sum_{j=1}^{k}\cos\big(\frac{2\pi (l-1)j}{N}\big).
\label{bp02_eq1}
\end{equation}
By a series expansion, we obtain
\begin{equation}
\lambda_2\simeq \frac{2\pi^2 k(k+1)(2k+1)}{3N^2},  \lambda_{N}= (2k+1)(1+2/3\pi ).
\label{bp02_eq2}
\end{equation}
and the eigenratio $R$ for $k\ll N$  can be approximated as
\begin{equation}
R\simeq \frac{(3\pi +2)N^2}{2\pi^3 k(k+1)}.
\label{bp02_eq22}
\end{equation}
These relations show that regular networks have a poor Type I and Type II synchronizability. The asymptotics
on the system size  $\lambda_2\sim 1/N^2$, makes
synchronization effectively impossible for both bounded and unbounded MSFs  of chaotic
oscillators with $\alpha_1>0$. In such regular networks, usually we have complicated pattern formation (waves) instead of
a stable spatially homogeneous (completely synchronized) state~\cite{mg98}.
A way to improve this situation is, for fixed $N$, adding connections to generate a higher range $k$, since $\lambda_2$ increases approximately as $\lambda_2\sim k^3$ the
the eigenratio  $R$ decreases as $R \sim 1/k^2$ for $k\ll N$, and synchronizability of Type I and II is enhanced.

\paragraph*{SW networks}

In \cite{bp02},  SW  networks are obtained by adding $NS$ links at random, to a regular network where each node is connected to its $2k$ nearest neighbors, so that the average number of shortcuts per node is $S$.
As shown in Fig.~\ref{bp02_fig2}, adding a small fraction of such random connections reduces the eigenratio $R$
so that the synchronizability is improved significantly.

The eigenvalues of the SW  networks have been obtained~\cite{bp02}
through a  perturbation analysis of the SW  Laplacian $L=L^0+L^r$.
Here  $L^0$ is  the deterministic Laplacian of the regular networks
and $L^r$ the Laplacian formed by the random shortcuts.
In the  stochastic Laplacian
matrix $L^r$  (symmetric, zero row-sum),  any of the
remaining entries  $N(N-2k-1)$ of $L^0$  takes the value 1 with probability $p_s=2S/(N-2k-1)$ and the value 0 with probability $(1-p_s)$.
For $p_s \ll 1$, and $ N^{1/3} <k\ll N$,  the perturbations of the eigenvalues are
\begin{equation}
\varepsilon \lambda_2^{(1)}\simeq Np_s-\sqrt{3\pi p_s/4}, \;\; \varepsilon \lambda_N^{(1)}\simeq Np_s+\sqrt{3\pi p_s/4}.
\label{bp02_eq4}
\end{equation}
The extreme eigenvalues are
\begin{equation}
\lambda_2=\lambda_2^{(0)}+ \varepsilon \lambda_2^{(1)}, \;\; \lambda_N=\lambda_N^{(0)}+ \varepsilon \lambda_N^{(1)},
\label{bp02_eq5}
\end{equation}
where $\lambda_2^{(0)}$ and $\lambda_N^{(0)}$ are the eigenvalues of the regular networks ($L^0$) as in
Eq.~(\ref{bp02_eq2}).
From this analysis, it follows that  for a  fixed small value of $S$, the minimal
non-zero eigenvalue $\lambda_2$ is driven away from $\lambda_2^{(0)}\approx 0$
to $\varepsilon \lambda_2^{(1)}\approx  Np_s \approx 2S$ for \emph{any} SW network with large $N$ and $N^{1/3}<k\ll N$,
while the maximal eigenvalue $\lambda_N$ is not affected very much for small $S$. This means that both types of
synchronizability are mainly determined by the average number of  shortcuts per node $S$.

\begin{figure}
\begin{center}
\epsfig{figure=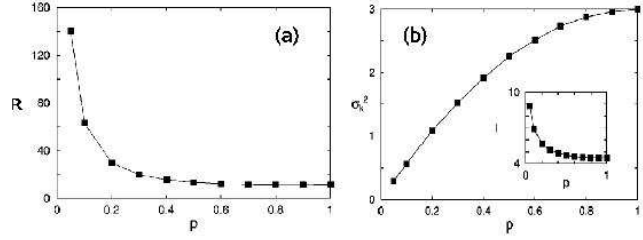,width=0.55\textwidth}
\end{center}
\caption{Synchronizability (eigenratio $R$, (a)) and heterogeneity of degrees
(variance $\sigma_k^2$ of the degree distribution (b))  of  SW
networks as a function of the rewiring probability $p$.  Inset of (b): average path length $\ell$ vs $p$.
The network has a size $N=2000$ and the range $k=3$. From \cite{hkcp04}.
}
\label{hkcp04_fig1}
\end{figure}

In SW networks,  the variance of the degree distribution raises as the regular  network is rewired with an increasing
probability $p$~\cite{hkcp04}  or when more shortcuts are added, see Fig.~\ref{hkcp04_fig1}(b).
This process results in an improvement of
the synchronizability  (the eigenvalue ratio, $R$, is reduced), as illustrated in Fig.~\ref{bp02_fig2} and Fig.~\ref{hkcp04_fig1}(a).
This is because $\lambda_2$ increases proportionally to the
number of shortcuts per node, i.e., $\lambda_2\approx 2S= 2 kp$.

\paragraph*{Random networks}

In purely random graphs,
in which a fraction $f$ of the $N(N-1)/2$ possible links is established at random, the  eigenratio is a function
of $f$ and $N$ ~\cite{bp02}. It reads
\begin{equation}
R\simeq \frac{Nf+\sqrt{2f(1-f)N \ln N}}{Nf-\sqrt{2f(1-f)N \ln N}}
\label{bp02_eq3}.
\end{equation}
Note that the networks are synchronizable only when $f\gtrsim 2\ln N /(N+2\ln N)$,
where  it is almost sure
that the networks are connected \cite{bollobas01}.  If this condition is verified
the synchronizability is improved with increasing $f$, as seen in
Fig.~\ref{bp02_fig2}.


Note that for small $f\lesssim \ln N /N$, the  purely random networks are almost surely disconnected  and thus
non-synchronizable. On the contrary, the  regular backbone of nearest connections with $k\ge 1$ can already make
the semirandom SW networks connected regardless of $N$ and thus synchronizable as a whole. In this sense,
semirandom SW networks  turn out to be much superior to the purely random networks in terms
of synchronizability, as can be seen from Fig.~\ref{bp02_fig2} for small $k$, $k\lesssim \ln N$
($k=1\sim 4$ in networks with $N=100$). The improvement
is even more pronounced for larger $N$.

\paragraph*{SF  networks}

\begin{figure}
\begin{center}
\epsfig{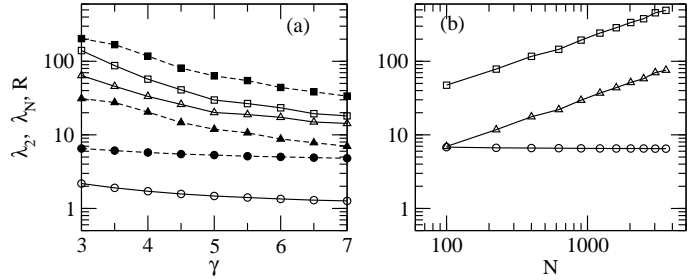}
\end{center}
\caption{Synchronizability of random SF networks. Plotted are  eigenvalues $\lambda_2$ (circles) and
$\lambda_N$ (squares) and eigenratio $R$ (triangles). (a) As functions of the exponent $\gamma$ for network size $N=2^{10}$;
open symbols for $k_{\min}=5$ and filled symbols for $k_{\min}=10$.  (b)  As a functions
of network size $N$ at $\gamma=3$ and $k_{\min}=10$. The results are averaged over 50 realizations.
}
\label{MSF_fig_cs2}
\end{figure}

Here we discuss  the synchronizability of SF  networks for different values of the degree distribution exponent $\gamma$. To this end, we consider a random  model introduced in \cite{nsw01} to construct SF networks. The algorithm works as follows, first, a degree $k_i$ is assigned to each node $i$
according to the probability distribution $P(k)\sim
k^{-\gamma}$ and $k\ge k_{\min}$.  The network is generated by randomly connecting the nodes so that  node $i$ has exactly the prescribed $k_i$ links to other nodes, prohibiting self- and repeated links.

The dependence of the eigenvalues $\lambda_2$ and $\lambda_N$ and the eigenratio $R$ on $\gamma$  and $N$ are shown in Fig.~\ref{MSF_fig_cs2}. We observe that $\lambda_2$ has no noticeable dependence on $\gamma$  and $N$. However, $\lambda_N$ becomes larger as the degree
heterogeneity is increased. So the changes of the eigenratio $R$ follow the trend of $\lambda_N$ closely. This result is somehow expected given that the largest eigenvalue $\lambda_N$ is intimately related to the degree of hubs, and this is the essential fingerprint of SF networks with different exponents $\gamma$.
On the other hand, $\lambda_2$ increases  with
$k_{\min}$, and $R$ is larger at smaller $k_{\min}$ for the same $\gamma$ and $N$. The variation of $R$ as a function of $\gamma$ was
reported in \cite{nmlh03}. The dependence of the synchronizability  on $k_{\min}$ and $N$ in this random SF network model  turns  out to be very similar in the BA growing model, as shown in \cite{pb05}. The conclusion is that in SF networks, the two types of synchronizability  (I and II) associated to $R$ (bounded MSF) and $\lambda_2$ (unbounded MSF)  are very different.

In an early paper ~\cite{wc02} that studied robustness and fragility of synchronization of SF networks,
it was reported that  $\lambda_2$ is a constant almost unrelated with $k_{\min}$ ($k_{\min}=3,5,7,9,11$).
This  result in \cite{wc02} is inconsistent with the observation in Fig.~\ref{MSF_fig_cs2}, with the work in \cite{pb05}, and with graph theoretical
analysis in \cite{w03}, which we will discuss in more detail later on.
Thus  the observation of robustness and fragility of synchronizability
in \cite{wc02}  (changes of $\lambda_2$ due to random or deliberate attack of the nodes)
should be taken cautiously, and a detailed reexamination of this issue is mandatory before assessing conclusions.

\subsubsection{Synchronizability and structural characteristics of networks}

The relationship
between structural characteristics of networks and synchronizability has been explored intensively
in the literature, mainly based on numerical experiments on various network models. The observations, which are summarized in
what follows, are quite confusing. The main problem  is that many works have made a naive use of complex network models to assess synchronizability. Very often network models do not allow us to isolate one structural characteristic while keeping the other properties fixed. For this reason many results have been misinterpreted.
As we will see in the next section, a more objective graph theoretical analysis sheds some light to the whole problem. Unfortunately, the complete understanding of the structure-synchronizability relationship is still missing.

\paragraph*{Synchronizability dependence on $\ell$}

The average shortest path length $\ell$ is a property of the network closely related to the efficiency of information processing.
Most real-world complex networks are characterized by a small $\ell  \lesssim \ln N$~\cite{dgm07}.
Indeed, it has been conjectured and rationalized that in biological neuronal networks, $\ell$ has
been minimized by evolution~\cite{shboyk00, kh06}.
Generally speaking, $\ell$ is lower in SF networks than in ER networks due to the presence of hubs~\cite{ch03}, and $\ell$ is lower in SW networks than in regular lattices due to the presence of shortcuts.

In \cite{ws98} it is suggested that the decrease in the distance in the WS  network would lead to more efficient
coupling and thus enhanced synchronization of the oscillators.
Investigation of phase oscillators~\cite{hck02} or circle maps~\cite{bpvl03} on WS networks has
shown that when more and more shortcuts are created at larger
rewiring probability $p$, the transition to the synchronization regime
becomes easier.
On the other hand, the synchronizability of identical oscillators
follows the same trend of $\ell$, in networks with fixed  $N$ and $k$, as $p$ is increased
(Fig.~\ref{hkcp04_fig1}).
A similar type of behavior is observed if shortcuts
are added to the regular networks (Fig.~\ref{bp02_fig2}).
From these observations, the generalized conclusion that smaller distances will always be correlated to enhanced
synchronization, has been intuitively used by many authors, and is not so.
Indeed, a more detailed analyses of various network models have shown that
there is no direct relationship between $\ell$ and the synchronizability of the networks.
The reason is that the transition to the small-world regime occurs at a value of the rewiring probability
for which there is no significant effect on $\lambda_2$.

In fact, in WS networks, $\ell $ is a function of the network size $N$,
the degree of the nodes in the original regular network, $k$, and the randomness parameter $p$ \cite{nmw00}
\begin{equation}
\ell (N,k,p)\sim \frac{N}{k}f(pkN),
\label{nmw00_eq1}
\end{equation}
where $f(u)$ is a universal scaling function, $f(u)=\hbox{const}$ if $u\ll 1$ and $f(u)=\ln(u)/u$ if $u\gg 1$.
From this result, it turns out that  $\ell$ begins to decrease with $p$, and consequently the SW behavior emerges, for $p\gtrsim p_{\mbox{\scriptsize{SW}}} = 1/Nk$.
At $p=p_{\mbox{\scriptsize{SW}}}$ the average number of shortcuts per node is $S\sim 1/N$, and then $\lambda_2\sim 1/N$ as well. This shows that at this point the synchronizability is not enhanced by the rewiring.
To achieve such an enhancement, the density of shortcuts has to be independent of $N$, which happens
for  $p\gtrsim p_{\mbox{\scriptsize{sync}}} = 1/k$, that is deep in the SW regime.
In other words, in the intermediate region $p_{\mbox{\scriptsize{SW}}} < p < p_{\mbox{\scriptsize{sync}}}$,
$\ell$ decreases while the synchronizability of the system remains roughly the same.

\begin{figure}
\begin{center}
\epsfig{figure=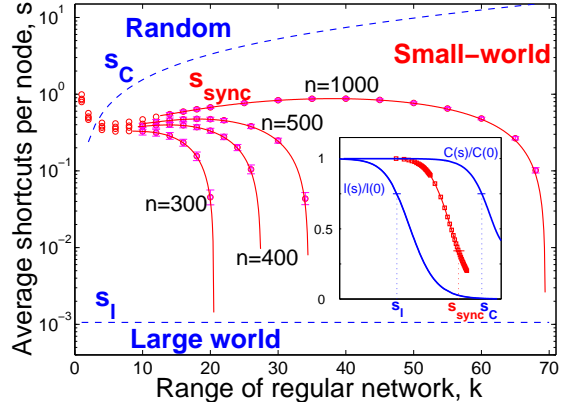,width=0.45\textwidth}
\end{center}
\caption{
(color online)
Synchronizability thresholds $S_{\mbox{\scriptsize{sync}}}$ ($\circ$)  for
graphs with $N$  nodes ($N=300, 400, 500, 1000$) and range $k \in [1, 70]$,
averaged over 1000 realizations. Solid lines are based on  an
analytical perturbation (Eqs.~(\ref{bp02_eq4},\ref{bp02_eq5}) valid in $N^{1/3} <k < k^0_{\mbox{\scriptsize{sync}}}$).
For most parameters, $S_{\mbox{\scriptsize{sync}}}$  lies within the SW region between the dashed
lines (depicted for $N=1000$), but it is distinct from its onset $S_\ell$.
Inset: decay of the average distance $\ell $, clustering $C$, and eigenratio
(squares) as shortcuts are added to a regular network of $n=500$ and $k=20$.
We define $S_\ell$ and $S_{\mbox{\scriptsize{C}}}$  as the points where $\ell$ and
$C$ are 75\% of the regular network  value. From \cite{bp02}.
}
\label{bp02_fig3}
\end{figure}

\citeauthor{bp02} \cite{bp02} showed the existence of two thresholds, one for the small-world transition and the other for the
enhancement of synchronizability, in SW networks using a more rigorous analysis. They obtained that the SW regime starts at $S_{\ell}\sim 1/N$ whereas the threshold beyond which the synchronizability is improved goes like $S_{\mbox{\scriptsize{sync}}} \sim k$. This means that
when $k$ is large, $\lambda_2^{(0)}$ of the underlying regular network contributes significantly to the
synchronizability, and then it can be enhanced  without additional shortcuts.
This is manifested
in Fig.~\ref{bp02_fig3} by the fast decrease of $S_{\mbox{\scriptsize{sync}}}$ when $k$  approaches the  critical value  $k^0_{\mbox{\scriptsize{sync}}}$.
On the other hand, at low $k$, the approximation in Eq.~(\ref{bp02_eq4}) is not valid. However, the results in
Fig.~\ref{bp02_fig3} show that  $S_{\mbox{\scriptsize{sync}}}\in [0.3,1]$ depends on $k$, but not noticeably on $N$.

\begin{figure}
\begin{center}
\epsfig{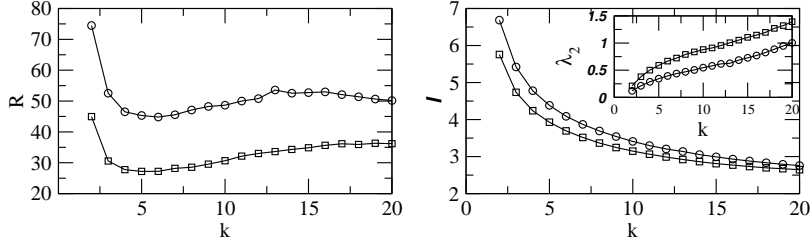}
\end{center}
\caption{Eigenratio $R$ (left),  distance $\ell$ (right) and eigenvalue $\lambda_2$ (inset)  as functions of $k$ (the range of  nearest neighbours
 in the WS model of SW networks). $N=500$ nodes, averaged over 100 realizations. The average number of shortcut per node $S$ is fixed, so that  the rewiring probability is $p=S/2k$;  $S=0.3$ (circle) and $S=0.5$ (square). }
\label{Fig_WS_k}
\end{figure}

One observation from Fig.~\ref{bp02_fig3} is that  smaller distances are not be necessarily correlated to enhanced synchronizability as intutively believed.
Indeed, if we keep the number of shortcut per node fixed, $S\sim ~ 1$, and increase $k$ as in Fig.~\ref{bp02_fig3}, we can see that the network
distance decreases monotonically, while the eigenratio does not always follow the same trend. There is a range of $k$ where the synchronizability
of Type I is reduced ($R$ increases)  while the network distance  becomes smaller, see Fig.~\ref{Fig_WS_k}. The Type II synchronizability ($\lambda_2$, unbounded MSF),
however, is enhanced.  In summary, it is not very meaningful to compare the synchronizability of two SW networks
(with different $N$, $k$ or $p$) considering only $\ell$.

\begin{figure}
\setlength{\unitlength}{5.3in}
\begin{picture}(1,0.52)
\put(0.99,0.27){\makebox(0,0)[r]{
\resizebox{0.93\unitlength}{!}{
\includegraphics{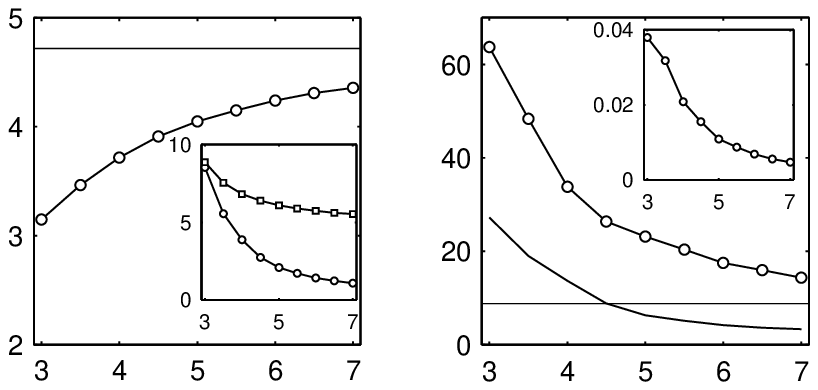}
}}}

\put(0.13,0.49){\makebox{(a)}}
\put(0.64,0.49){\makebox{(b)}}

\put(0.285,0.01){\makebox(0,0)[b]{$\gamma$}}
\put(0.415,0.1){\scalebox{0.7}{\makebox(0,0)[b]{$\gamma$}}}
\put(0.39,0.2){\scalebox{0.7}{\makebox(0,0)[b]{$s$}}}
\put(0.39,0.27){\scalebox{0.7}{\makebox(0,0)[b]{$\langle  k \rangle$  }}}
\put(0.02,0.28){\makebox(0,0)[l]{$\ell$}}
\put(0.12,0.395){\makebox(0,0)[l]{$\gamma = \infty$}}

\put(0.79,0.01){\makebox(0,0)[b]{$\gamma$}}
\put(0.48,0.28){\makebox(0,0)[l]{$\displaystyle\frac{\lambda_N}{\lambda_2}$}}
\put(0.88,0.39){\scalebox{0.7}{\makebox(0,0){$b_{\max}$}}}
\put(0.9,0.24){\scalebox{0.7}{\makebox(0,0){$\gamma$}}}
\put(0.61,0.12){\scalebox{0.7}{\makebox(0,0)[l]{$\gamma = \infty$}}}
\end{picture}
\caption{
Synchronizability of SF networks of size $N=1024$.
The average network distance (a) and the eigenratio (b)
for the random SF  model with $k_{\min}=5$. The inset of (a) shows the mean degree $\langle  k \rangle$
and the standard deviation $s$  of the
connectivity distribution.  Inset of (b): the maximum   normalized
load $b_{\max}$. The horizontal
lines in (a) and (b) indicate the values of $\ell$
and the eigenratio $R$  for the $\gamma=\infty$  case.
The solid curve in (b) is  the lower bound $k_{\max}/k_{\min}$ in Eq.~(\ref{nmlh03_eq1}).
The upper bounds of in Eq.~(\ref{nmlh03_eq1}) are above the limits, but follow
the same trend. All quantities are averaged over 100
realizations. From \cite{nmlh03}.
}
\label{nmlh03_fig1}
\end{figure}

The relationship  between  $\ell$ and the synchronizability of the system was also scrutinized for SF networks in \cite{nmlh03}, an important work that raised the interest of studies on the structure-synchronizability relation in complex networks.
As seen in Fig.~\ref{nmlh03_fig1}, for random SF networks, $\ell$
decreases when the degree distribution becomes more heterogeneous (decreasing of the exponent $\gamma$),
however, the network becomes less synchronizable, since $R$ increases.
In these simulations, the mean degree of the network
$\langle  k \ra=k_{\min} \frac{\gamma-1}{\gamma-2}$ also changes with $\gamma$, as well as the standard deviation of the degree distribution.
As will be clarified later on (e.g., Eq.~(\ref{tight_bounds})), in this case, the eigenratio $R$ is controlled by the heterogeneity of the  degree
distribution, being
$R \sim k_{\max}/k_{\min}$ if $k_{\min}$ is large enough, while the eigenvalue $\lambda_2$ has a dependence on the mean degree when the minimal degree is
fixed~\cite{zmk06,km07}.
In \cite{nmlh03} the authors observed similar results in a SW network model with hubs, and they considered them counterintuitive.
Similar to the SW networks with different range $k$ in Fig.~\ref{Fig_WS_k}, where the eigenratio $R$ is not
simply controlled by $\ell$, here it is also not surprising to observe that synchronizability can be reduced when $\ell$ becomes smaller~\cite{nmlh03}.

Besides, it should be possible to observe the situation that  $R$  decreases at smaller  $\ell$ in random SF network  for suitable combinations of parameters
$N$, $\gamma$ and $k_{\min}$, for example, when $k_{\min}$ increases at fixed $\gamma$ and $N$.
Therefore, the conclusion is that
the synchronizability of complex networks cannot be assessed
solely on the average shortest path length $\ell$.


\paragraph*{Synchronizability dependence on betweenness centrality}

In \cite{nmlh03} it is argued heuristically that this apparently surprising behavior
(smaller $\ell$, less synchronizability in SF networks) is due to
the fact that   a few central  oscillators interacting with a
large number of other oscillators tend to become overloaded.
When too many independent signals with different phases and
frequencies are going through a node at the same time,
they can cancel out each other, resulting in no effective
communication between oscillators.  Thus the authors were motivated to examine
the influence of the load (betweenness) of the nodes.
It was shown  (see the inset of Fig.~\ref{nmlh03_fig1} (b))that the synchronizability follows the same trend as the
maximum load $b_{\max}$  (normalized by the total load of the network).
However, the load of a node in SF networks is closely related to
the degree~\cite{gkk01,b04}, i.e., nodes with large degrees or links connecting nodes with large degrees
have, on average, a large load.
The correlation between reduced synchronizability and
heterogeneous load has also been observed in
a variant of the WS network model by adding $m$  shortcuts to the network from a randomly
selected node to one out of the $n_c$ center nodes~\cite{nmlh03}.  In this case, when $n_c$ is small,  the  $m$ shortcuts are connected to a few hubs,
and  the degree becomes more heterogeneous and the maximum  load $b_{\max}$ increases, while the synchronizability decreases.
As before,
in these two examples,
it is not very clear whether the change of synchronizability is mainly influenced by the degree heterogeneity or
by the load itself, because these two properties are closely related.

In the original WS network ~\cite{ws98}, the maximum load $b_{\max}$ decreases
when the degree distribution becomes more heterogenous as $p$ is increased. Based on this observation, it was claimed that
more homogeneous load predicts better synchronizability on complex networks~\cite{hkcp04}.
However, a direct relationship between load and synchronizability can not be clearly established.
In fact,  in\cite{zzwyyb06} it is shown an example where the network displays improved synchronizability while
the load becomes more heterogeneous when an original random SF  is rewired to obtain
nontrivial clustering.

The heuristic argument in \cite{nmlh03} is not clearly justified.
If the picture described is correct, i.e., signals  can cancel out each other at the central nodes, resulting
in no effective communication, then the Type  II synchronizability ($\lambda_2$, unbounded MSFs) should also be reduced
significantly.
However, as seen in Fig.~\ref{MSF_fig_cs2}, $\lambda_2$
in SF networks is mainly determined  by the minimal degree $k_{\min}$, without a noticeable dependence on the degree distribution exponent or load.
Moreover, when the minimal degrees becomes larger at fixed $\gamma$, the maximal load increases, however, both types of synchronizability are
enhanced (smaller $R$ and larger $\lambda_2$), in contrast to the heuristics. In fact,  several further  investigations have shown that the highly connected  oscillators synchronize faster among them and form
synchronization clusters~\cite{l05,zk06a,zk06b,gma07a}, which is in contrast with the
argument provided in \cite{nmlh03}.

\paragraph*{Synchronizability dependence on the clustering coefficient}

In \cite{mzk05b} it is pointed out that, in general,  the eigenratio $R$ increases with increasing clustering
in  a modified version of the  BA model ~\cite{st01}. Unlike the original
one,
in this model, motivated by the evolution of language, a new node is first linked  to an existing node according to the preferential attachment rule, and also linked to the neighbors of this target node.
Thus this model displays nontrivial clustering while keeping the same SF degree distribution.
Besides, the eigenratio $R$ is larger than that of the BA model.

A recent work demonstrated that for both SW and SF networks, large values of
clustering hinders global synchronization of phase oscillators, since the network splits into dynamical clusters
that oscillate at different  frequencies~\cite{mm05,gm07}.
In \cite{zzwyyb06}, using the rewiring scheme proposed in \cite{k04},
that changes the clustering but keeps the degree sequence,
it was shown that the eigenratio $R$ also increases with $C$, for both SW and SF networks.
This indicates  that synchronizability is reduced when clustering $C$ increases.

\begin{figure}
\epsfig{figure=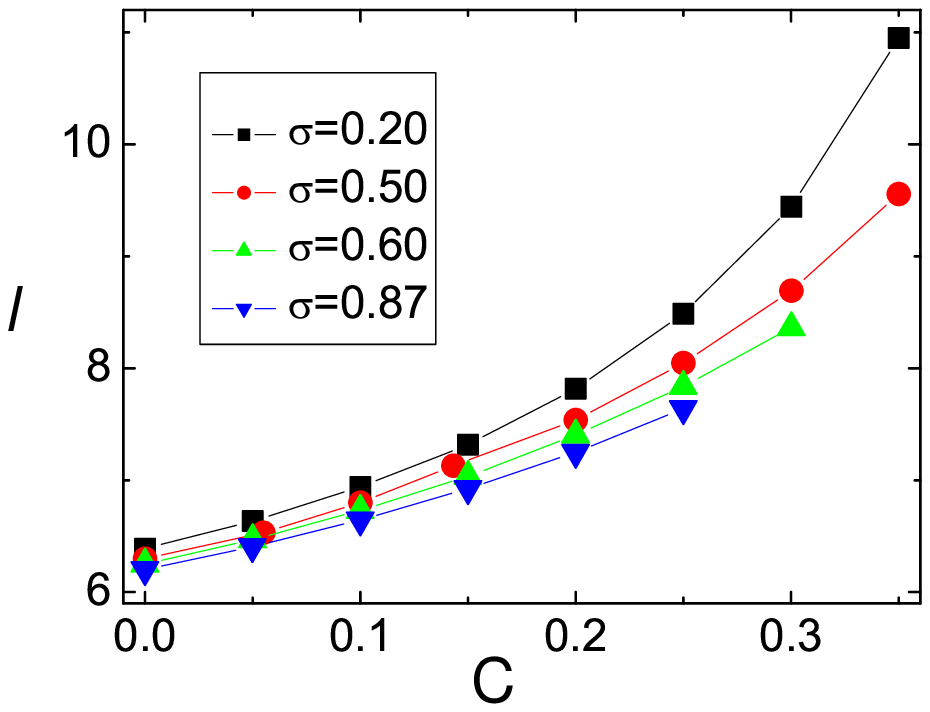,width=0.5\textwidth}
\epsfig{figure=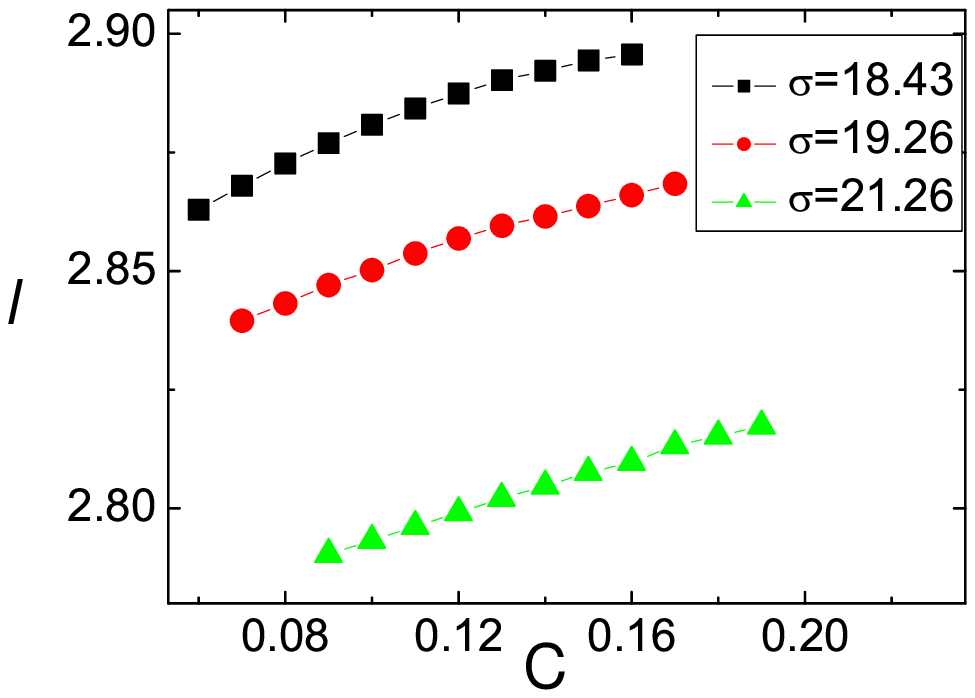,width=0.5\textwidth}
\caption{ (color online) The relationship between graph distance $\ell$  and clustering
coefficient $C$.  Left: the original networks are the SW networks.  Right: the original networks are the extensional BA networks.
The different symbols are for different heterogeneity of degrees, measured by the standard
deviation $\sigma$ of the degree distribution.
All the data are the average over 20 different realizations. From \cite{zzwyyb06}.
}
\label{zzwyyb06_fig1}
\end{figure}

Note that in the above investigations different
structural properties change at the same time~\cite{bp02,hkcp04,nmlh03} when the parameters characterizing the original networks are modified.
Once again, it is difficult to draw conclusions about the relationship between one single statistical descriptor
of the network
and its synchronizability.
Special attention was paid to this problem in \cite{zzwyyb06}, showing
that in the rewiring scheme of \cite{k04}, $\ell$ is correlated with the clustering coefficient $C$ (Fig.~\ref{zzwyyb06_fig1}).

\paragraph*{Synchronizability dependence on degree correlations}

In many real-world networks, the degree of a node is often correlated with the degree of the neighboring nodes.
Correlated networks show \emph{assortative (disassortative) mixing} when  high degree nodes are mostly attached to nodes with high (low) degree ~\cite{n02}.
In practice, the  degree-degree correlation of a network  can be calculated as the Pearson correlation coefficient between  degrees ($j_l, k_l$) of the nodes  at the ends of
the $l$th link, i.e.,
\begin{equation}
r_k=\frac{\langle   j_l k_l \rangle -\langle   k_l\rangle^2}{\langle   k^2_l\rangle- \langle   k_l\rangle^2},
\end{equation}
where $\langle   \cdot \rangle$ denotes  average over the total number of links  in the network.
Positive values of $r_k$  indicate assortative mixing, while negative values refers to disassortative
networks. A specific correlation $r_k$ can be obtained by rewiring the links while keeping the degree sequences unchanged~\cite{n03b}. However, one can expect  that $\ell$ and  $b_{\max}$ change also when the networks are rewired.

\begin{figure}
 	
\begin{center}
\epsfig{width=0.55\textwidth, file=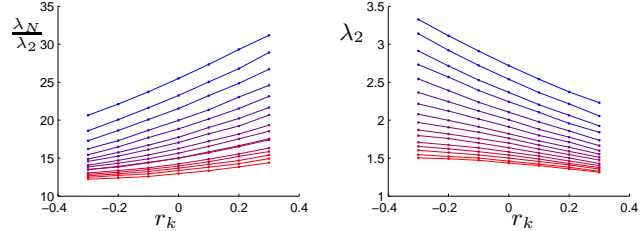}
\begin{picture}(0,0)(0,0)
\put(-200,-20){\small $r_{k}$}
\put(-255,50){\small $\frac{\lambda_N}{\lambda_2}$}
\put(-65,-20){\small $r_{k}$}
\put(-130,50){\small $\lambda_2$}
\end{picture}
\end{center}
\caption{ (color online)
Synchronizability of degree correlated SF networks of size $N=1000$,
$k_{\min} = 5$ and $\gamma=2.5$. Left: Behavior of the eigenratio $R$. Right: behavior of the second lowest eigenvalue $\lambda_2$.
Both as functions of the correlation coefficient $r_{k}$
defined in \cite{n02}, for $\gamma$ varying from 2 (blue line) to 5 (red line) in
steps of 0.2.  The results have been averaged over 100 different realizations. From \cite{dgs06}.
}
\label{dgs06_fig1}
\end{figure}

The influence of degree correlations on synchronizability was addressed in \cite{mzk05b}, showing that
the eigenratio $R$ increases when increasing the assortativity of the network.
A more detailed and systematic analysis was later on carried out in \cite{dgs06}.
They generated random
SF  network as in \cite{nsw01} with a minimal degree $k_{\min}=5$  and obtained the desired degree correlation $r_k$ using the rewiring
procedure of Refs. \cite{n03b,mzk05b}.
The dependence of the synchronizability on  $r_k$ is shown
in Fig.~\ref{dgs06_fig1}. The main effect on the eigenratio $R$ comes from the fact that  $\lambda_2$ decreases when
$r_k$ grows,
while $\lambda_N$ remains roughly constant~\cite{dgs06}.
As it happens for the other parameters,
the  dependence of $\lambda_2$ on $r_k$ can be
obtained from graph theoretical analysis, which is the subject we are going to discuss next.

\subsubsection{Graph theoretical bounds to synchronizability}
\label{sect_graph}

Previously we have seen that
many structural properties can influence
the synchronizability of the networks, but none of them can be regarded as the exclusive factor in the
observed dependencies.
The moral is that all works described in the preceding paragraphs seem not to be on the most appropriate way to elucidate the dependence of synchronizability on the network characteristics.
Since the synchronizability depends on the two extreme eigenvalues of the Laplacian matrix,
a sound analysis
must attack the raw problem of the spectral properties of networks from a mathematical point of view,
given that the simulation experiments are far from being conclusive.

Graph theoretical analyses of the Laplacian matrix $L$, in the context of synchronizability, mainly
focus on the bounds of its extreme eigenvalues. The implications of these bounds on different types of complex networks
have been discussed in several works~\cite{clv03,w03,w05,abj06a,abj06b,nmlh03,pb05,dhm07}.
Here we summarize the main results and discuss how they help to understand  the synchronizability of complex networks.
Some detailed proof of the results can be found in the corresponding references and in
monographs on graph theory, e.g. \cite{fiedler73,am85,m91,chung97}.

First, we  discuss the bounds for networks with prescribed degree sequences
$k_{\min}=k_1\le k_2 \le \cdots \le k_N=k_{\max}$.
The bounds satisfy the following relations (see \cite{fiedler73,am85})
\begin{equation}
2 \left( 1-\cos(\frac{\pi}{N}) \right) e(G) \le \lambda_2 \le \frac{N}{N-1} k_{\min},
\label{w05_eq1}
\end{equation}
and
\begin{equation}
\frac{N}{N-1}  k_{\max} \le \lambda_N \le 2 k_{\max},
\label{lambdaN_bound}
\end{equation}
where $e(G)$ is the edge connectivity of the graph, i.e. the minimal number of edges whose removal would
result in losing connectivity of the graph.
It follows that
\begin{equation}
 \frac{k_{\max}}{k_{\min}} \le R \le  \frac{k_{\max}}{\left( 1-\cos(\frac{\pi}{N}) \right) e(G)}.
\label{boundR1}
\end{equation}
Note that the edge connectivity $e(G)\le k_{\min}$. The equality holds only
for special cases, in particular for networks which are homogeneous in degree.
Then for a network with a fixed number of nodes and edges the edge connectivity
ranges from 1 (a structured network with communities connected through single links)
to a maximum value for a homogeneous network, and this turns out in the large variability of
the lower bound for $\lambda_2$.

The upper bound in Eq.~(\ref{w05_eq1})  is approached when the network is random. As shown in \cite{w03,w05},
in a $k$-regular  random network, where each node  is randomly connected to other $k=k_{\min}$ nodes in the network,
$\lambda_2=k-O(\sqrt{k})$ as $N\to \infty$. In fact, $\lambda_2=c_k k$, where  $c_k \to 1 $ as $k\to \infty$~
both for  $k$-regular networks and random SF networks~\cite{w03}. In this sense, the observation in \cite{wc02} that
$\lambda_2$ is almost constant, practically unrelated to $k_{\min}$, is incorrect.
So for large random networks with large enough minimal degree, $\lambda_2\approx k_{\min}$ is a good approximation.

Another general lower bound for $\lambda_2$ is \cite{mohar91}
\begin{equation}
\frac{4}{ND} \le \lambda_2  \le \frac{N}{N-1} k_{\min},
\label{pb05_eq1}
\end{equation}
so that
\begin{equation}
 \frac{k_{\max}}{k_{\min}} \le R \le   \frac{ND k_{\max}}{2},
\label{boundR2}
\end{equation}
where $D$ is the diameter of the graph, i.e. the maximum value of the shortest path lengths between any two nodes.

Let us discuss the implications of the above bounds for WS networks.
In this model, when the probability of having shortcuts is very low, $pkN \ll 1$, then  $D \sim N/k$
and the lower bound in Eq.~(\ref{pb05_eq1}) approaches zero as $1/N^2$ (regular  network).
Beyond the onset of the SW regime $(pkN\sim 1)$,  $D$ decreases and  approaches $D \sim  \ln (N)$,
and $\lambda_2$ increases for fixed $N$. Thus  this bound allows us to understand why the network
synchronizability is inversely proportional to $\ell$,
as observed numerically  in \cite{bp02,hkcp04} (Figs.~\ref{bp02_fig2} and \ref{hkcp04_fig1}).
However, $\lambda_2$ is not immediately bounded away from zero, since it approaches zero faster than $1/N$ just after
the onset of the SW regime.
When moving  deeper into the SW region, $pkN\sim N$ so that
each node has $S\sim 1$ shortcuts, and $\lambda_2$ is already bounded away from the lower bound
$4/ND$ and approaching the upper bound $k_{\min}$. In this regime, $\lambda_2$ will not be sensitive to changes in the diameter $D$. This helps  to understand why the synchronization threshold is different from the onset of the SW  behavior in Fig.~\ref{bp02_fig3}.

Thus, in WS networks with high $p$, $k_{\max}/k_{\min}$ provides a good lower bound to the eigenratio $R$. The upper bound,  $NDk_{\max}/2$,
still increases with $N$, and can be several orders of magnitude larger than the actual value of $R$ even for small random networks.
In fact, $R$ may not follow the variation of the distant upper bound when the diameter $D$ or the size $N$ change.
Thus, in complex networks with both local and random connections, a close relationship between the synchronizability
and $\ell$ is not expected.

In \cite{nmlh03} similar  bounds are obtained but as a function of  $\ell$ and  $b_{\max}$ as
\begin{equation}
\frac{k_{\max}}{k_{\min}} \le R \le Nk_{\max}b_{\max} D \ell.
\label{nmlh03_eq1}
\end{equation}
These authors pointed out that experimental values of $R$ are closer to the lower bound, and
far away from the upper bound (Fig.~\ref{nmlh03_fig1}).
Thus quite probably such upper bound  \label{nmlh03} does not provide
meaningful understanding of the relationship between  synchronizability and $\ell$, $D$
or $b_{\max}$, since the change of the upper bound by these structural measures is likely not to be reflected in the actual value of  $R$.
Indeed, for the SF model considered in this paper and arbitrary random enough networks with suffiently large mimimal degrees,
recent results\cite{zmk06}  showed that the eigenvalues are bounded as
\begin{eqnarray}
k_{\min}(1-\frac{2}{\sqrt{\langle  k \ra}}) \lesssim \lambda_2 \lesssim  k_{\min} \nonumber, \\
k_{\max}\le \lambda_N \le k_{\max} (1+\frac{2}{\sqrt{\langle  k \ra}}),
\label{tight_bounds}
\end{eqnarray}
which can explain more clearly how the eigenvalues in  Fig.~\ref{MSF_fig_cs2} and in \cite{nmlh03} depend precisely on the maximal degree $k_{\max}$ and the mean degree $\langle  k \rangle$  . A recent advance \cite{km07} in spectral analysis of random SF networks
shows that the maximal eigenvalue is very close to the lower bound in Eq.~(\ref{lambdaN_bound}), $\lambda_N \simeq k_{\max}+1$.
The observation in \cite{nmlh03} that the synchronizability (Type I) is anti-correlated with a more heterogeneous load
distribution, and smaller distance, is a consequence of the positive correlations between the load and the degree, and
negative correlation between the distance and the degree, in the models considered.
Indeed, when keeping the degree sequence unchanged, it is observed  that more heterogeneous
load distributions can be positively correlated with synchronizability of Type I~\cite{zzwyyb06}.

\begin{figure}
\begin{center}
\epsfig{figure=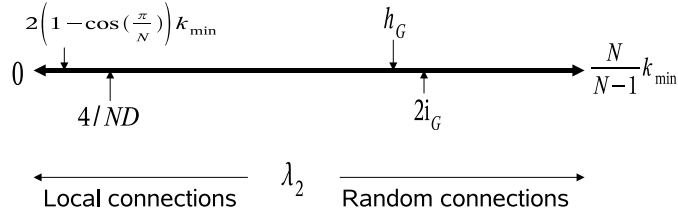,width=0.55\textwidth}
\end{center}
\caption{ A schematic plot of the bounds of $\lambda_2$ for  networks with minimal degree $k_{\min}$.
}
\label{MSF_fig_cs4}
\end{figure}

The above analysis of bounds provides justification about why we can observe different $R$ for
increased or decreased heterogeneity, distances or loads.
Moreover, the bounds expressed by these quantities are not
tight at all in the particular examples, e.g., in Eq. (\ref{nmlh03_eq1}). In general, the graph theoretical analysis
states that randomness improves  the synchronizability, since $\lambda_2$
is well  bounded away from $0$, while in networks with dominantly  local connections, $\lambda_2$ approaches to  0 in
large networks. In other words, for a prescribed degree sequence, the eigenratio $R$ changes mainly because of $\lambda_2$.
A schematic plot of the bounds of $\lambda_2$ is shown in Fig.~\ref{MSF_fig_cs4}.

On the other hand, the local organization of the connections even within
a small part of the network, can make the eigenvalue $\lambda_2$ deviate significantly from the upper bound $k_{\min}N/(N-1)$. For an arbitrary network, there is a better upper bound for $\lambda_2$ as
\begin{equation}
\lambda_2 \le 2i_\mathcal{G},
\label{boundiG}
\end{equation}
where $i_\mathcal{G}$ is the isoperimetric number of a graph~\cite{m89,abj06a,abj06b},  which is defined as
\begin{equation}
i_\mathcal{G}=\min_\mathcal{S} \frac{|\partial \mathcal{S}|}{|\mathcal{S}|}.
\end{equation}
Here  ${\mathcal S} \subset {\mathcal G}$ is a subset of the nodes, with ${\mathcal G-S}$ denoting its complement, and $|\mathcal{S}|$ is the  number
of nodes within $\mathcal{S}$, with $0<|\mathcal{S}| <N/2$.  Besides, $|\partial \mathcal{S}|$ is the number of connections
between $\mathcal{S}$ and its complement,
namely,
\begin{equation}
|\partial \mathcal{S}|=\sum_{i\in \mathcal{S}} \sum_{j\in (\mathcal{G-S})} a_{ij}.
\end{equation}
For an arbitrary partition of the network into $\mathcal{S}$ and $\mathcal{G}$, we  have~\cite{abj06b}
\begin{equation}
\lambda_2 \le 2\frac{|\partial \mathcal{S}|}{|\mathcal{S}|}.
\end{equation}
This  means that the synchronizability of the network is determined
by the  sparse connections between the two subnetworks.
For instance, if a small set $\mathcal{S}$ made up of, say, 20 nodes, is connected to $\mathcal{G}-\mathcal{S}$ with just one link, then
$\lambda_2<0.1$, regardless of how the nodes are connected within $\mathcal{S}$ and within the large complement $\mathcal{G}-\mathcal{S}$.
It follows that
the statistical properties of the network $\mathcal{G}$ are
mainly determined by the huge part $\mathcal{G}-\mathcal{S}$, while $\lambda_2$ is independently constrained by the small subgraph. This result is in complete agreement with the path towards synchronization in modular networks presented in section \ref{kmsmn}, where the community structure has been demonstrated to impose scales in the synchronization process.

Everything up to now indicates that statistical properties, such as degree distribution, $\ell$, etc., may not be always
correlated with the synchronizability of the network. In fact, it was particularly shown in \cite{w05,abj06a,abj06b},
that networks with very different synchronizability can be constructed for the same prescribed degree sequences, because $i_\mathcal{G}$
can be at any place in a broad range between the lower and upper bounds in Eq.~(\ref{w05_eq1}), see also Fig.~\ref{MSF_fig_cs4}.


Another similar bound in graph theory is based on the Cheeger
inequality~\cite{c70,dgs06},
\begin{equation}
\lambda_2\le  h_{\mathcal{G}},
\label{boundhG}
\end{equation}
where $h_{\mathcal{G}} =\min_{\mathcal{S}} h_{\mathcal{G}}(\mathcal{S})$ and
\begin{equation}
h_{\mathcal{G}}(\mathcal{S})=\frac{|\partial \mathcal{S}| N}{|\mathcal{S}| (N-|\mathcal{S}|)}.
\end{equation}

As for correlated networks, in \cite{dgs06} the authors showed that in random networks with degree correlations $r_k$
\begin{equation}
\frac{\partial h_{\mathcal{G}}(\mathcal{S})}{\partial r_k}<0,
\end{equation}
which explains why the synchronizability is reduced in networks with assortative connections, as
shown in Fig.~\ref{dgs06_fig1}. Along the same line, the dependence of the minimum nonzero eigenvalue on the topological properties of the network and its degree-degree correlation coefficient $r$ was also analyzed in \cite{dhm07}. The authors derived a rigorous upper bound for $\lambda_2$ as,
\begin{equation}
\lambda_2\le(1-r)\frac{\langle   k \rangle \langle   k^3 \rangle-\langle   k^3 \rangle \langle   k^2 \rangle}{\langle   k \rangle(\langle   k^2 \rangle-\langle   k \rangle^2)}.
\end{equation}

It is worth mentioning also the inequality presented in~\cite{m89,dgs06},
\begin{equation}
\lambda_2\ge k_{\max}-\sqrt{k_{\max}^2-i_{\mathcal{G}}^2}.
\label{dgs06_eq1}
\end{equation}
Note that  $k_{\max}-\sqrt{k_{\max}^2-i_\mathcal{G}^2}$ is a decreasing function of $k_{\max}$ if $i_\mathcal{G}$ is fixed.
Based on this bound, we can expect  some apparently   counterintuitive effects on the synchronizability of complex networks.
Suppose we put more links to the graph $\mathcal{G}$, but only add them to the nodes within
$\mathcal{G-S}$ and $\mathcal{S}$, but not to the nodes between $\mathcal{S}$ and $\mathcal{G-S}$ so that $i_\mathcal{G}$ does not change.
For simplicity, we  can further  assume  that the nodes within $\mathcal{S}$ and $\mathcal{G-S}$ are connected with a uniform probability $f$ (random networks), so that
$k_{\max}$  and the mean degree $\langle  k \rangle$   increase when more and more links are added.
In this case, the synchronizability of the subnetworks $\mathcal{S}$ and
$\mathcal{G-S}$ is enhanced at larger $f$, see Fig.~\ref{bp02_fig2}. However, $\lambda_2$ of the whole network $\mathcal{G}$ is
reduced according to Eq.~(\ref{dgs06_eq1}). Therefore,  in the case of two coupled networks, enhancing the synchronizability
of the subnetworks may actually  reduce  the synchronizability of the whole network. Phenomenologically, this is intuitively
expected, because the subnetworks tend to form distinct synchronized clusters.

Based on the above arguments (Eqs. (\ref{boundiG},\ref{boundhG},\ref{dgs06_eq1})),
networks possessing a clear community organization display a small synchronizability,
since the density of connection between
different communities can be much smaller than the density  within the
communities~\cite{hplyy06,zzcyw06,ad07}.

Recapitulating, we have seen that for a prescribed degree sequence, it is possible to construct a very large
number
of networks ranging from fully local connections to fully random networks~\cite{w05}, with many possible
structures in between.
However, the degree sequence by itself is not sufficient to determine the synchronizability.
On the other hand, we have seen that the synchronizability is not directly related to graph measures, such as distance, clustering or
maximal betweenness.
Admittedly, the weak connections
between two subnetworks
(characterized by the isoperimetric number $i_\mathcal{G}$)
determine the behavior of the eigenvalue $\lambda_2$,
and hence that of the synchronizability of the whole network.
The bounds discussed in this section are valid for any network. However, we
would like to  point out that one needs to be careful in the interpretation of these  general analytical results.
A linear relationship between the bound and some network descriptor does not mean that we can always expect
the same relation to hold between these descriptors and the actual eigenvalues for particular types of networks.
The reason is that the bounds, the network descriptors and the eigenvalues of the Laplacian can follow totally different scaling laws
in particular types of networks.  For example, for random SF networks with large enough minimal degree, $R\approx k_{\max}/k_{\min} \sim N^{1/(\gamma-1)}$ from Eq.~\ref{tight_bounds},
while the upper bound in Eqs.~(\ref{boundR2},\ref{nmlh03_eq1}) increase faster than $N\cdot N^{1/(\gamma-1)}$.   The numerical results in \cite{nmlh03}
showed  that the eigenratio $R$ and the load follow different scaling laws also. This indicates that one must be cautious not to generalize the observation of such correlations in one type of network, not to interpret such observed correlations as the ultimate responsible for the synchronizability without a deep analysis of their constrains.

\subsubsection{Synchronizability of weighted networks}

Up to now, we have considered the influence of the network topology on synchronization,
assuming that the connection weights are the same for all the links in the network, i.e., the networks are unweighted.
However, this is not the case for many real-world networks.  Indeed, many complex networks where
synchronization is relevant are actually  weighted and display a
highly heterogeneous distribution of both degrees and
weights~\cite{bbpv04,yjbt01,n01,bbchs03}. Examples include neural
networks~\cite{fv91, sbhoy99},  airport networks~\cite{bbpv04} and the structure of the networks characterizing epidemic outbreaks in different cities~\cite{gbk01,gfg05}.
Furthermore, it has been observed that in many cases, the connection strength is not an independent
parameter, but  it is correlated to the network topology.  The
analysis of some real networks~\cite{bbpv04} yields the following
main properties:

(i) the weight $w_{ij}$ of a connection between nodes $j$ and $i$ is
strongly correlated with the product of the corresponding degrees as
$\langle   w_{ij}\rangle \sim (k_ik_j)^\theta$;

(ii) the average intensity $S(k)$ of nodes with degree $k$ increases as
$S(k)\sim k^{\beta}$.
Here the intensity $S_i$  of a node $i$ is defined as the total input weight of the node:
\begin{equation}
S_i=\sum_{j=1}^N a_{ij}w_{ij}.
\label{S_i}
\end{equation}

Note that the inclusion of a distribution of weights in the network affects directly its classification within topological
homogeneity or heterogeneity. For example, a regular lattice with a very skewed distribution of weights can eventually  represent a SF topology. From a mathematical point of view, the adjacency matrix is in this case simply substituted by the weight matrix. On the contrary, from a physical perspective, it is still interesting to keep separated the topology of interactions from the distribution of weights, and answer questions, whenever possible, discriminating these two topological aspects.

The first works on synchronization in  weighted networks considered that the weighted input of  a
node $i$ from a node $j$ depends on the degree $k_i$  of node $i$~\cite{mzk05a,mzk05b}, with a model
of weighted coupling as $w_{ij}=k_i^{\theta}$, so that the matrix $G=(G_{ij})$  in Eq. (\ref{MSF_eq_2}) reads
\begin{equation}
G=D^{\theta} (D-A)=D^{\theta} L.
\label{mzk05_eq1}
\end{equation}
Here $D_{ij}=\delta_{ij}k_i$ is the diagonal matrix of degrees.
$\theta$ is a tunable parameter that keeps the network topology unchanged, but varies the distribution
of the weights of the links.

Within this scheme, the weights between a pair of nodes $i$ and $j$
are in general asymmetric, because $w_{ij}=k_i^{\theta}$ and $w_{ji}=k_j^{\theta}$.  However,
since
\begin{equation}
\det (D^{\theta}L-\lambda I)=\det
(D^{\theta/2}LD^{\theta/2}-\lambda I),
\end{equation}
is valid for any $\lambda$,  the spectrum of
eigenvalues of the matrix $G$ is equal to the spectrum of a
symmetric matrix defined as $H=D^{\theta/2}LD^{\theta/2}$.  As a
result, all the eigenvalues of  $G$ are real, and the synchronizability can still be characterized in the framework of the MSF.

\begin{figure}
\begin{center}
\epsfig{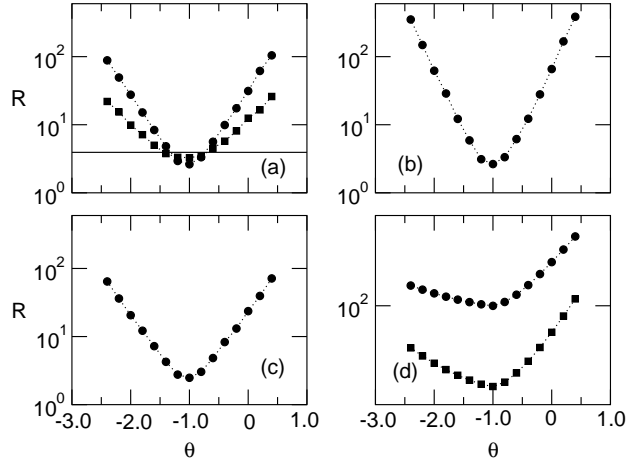}
\end{center}
\caption{Eigenratio $R$ as function of $\theta$ in Eq.~(\ref{mzk05_eq1}): (a) random SF networks with
$\gamma=3$ $(\bullet)$, $\gamma=5$ ($\blacksquare$) and $\gamma=\infty$ (solid line), for $k_{\min}=10$; (b) random networks
with expected SF sequence for $\gamma=3$ and $\tilde{k}_{\min}=10$; (c)
growing SF networks for $\gamma=3$ and $m=10$; (d) SW networks with $M=256$ ($\bullet$)
and $M=512$ ($\blacksquare$), for $k=1$. Each curve is the result of an average
over 50 realizations for $N=1024$. Modified after \cite{mzk05b}.
}
\label{mzk05_PRE_fig1}
\end{figure}

The synchronizability of various complex networks as a function of the parameter $\theta$ is
shown in Fig.~\ref{mzk05_PRE_fig1}. Except for k-regular networks, in all other cases, including
the  SW networks, the eigenratio $R$  exhibits a pronounced minimum at $\theta=-1$. Here the SW networks are obtained by adding $M\le N(N-2k -1)/2$ new links between randomly
chosen pairs of nodes on the basic regular array  where each node is connected to its $2k$
first neighbors.

\begin{figure}
\begin{center}
\epsfig{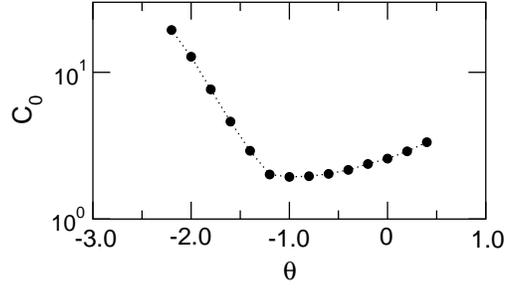}
\end{center}
\caption{Normalized cost $C_0$ as a function of $\theta$ in Eq.~(\ref{mzk05_eq1}) for random SF networks  with
$\gamma=3$,  $k_{\min}=10$ and $N=1024$. Modified after \cite{mzk05a}.
}
\label{mzk05_PRE_fig5}
\end{figure}

In \cite{mzk05a,mzk05b} the authors also characterize the synchronizability of the network,
related to $\lambda_2$, using the notion of the {\em cost} of the network.  When Eq.~(\ref{MSF_R})
is satisfied, the fully synchronized state is linearly stable for
$\sigma> \sigma_{\min}\equiv\alpha_1/\lambda_2$. The cost is defined as the total input strength of the connections of all
nodes at the synchronizability threshold:
$\sigma_{\min}\sum_{i,j} w_{ij}a_{ij}=\sigma_{\min}\sum_{i=1}^NS_i$.
A more convenient definition for comparisons is obtained normalizing by the number of nodes, such that
\begin{equation}
C_0\equiv \frac{\sigma_{\min}\sum_{i=1}^NS_i}{N\alpha_1}=\langle  S \rangle /\lambda_2,
\label{cost}
\end{equation}
where $\langle  S \rangle $ is the average intensity of nodes in the network.
Similar to $R$, $C_0$ is also minimal at $\theta=-1$ (Fig.~\ref{mzk05_PRE_fig5}).

\begin{figure}
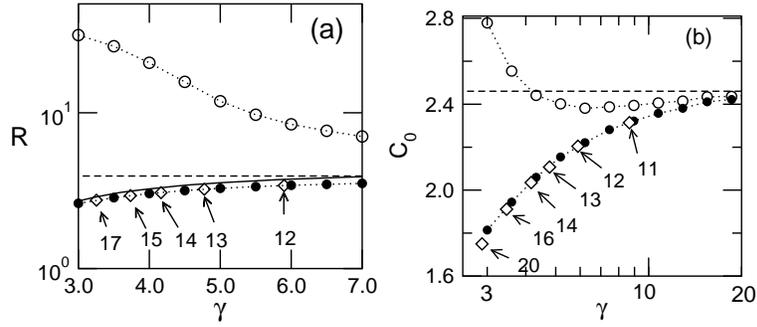

\begin{center}
\epsfig{figure=fig19a.eps,width=0.3\textwidth,clip=,}
\epsfig{figure=fig19b.eps,width=0.3\textwidth,clip=,}
\end{center}
\caption{Eigenratio $R$ (a) and normalized cost $C_0$ (b) as a function of $\gamma$  for random SF networks  with
with   $\theta=0$ ($\circ$) and $\theta=-1$ ($\bullet$) and for random homogeneous networks with the same average degree $\langle  k \rangle$
($\diamond$).  The other parameters are $k_{\min}=10$ and $N=1024$.
The dashed line corresponds to $\gamma=\infty$ ($\langle  k \rangle =10)$. The solid line in (a)  is the approximation given by Eq.~\ref{eq25}.
Modified from \cite{mzk05b}.
 }
\label{mzk05_PRE_fig2}
\end{figure}

Interestingly enough, in \cite{mzk05a,mzk05b} it was obtained that in SF networks with fixed minimal degrees $k_{\min}$, the weighted
versions ($\theta=-1$) behave differently to the unweighted networks when one looks at the dependence of both the eigenratio $R$ and the cost $C_0$ on the scaling exponent $\gamma$,  as shown in Fig.~\ref{mzk05_PRE_fig2} .

$\theta=-1$ is a special case.  The coupling matrix is now $G=D^{-1}L$, and
all the diagonal elements  $G_{ii}\equiv 1$. It is usually called
the {\it normalized Laplacian}  of a graph.
Based on graph spectral analysis results in \cite{clv03} for random networks with arbitrary
given degrees, it can be shown that the spectrum of the normalized Laplacian tends to the semicircle law for large
networks. In particular, for $k_{\min}\gg \sqrt{\langle  k \ra}\ln^3N$, one has
\begin{equation}
\max \{1-\lambda_2,\lambda_N-1\}= [1+\mathcal{O}(1)]\frac{2}{\sqrt{\langle  k \ra}}.
\label{eq24}
\end{equation}
This result is rigorous for ensembles of networks with a given expected
degree sequence and sufficiently large minimum degree $k_{\min}$, but the  numerical
results reported in Fig.~\ref{mzk05_PRE_fig2} support the hypothesis
that the approximate relations
\begin{equation}
\lambda_2 \approx 1 - 2/\sqrt{\langle  k \ra}, \;\;\;
\lambda_N \approx 1 + 2/\sqrt{\langle  k \ra}
\label{eq25}
\end{equation}
hold under much milder conditions. In particular, the relations (\ref{eq25}) are
expected to hold for any large network with a sufficient number of random connections, $k_{\min}\gg 1$.

Furthermore, the synchronizability in this case seems to be independent of the degree distribution.
It is only  controlled by the average degree $\langle  k \rangle$  , since the synchronizability of the weighted SF networks is
almost identical to that of a regular random network where each node has the same degree $k_i=\langle  k \rangle$  .
These results demonstrate that the topological degree of networks is not the only determinant of the synchronizability
of the networks; having a heterogeneous distribution for the connection strengths can significantly influence the synchronizability.

\subsubsection{Universal parameters controlling the synchronizability}

What are the leading parameters governing the synchronizability for
the more general case in which weighted networks are considered?
It \cite{zmk06} it has been  proved analytically and verified numerically what
controls  the synchronizability of sufficiently random networks with large enough
minimal degree ($k_{\min}\gg 1$). It is the distribution of the intensities $S_i$ defined in Eq.~(\ref{S_i}). The intensity of a node incorporates
both the information about the topology and  the weights of the connections  in  the networks.  The main finding is  that
the synchronizability is  sensitively controlled by the heterogeneity of the intensity $S_i$.
The eigenratio $R$  and the normalized cost $C_0$ can be expressed as
\begin{equation}
R=A_{R} \frac{S_{\max}}{S_{\min}}, \;\;\;\;\; C_0=A_{C} \frac{\langle  S \ra}{S_{\min}},
\label{fit_all}
\end{equation}
where $S_{\min}$,  $S_{\max}$, and $\langle  S \rangle$   are the minimum, maximum and
average intensities, respectively.
The  pre-factors $A_R$ and $A_C$  are expected to approach 1 for large average degree $\langle  k \rangle$  .
Equations (\ref{fit_all}) are universal in the sense that they apply to many random networks with arbitrary degree and weight
distributions provided that the minimal degree is sufficiently large. The main hypothesis behind this result is the assumption that the
local mean fields ${\bf \tilde H}_i=(1/k_i)\sum_j a_{ij} {\bf H(x_j)}$ can be substituted by the global mean field ${\bf \tilde H}_i \approx {\bf \tilde H}=(1/N)\sum_j {\bf H(x_j)}$.

\begin{figure}
\begin{center}
\epsfig{figure=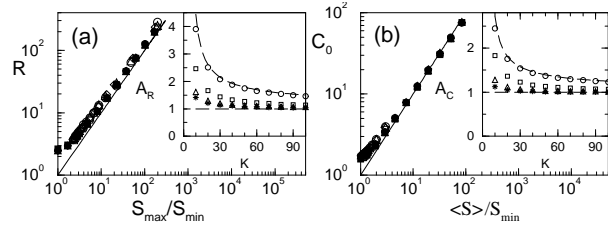,width=0.48\textwidth, clip=,}
\caption{(a) $R$ as a  function of $S_{\max}/S_{\min}$ and (b)
$C_0$ as a function of $\langle  S \ra/S_{\min}$, averaged over 50 realizations of the networks.
Filled symbols: uniform distribution of $S_i \in [S_{\min},S_{\max}]$.
Open symbols:  power-law  distribution of $S_i$, $P(S)\sim S^{-\Gamma}$
for $2.5 \le\Gamma\le 10$.
Different symbols are for networks with  different topologies:
BA networks ($\circ$), growing SF network  with aging exponent $\alpha=-3$ ($\Box$),
 random SF network with $\gamma=3$ ($\diamond$),
and $k$-regular random networks ($\triangle$). The number of nodes is $N=2^{10}$ and the average degree is $\langle  k \ra=20$.
Insets of (a) and (b): $A_R$ and $A_C$ as functions of $\langle  k \rangle$   for
$S_{\max}/S_{\min}=1$ $(\circ)$, $2$ $(\Box)$, $10$ $(\triangle)$,  and  $100$ $(\ast)$, obtained
with  uniform distribution of $S_i$ in $k$-regular networks.
The dashed lines are  the  bounds.
Solid lines in (a) and (b): Eqs.~(\ref{fit_all}) with $A_R=A_C=1$.
From \cite{zmk06}.
}
\label{zmk06_fig1}
\end{center}
\end{figure}

In Fig.~\ref{zmk06_fig1}, Eq.~(\ref{fit_all}) is
corroborated by numerical results of $R$ and $C_0$ for networks with several degree and intensity distributions of degrees, using the weighted coupling scheme
\begin{equation}
w_{ij}=S_i/k_i.
\label{zmk06_eq14}
\end{equation}
This coupling scheme means that the intensity of the nodes, which is not necessarily correlated with the degrees, is uniformly
distributed into the input links of the nodes. It covers the coupling scheme in Eq. (\ref{mzk05_eq1}) as a special example: $S_i=k_i^{1+\theta}$. Recent analysis in \cite{kk07} on the spectral density of SF networks with a weighted Laplacian matrix
similar to Eq.~(\ref{mzk05_eq1}),  also confirms that Eq.~(\ref{fit_all}) holds.

\begin{figure}
\begin{center}
\epsfig{figure=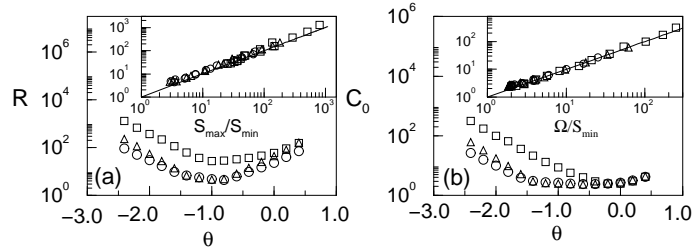,width=0.55\textwidth, clip=,}
\caption{ (a) Eigenratio $R$ and (b) cost $C_0$ as functions of $\theta$
for BA networks  ($\circ$), growing SF networks with aging exponent $\alpha=-3$ ($\Box$)
and random SF networks  with $\gamma=3$ ($\triangle$).
Each symbol is an average over 50 realizations of the networks with  $\langle  k \rangle =20$ and $N=2^{10}$.
Inset of (a): the same data for  $R$ as a function of $S_{\max}/S_{\min}$.
Inset of (b): the same data for  $C_0$ as a function of $\langle  S \ra/S_{\min}$.
Solid lines:  Eqs.~(\ref{fit_all}) with $A_R=A_C=1$. From \cite{zmk06}.
}
\label{zmk06_fig2}
\end{center}
\end{figure}

In \cite{zmk06} the authors also presented results for the
following coupling scheme
\begin{equation}
w_{ij}=(k_ik_j)^{\theta},
\label{real_weight}
\end{equation}
that describes the relationship between the weights and the degrees in some real networks~\cite{bbpv04,mab05}. The tunable parameter $\theta$ controls  the heterogeneity of  the intensity $S_i$ and  the
correlation between $S_i$ and $k_i$,
since $S_i=k_i^{1+\theta} \langle   k_j^\theta\rangle_i$,
where $\langle   k_j^\theta\rangle_i =(1/k_i)\sum k_j^\theta$ is  approximately
constant for $k_i\gg 1$ when the degree correlations can be neglected.
Variations of  $\theta$ have a significant impact on the synchronizability of
networks which are heterogeneous in degree(Fig.~\ref{zmk06_fig2}).
Note that in heterogeneous in degree networks, the weighted coupling in Eq.~(\ref{real_weight}) may result
in a broad distribution of the  input  weights $w_{ij}$ among the $k_i$ links of the node $i$, especially when
$\theta$ is not close to $0$.
However, as shown in the insets of  Fig.~\ref{zmk06_fig2} for various networks
and $\theta$ values, $R$ and $C_0$  collapse again to the
universal curves when regarded as functions of $S_{\max}/S_{\min}$ and $\langle  S \ra/S_{\min}$,  respectively.
The fact that the universal formula holds for a broad range of $\theta$ values shows that the mean field approximation used to obtain Eq.~(\ref{fit_all}) often remains valid under milder conditions.

\begin{figure}
\begin{center}
\epsfig{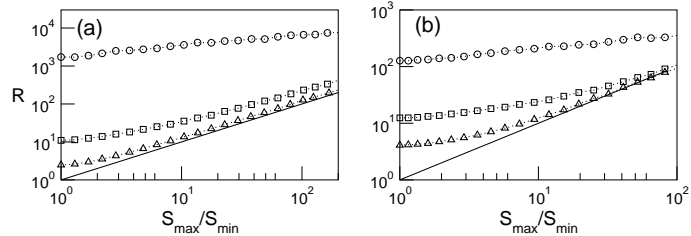}
\caption{ (a) Eigenratio $R$ as a function of $S_{\max}/S_{\min}$ for BA growing networks
with $m=1 (\langle  k \rangle =2)$ $(\circ)$, $m=2 (\langle  k \ra=4)$ $(\Box)$ and $m=10 (\langle  k \rangle =20)$ ($\bigtriangleup)$. (b)
$R$ for SW networks with $\langle  k \ra=20$ and rewiring probability $p=0.01$ ($\circ)$, $p=0.1$ $(\Box)$ and $p=0.3$ ($\bigtriangleup)$.
Solid lines:  Eqs.~(\ref{fit_all}) with $A_R=A_C=1$.
}
\label{Fig_SF_SW}
\end{center}
\end{figure}

It is important to stress that
these  results also hold for unweighted random networks.
In this case $S_i=k_i$ for all the nodes and one gets the results in Eq.~(\ref{tight_bounds})
for large random networks with  minimal degree $k_{\min}\gg 1$.
These results provide much tighter bounds  than those discussed in the previous
sections (e.g.,  cf. Eqs.~(\ref{boundR1},\ref{boundR2}, \ref{nmlh03_eq1})) which depend on the system size $N$. Interestingly, Eqs.~(\ref{fit_all}) also provide meaningful insights into the problem for other special networks. For example, consider the class of SF networks generated using the BA model. When $m=1$, the network is a tree,
and $R$ is much larger than $S_{\max}/S_{\min}$. However, it is an increasing function of $S_{\max}/S_{\min}$, showing that the heterogenity of the intensity
is still an important parameter. But for $m=2$ ($\langle  k \rangle =4$) it approaches the universal curve quickly
(Fig.~\ref{Fig_SF_SW} (a)). The drastic change of  synchronizability from $m=1$ to $m=2$
can be  attributed to the appearance of loops~\cite{ym06}.
In  WS networks \cite{ws98} with $N=2^{10}$ and  $\langle  k \ra=20$,
$R$ collapses to the universal curve
even when the networks are dominated by local connections, e.g.
for a rewiring probability $p= 0.3$ with intensities $S_{\max}/S_{\min}\gtrsim 10$,
see Fig.~\ref{Fig_SF_SW} (b).

\subsection{Design of synchronizable networks}

An interesting subject related to the impact of network structure on synchronization dynamics is the design of
synchronizable networks. Here we review several ideas exploring this issue: weighting
the couplings leaving the topology unchanged, perturbing  part of the network topology,
and finally searching  for optimal topologies with respect to synchronizability. Note that the following theoretical schemes
may not directly apply to real complex networks. It is difficult to conceive real systems where the weights can be tuned
at discretion, or where the topological substrate of interactions can be changed accordingly. Nevertheless, the insights given by these
works allow for a deeper understanding of the synchronizability of networks.

\subsubsection{Weighted couplings for enhancing synchronizability}

The previous analysis shows that for networks that are heterogeneous in degree, synchronizability can be enhanced by balancing
the heterogeneity in the degree distribution with suitable weighted couplings, towards the obtention of a homogeneous
distribution of the intensities $S_i$.

\begin{figure}
\begin{center}
\epsfig{figure=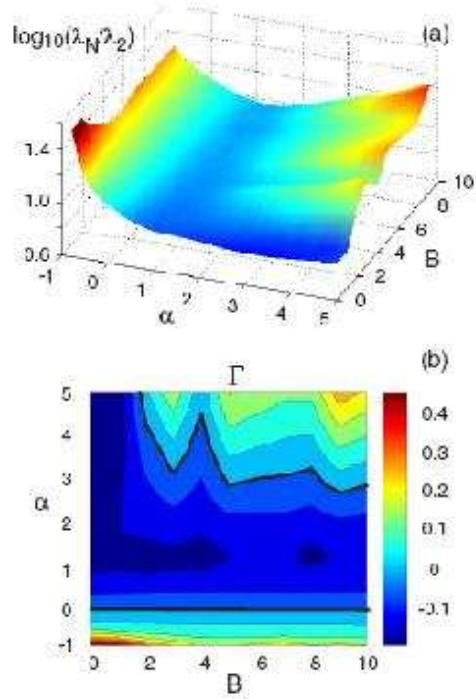,width=0.4\textwidth, clip=,}
\caption{(color online) (a)  eigenratio $\lambda_N/\lambda_2$
 (in logarithmic scale) for SF networks in the parameter space ($\alpha,  B$). (b)The relative synchronizability
$\Gamma=\log(\lambda_N/\lambda_2)-[\log(\lambda_N/\lambda_2)]_{\alpha=0}$
vs ($\alpha,  B$). In all cases $m=2$, and the graphs
refer to averaging over 10 realizations of networks with $N=1000$.
The domain with with $\Gamma < 0$ is outlined by the black contours
drawn on the figure. From  \cite{chahb05}.
}
\label{Fig_chahb05_fig2}
\end{center}
\end{figure}

A different scheme is presented in \cite{chahb05}, where it is assumed that the weight of a link is related to its betweenness $b_{ij}$ as

\begin{equation}
G_{ij}=-\frac{b_{ij}^{\alpha}}{\sum_{j\in \Gamma_i} b_{ij}^{\alpha}},
\label{chahb05_eq}
\end{equation}
where $\alpha$ is a tunable parameter that controls the dependence of the weights $w_{ij}$ on the loads $b_{ij}$.
The zero-sum requirement of the matrix $G$ implies that $G_{ii}=1$ for all $i$. Note that $\alpha=0$ corresponds to  the weighted coupling scheme
in Eq.~(\ref{mzk05_eq1}) at the optimal point $\theta=-1$. As seen in Fig.~\ref{Fig_chahb05_fig2}, the eigenratio $R$
depends  on $\alpha$, reaching a minimum at a value $ 0<\alpha \lesssim 1$, showing that the synchronizability in SF networks can be
slightly enhanced compared to the optimal case of the  weighted coupling scheme \cite{mzk05a,mzk05b,zmk06}.
The SF network  considered in \cite{chahb05} is a generalized BA model with a preferential attachment probability
$\pi_i \sim k_i+B$~\cite{dms00},
where the parameter $B$ controls the exponent
$\gamma=3+B/m$ of the power law degree distribution, and $m=k_{\min}$.
At large $\alpha$ values, only the links with the largest loads $b_{ij}$  are  significant,
which can lead to effectively disconnected nodes, so that synchronizability
is reduced. In \cite{chahb05} it is also pointed out that for large minimal degrees,
the regimes corresponding to enhanced synchronizability are
reduced so that the minimum approaches to  $\alpha=0$.  This demonstrates again that
for random networks with large enough minimal  degree, Eq.~(\ref{eq25}) is  asymptotically valid  regardless of the
detailed weighted scheme, as claimed in \cite{zmk06}.

\begin{figure}
\begin{center}
\epsfig{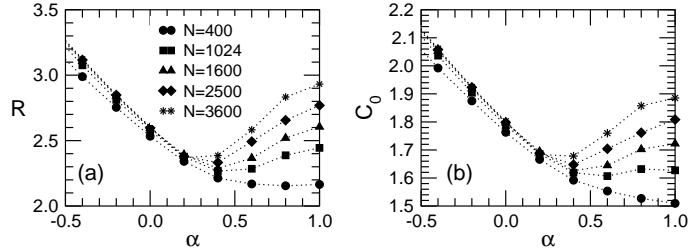}
\caption{Eigenratio $R$  (a) and normalized cost $C_0$ (b) as a function of the parameter $\alpha$
in Eq.~(\ref{AIP_eq}), for random SF networks  with $\gamma=3$ and $k_{\min}=10$. From \cite{mzk05c}
}
\label{Fig_AIP_MSF}
\end{center}
\end{figure}

The explanation of the observed enhanced synchronizability proposed in \cite{chahb05} is that the load $b_{ij}$
reflects the global information of the network, while at $\alpha=0$ only the local information (degree)
is employed. Such a heuristic explanation, however,  is not supported by several further investigations.
In fact, only the local information can also lead to similar enhanced synchronizability.
For example, consider that the weights depend
on the degrees following Eq.~(\ref{real_weight}), and then normalize to allow fully uniform intensity $S_i=1$, namely \cite{mzk05c}
\begin{equation}
G_{ij}=-\frac{(k_ik_j)^{\alpha}}{\sum_{j\in \Gamma_i} (k_ik_j)^{\alpha}}=
-\frac{k_j^{\alpha}}{\sum_{j\in \Gamma_i} k_j^{\alpha}}.
\label{AIP_eq}
\end{equation}
Again, $\alpha=0$ corresponds to the optimal case (Eq.~(\ref{mzk05_eq1}), $\theta=-1$) of
the  weighted coupling scheme \cite{mzk05a,mzk05b,zmk06}.
Similar to \cite{chahb05}, the synchronizability can be further enhanced in a range $\alpha>0$ (Fig.~\ref{Fig_AIP_MSF}).
However, the minimum moves closer to $\alpha=0$ when the  networks are larger, indicating that the
synchronizability in large random networks with $k_{\min}\gg 1$
can hardly be  enhanced further with other weighted coupling schemes different from  the optimal
case in Eq.~(\ref{mzk05_eq1}).
This coupling scheme was also considered in \cite{lgh04} where synchronizability in SF networks
of maps is enhanced for $\alpha>0$.
Additionally, we note that the coupling form in Eq.~(\ref{AIP_eq}) has been recently revisited from the
viewpoint of gradient-network~\cite{wll07}.

Another work ~\cite{zzwor06} introduced an additional parameter $\beta$ into the coupling scheme in Eq.~(\ref{AIP_eq}),  as
\begin{equation}
G_{ij}= -\frac{k_j^{\alpha}}{\big(\sum_{j\in \Gamma_i} k_j^{\alpha} \big)^{\beta}}.
\label{zzwor06_eq}
\end{equation}
In this case, the intensity of the node is $S_i=\big( \sum_{j\in \Gamma_i} k_j^{\alpha}\big)^{1-\beta}$.
If $\beta \ne 1$, the intensity is not uniform and becomes more  heterogeneous when $|1-\beta|$ increases.
For any given $\alpha$, the synchronizability is optimal at $\beta=1$
where the intensity is fully uniform. Besides, for a fixed $\beta$, there is also a value of $\alpha$ for which the
best synchronizability is achieved.

More weighted coupling methods have been proposed. In \cite{hcab05} the authors use the information about the age of the nodes in growing
networks and introduce asymmetrical coupling  between old and young nodes (first and latest nodes to join the network, respectively).  Old nodes in general have large degree and young nodes small degree.
The authors propose a connectivity matrix
\begin{equation}
G_{ij}= -\frac{a_{ij}\Theta_{ij}}{\big(\sum_{j\in \Gamma_i} \Theta_{ij} \big)},
\label{hcab05}
\end{equation}
where $\Theta_{ij}=(1-\theta)/2$ if the connection is from old to young nodes ($i>j$) and
$\Theta_{ij}=(1+\theta)/2$ if the connection is from young to old  nodes ($i<j$).
The limit $\theta=-1$ ($\theta=1$)  gives a unidirectional coupling where the old (young) nodes drive the
young (old) ones.  It was shown that in SF networks, synchronizability is enhanced
when couplings from older to younger nodes are dominant ($\theta<0$).
When large heterogeneous degrees between the old and young nodes occur, this scheme is quite similar to those in
Eqs.~(\ref{chahb05_eq},\ref{AIP_eq},\ref{zzwor06_eq}). However, in this case the eigenvalues are complex, and both, the ratios of
the real and imaginary parts of the eigenvalues were employed simultaneously to characterize the synchronizability.

In spite of these numerical observations, a clear understanding about why the synchronizability is further enhanced with the various weighted  coupling schemes (Eqs.~(\ref{chahb05_eq},\ref{AIP_eq},\ref{zzwor06_eq})) has not been obtained. The same scheme as in ~(\ref{hcab05}), but \emph{without} the normalization ($G_{ij}=-a_{ij}\Theta_{ij})$ was considered in \cite{zzc06},
and again the synchronizability is enhanced for $\theta<0$. In this case, the change of synchronizability
is mainly due to the heterogeneity of the intensity distribution $S_i$, which becomes more homogeneous
for $\theta<0$ because the old nodes are hubs  and have most of the connections to young nodes.
In this case, the universality in Eq.~(\ref{fit_all}) should apply when the minimal degree is large enough. Finally, it has also been shown that a random distribution of the weights of the connections of regular networks,
with only nearest neighbors, can also enhance synchronizability. This fact  is related to the
effective presence of short cuts in terms of weights~\cite{llwdf07}.

As it has been already pointed out, the particular scientific course of action taken by proposing different weighting schemes to enhance synchronizability is unfinished, and probably an unfruitful quest given the many possibilities of inventing new weights. A more rigorous analysis of the eigenspectra of general graphs beyond the already obtained bounds is absolutely required to boost this line of research.



On the other hand, the above weighted coupling schemes are static. In many real-world systems, the network structure evolves and changes with time. In \cite{zk06b} the authors proposed a scheme that can {\it adaptively} tune
the correlation between the degrees of the nodes and the weights of the links as in Eq.~(\ref{mzk05_eq1}), with $\theta\approx -0.5$,
so that synchronizability can be significantly enhanced compared to the unweighted counterpart.
The adaptation scheme  is based on the local synchronization between  a node  and its $k_i$ direct
neighbors in the network. Each node tries  to synchronize to its neighbors
by increasing the connection strength among them. By doing this, the coupling strength of the node $i$ with its neighbors increases uniformly trying to suppress the difference $\Delta_i$ with the mean activity of its neighbors, namely,
\begin{equation}
G_{ij}(t)=a_{ij}V_i(t),  \;\;\;\dot{V}_i=\rho \Delta_i/(1+\Delta_i),
\label{vi}
\end{equation}
where $\Delta_i=|{\bf H}({\bf x}_i)-(1/k_i)\sum_j a_{ij}{\bf H}({\bf x}_j)|$, and
$\rho>0$ is the tuning parameter.
Note that with this adaptation scheme, the input weight
($w_{ij}=V_i$) and the output weight ($w_{ji}=V_j$) of a node $i$ are
in general asymmetrical.

\begin{figure}
\begin{center}
\epsfig{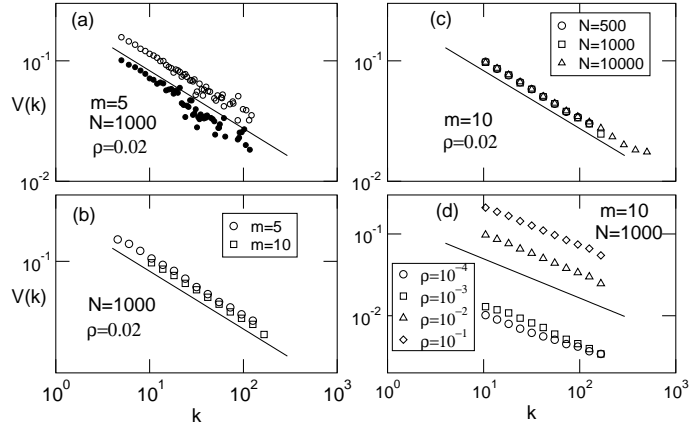}
\caption{The average input weight $V(k)$ of nodes
with degree $k$ as a function of $k$ for the R\"ossler oscillators ($\circ$) and
the food-web model ($\bullet$) (a)
 and its
dependence on various parameters: $m$ (b), $N$ (c) and $\rho$ (d).
Results in (b-d) are averaged over 10 realizations  of the networks
with  random initial conditions.
For clarity, only  the results for the R\"ossler oscillators are shown
in (b-d) and  are logarithmically binned.
The solid lines in (a-d) have a slope $-0.48$. From  \cite{zk06b}.
}
\label{zk06_fig2}
\end{center}
\end{figure}

This adaptive process was simulated using R\"ossler oscillators and a chaotic food web model
on BA  networks, and both the unbounded and bounded MSFs were considered.
For the unbounded case, the system approaches  complete synchronization when $\rho>0$, while
for the bounded case, one has that this happens for $0<\rho<\rho_c$, where $\rho_c$ depends on the particular oscillators,
and on the system size $N$. In both situations, when synchronization is achieved, the adaptation
process will lead to a weighted coupling structure where the input strength of the links
of a node displays a power law dependence on the degree as
\begin{equation}
V(k) \sim k^{\theta},
\label{vk}
\end{equation}
with $\theta\simeq -0.5$. The results for the unbounded MSF, which are
rather robust to variations in network models,  parameters and  oscillator models,
are shown in Fig.~\ref{zk06_fig2}.

The mechanism underlying such a self-organization of the weighted structure is due to the
degree-dependent synchronization difference $\Delta_i$ ~\cite{zk06a,zk06b}.
 Starting from  random initial conditions on  the chaotic
attractors, both the local synchronization difference $\Delta_i\gg 1$ and the input weights
for each node increase rather homogeneously in the whole network,
i.e., $w_{ij}=V_i(t)\approx \gamma t$.
Now, the intensity of the node $S_i(t)=V_i(t)k_i=\rho k_i t$. Hence, nodes with large degrees are coupled stronger to the mean activity of their
neighbors. As a consequence, after a short period of time the synchronization difference $\Delta_i$ for those highly connected nodes decreases,  and the weights $V_i$ of different  nodes
evolve at different  rates and converge to different values. Once synchronization is achieved, the
input strength $w_{ij}=V_i$ is small for nodes
with large degrees.

\begin{figure}
\begin{center}
\epsfig{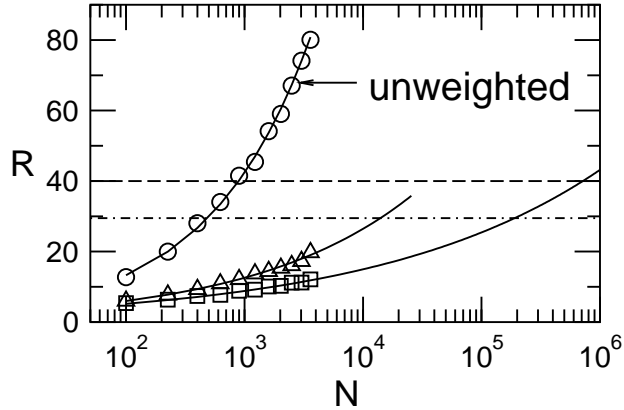}
\caption{
The eigenratio $R$  as a function of $N$ averaged
over 20 realizations of the networks.
The solid lines are  power-law fitting.
The weighted networks are obtained
by the adaptive process with the conditions:
$M=5$, $\gamma=0.002$ with ${\bf H} ({\bf x)}=(x,0,0)$ for the R\"ossler oscillators ($\Box$)  and ${\bf H} ({\bf x)}=(0,y,z)$
for the food-web model ($\bigtriangleup$).
The networks are synchronizable if $R<R_{\alpha}$:  R\"ossler oscillators, $R_{\alpha}=40$ (dashed line),
food-web model, $R_{\alpha}=29$ (dashed-dotted line). From \cite{zk06b}
}\label{zk06_fig5}
\end{center}
\end{figure}

The adaptation process makes the intensity more homogeneous, so that it is expected that the synchronizability
is enhanced.
In Fig.~\ref{zk06_fig5} we show
the eigenratio $R$, as a function of the network size $N$.
There we compare the original unweighted network with two weighted networks after the adaptation.
 Suppose that  the largest  network size synchronizable
for the bounded MSF ($R=\alpha_2/\alpha_1$) is $N_1$ for unweighted networks and $N_2$ for weighted networks
obtained from adaptation. It then follows that $N_2/N_1 \sim (\alpha_2/\alpha_1)^{1/(1+\theta)} \sim (\alpha_2/\alpha_1)^2$ for power law degree distributions regardless of the exponent $\gamma$ \cite{zk06b}.


\subsubsection{Topological modification for enhancing synchronizability}

Some authors have proposed to enhance synchronizability
by perturbing  the network topology. Based on the argument that heterogeneity
in the  betweenness distribution is related to poor
synchronizability, in \cite{zzww05,ywcw06} it is proposed to modify the nodes or links with the highest maximal betweenness.
As already noticed, in SF networks, the betweenness of a node and a link is  strongly
correlated with the degree $k_i$, and the product of degrees $k_ik_j$ of the two nodes
at the ends, respectively.
The perturbation proposed
in \cite{zzww05} consists of dividing the node with the highest degree into a  group of
several fully connected  nodes and redistribute  the $k_i$
 links equally over the new nodes. Following this scheme, the synchronizability
can be substantially enhanced by modifying
a very small portion of the  nodes. The enhanced synchronizability follows
closely the  reduced  maximal degree in the networks.
It was also shown that the average distance actually increases when the
hubs are divided. In \cite{ywcw06} the authors propose that the connections with the largest
$k_ik_j$ are broken, and again the synchronizability can be enhanced by
 cutting  a small fraction of links with high betweenness. These ideas can plausibly be implemented in technological networks, where the substitution
of hubs by a core of nodes is possible. In this way, the redistribution of load will improve traffic, and
as a by-product, the synchronizability.

\subsubsection{Optimization of synchronizability}

A more straightforward approach to the design is that of asking which are the best network architectures to get an optimal synchronizability. In \cite{dhm05} an optimization
scheme (e.g. simulated annealing) combined with a network rewiring algorithm to minimize the
eigenratio $R$ is applied. In this case, the total number of nodes and links are preserved. The resulting networks, with optimal synchronizability,
called \emph{entangled networks},  are found to be very homogeneous in many topological
measures, such as degrees, distance between nodes, betweenness etc. This result is quite relevant because it provides a null model that allows to compare the synchronizability of networks directly with its optimal counterpart.

A similar optimization scheme was applied to study the optimal synchronizability in
networks with a preserved SF  degree sequence~\cite{ldwgzw07}. In this case, the synchronizability
can only be slightly enhanced.  An interesting finding is that the optimized networks
become disassortative and the clustering and the maximal betweenness is reduced, which is
consistent with the observed enhanced synchronizability obtained by changing these features.

Thus,  optimization schemes are helpful to identify meaningful topological features
that correlate with the synchronizability of complex networks.
However, such  optimization schemes  are computationally demanding when we deal with large networks.
The development of analytical tools to attack this optimization problem is currently a major challenge in the subject.

\begin{figure}
\begin{center}
\epsfig{figure=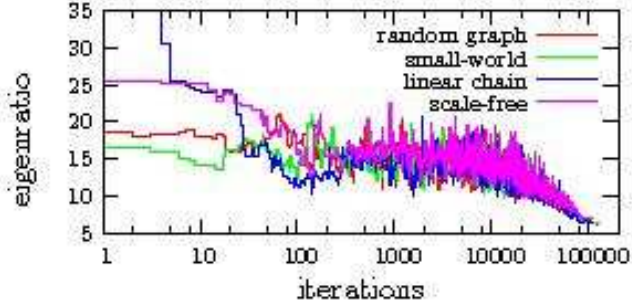,width=0.55\textwidth, clip=,}
\caption{
(color online) Eigenvalue ratio $R$ as a function of the
number of algorithmic iterations. Starting from different initial
conditions, with $N=50$, and $\langle   k \rangle= 4$, the algorithm converges
to entangled networks with very similar values
of $R$. From \cite{dhm05}.
} \label{dhm05fig}
\end{center}
\end{figure}

Furthermore, in these two studies, the underlying network topology is bi-directional. In many of the weighted
coupling  schemes previously mentioned,  the connections are effectively directed, since the coupling strength is asymmetrical
for the input and output links.
In  \cite{nm06b}  the authors went   to the extreme by imposing an unidirectional information flow
so that the networks become optimally synchronizable with $R=1$.
For any topology, the maximally synchronizable network
can be achieved by imposing that the network: (i) embeds a directed spanning tree, (ii)
has no directed loop, and (iii) has normalized input strengths. With these conditions,
the original networks are changed into  feedforward networks without any feedback, and
$0=\lambda_1<\lambda_2=\cdots=\lambda_N=\lambda$. Furthermore, the synchronization of the whole network is achieved in a hierarchical way. The conditions (i)-(iii) lead to the following path towards synchrony. There is a node in the directed spanning tree that has no input and acts as a master oscillator driving the network dynamics. If $\alpha=\sigma \lambda$, the oscillators that are just one hierarchical level below the master oscillator will synchronize. Then, the next lower level oscillators will also get synchronized and so on until the whole network reaches complete synchronization. Note that this happens for the entire range of the coupling strength where $\alpha=\sigma \lambda$.

Note that the optimization schemes discussed above  have considered maximizing the Type I synchronizability by minimizing
the eigenratio $R$ under certain constrains. Such optimized networks, however, may not have enhanced synchronizability of Type II
and smaller cost of synchronization  that are associated to the eigenvalue $\lambda_2$.
Furthermore, in these  studies, the underlying network topology is un-directed.

Is it possible to find the network structures that have the optimal synchronizability of both Type I and Type II and the smallest
cost, among all possible network configurations? In \cite{nm06a,nm06b} an elegant answer is provided. They assume that  MSF has a bounded convex region
in the complex plane and denote $\alpha_1$ and $\alpha_2$ as the thresholds along the real axis.
For any networks of oscillators, if synchronization manifold is stable for the coupling strength $\sigma_{\min}\le \sigma \le \sigma_{\max}$,
the synchronizability and cost of synchronization can be defined as
$$S_{syn}=\frac{\sigma_{\max}}{\sigma_{\min}}, \;\; C_{syn}=\sigma_{\min}\sum\limits_{i,j}^{n} w_{ij}.$$
For real eigenvalues ordered as in Eq.~\ref{eigenvalue_order}, we have
$$S_{syn}=\frac{\sigma_{\max}}{\sigma_{\min}} \cdot \frac{\lambda_2}{\lambda_N} , \;\; C_{syn}=\frac{\alpha_1}{\lambda_2}\sum\limits_{i=1}^{n} \lambda_i.$$
As in Eq.~\ref{cost}, the cost is the minimal total coupling strength when the network is just able to synchronize.
In  \cite{nm06a} it is proven that for any network
 $$S_{syn}\le \frac{\alpha_2}{\alpha_1}, \;\; C_{syn}\ge\alpha_1 (N-1).$$
If the spectum of the coupling matrix satisfys
\begin{equation}
0=\lambda_1<\lambda_2=\cdots=\lambda_N=1
\end{equation}
then the  network will achieve the maximal synchronizability \emph{and} the minimal cost, i.e.,
$$S_{syn}= \frac{\alpha_2}{\alpha_1}, \;\; C_{syn}=\alpha_1 (N-1).$$
In \cite{nm06a,nm06b} such optimal networks are obtained by imposing an unidirectional information flow.
For any topology, the maximally synchronizable network
can be achieved by imposing that the network: (i) embeds a directed spanning tree, (ii)
has no directed loop, and (iii) has normalized input strengths. With these conditions,
the original networks are changed into  feedforward networks without any feedback (reciprocial links or loops).
 Furthermore, the synchronization of the whole network is achieved in a hierarchical way.
The conditions (i)-(iii) lead to the following path towards synchrony. There is a node in the directed spanning
tree that has no input and acts as a master oscillator driving the network dynamics. If $\sigma \in (\alpha_1, \alpha_2)$,
the oscillators that are just one hierarchical level below the master oscillator will synchronize.
Then, the next lower level oscillators will also get synchronized and so on until the whole network reaches
complete synchronization.
Finally, we note that it could take a very long time to achieve complete synchronization when
the number of  effective layers is large.  It would be interesting to see how the topology of the original
networks is related to the depth of such effective directed trees and how it influences the
transient time towards synchronization. Very recently, it was shown \cite{lzzw07} that an age-based coupling similar to Eq.~\ref{hcab05}, but with $\Theta_{ij}=e^{-\alpha (i-j)/N}$, will lead to such an effective directed tree with $R=1$ at large values of $\alpha$ \cite{lzzw07}.

The findings of \cite{nm06a,nm06b} have important implications on the structural properties and dynamical processes
in real networks. Although most analysis of complex networks have been developed for undirected networks, many
real networks are directed. Recent studies about the local organization of directed networks found
that reciprocity (percentage of bi-directional links) in real networks differ clearly from models \cite{gl04,zzzsk08}, and suprisingly  many real directed
networks have very few short loops as compared to random networks \cite{bgm08}.
These properties seem to be constrained by the degree correlations \cite{zzzsk08,bgm08}, and therefore it will be interesting to study the
impact of these properties on the synchronizability of directed networks in the future.

Optimizing networks for synchronization has also been considered for networks of phase oscillators with
natural frequencies uncorrelated with the initial random network configuration\cite{gz06,brede08}. If the network is rewired
with a bias towards connecting oscillators with similar average frequencies, synchronization is enhanced and
for intermediate coupling strength non trivial network properties, such as high number of cliques and large
average distance, emerge \cite{gz06}. When a  measure that combines the local \cite{gma07a} and global synchronization is optimized
by network rewiring, communities having similar  natural frequencies are obtained at intermediate coupling,
while strong coupling  between dissimilar oscillators leads to highly synchronizale networks at large coupling strength \cite{brede08}.

Optimization approaches in networks thus are very useful to explore the structure-dynamics relation in model and
real networked systems and also in various  applications aiming at enhancing the performance of the networks, see \cite{mt07b}
for a recent focus issue on this interesting topic.

\subsection{Beyond the Master Stability Function formalism}
\label{subsect_beyond_MSF}

We have discussed how the impact of network topology on synchronizability
can be addressed using the MSF and graph theory. Away from the complete
synchronization regime, the linear stability does not strictly apply. However, it is still possible to go one step forward to
further understand some aspects of the dynamical synchronization patterns.

\begin{figure}
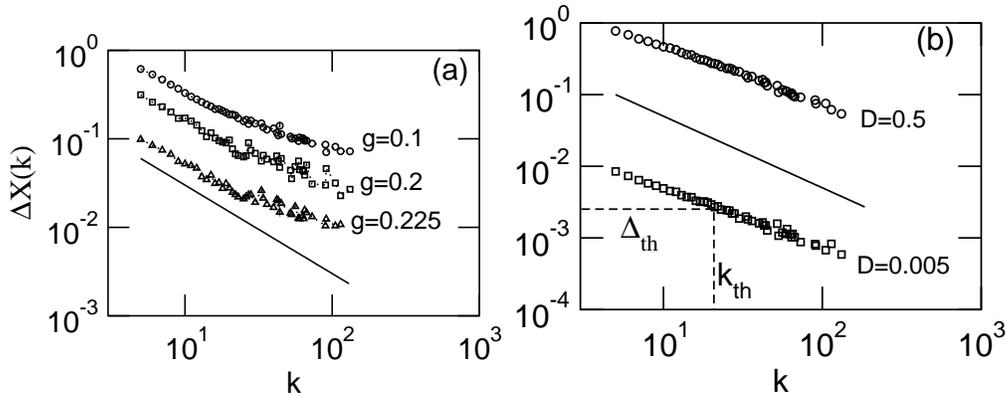

\begin{center}
\epsfig{figure=fig28a.eps,width=0.4\textwidth, clip=,}
\epsfig{figure=fig28b.eps,width=0.4\textwidth, clip=,}
\caption{
The average values $\Delta X(k)$ as a function of $k$ in
SF networks of R\"ossler oscillators outside the complete synchronization
regime. (a) At various  coupling strengths  $g=\sigma \langle   k \rangle$
below the threshold for complete synchronization   and (b) the synchronized state is perturbed
by noise of different intensities $D$. Qualitatively, the same behavior is observed when the oscillators
are nonidentical.   From  \cite{zk06a}.
} \label{zk06_fig}
\end{center}
\end{figure}

In \cite{zk06a}  the authors studied effective synchronization patterns in unweighted
SF networks of chaotic oscillators
(with the coupling function ${\bf H} ({\bf x})={\bf x})$  in several situations: away from the complete synchronization regime, when the coupling
strength is smaller than the threshold for complete synchronization, when the oscillators have
mismatches in parameters, and when there are noise perturbations. They considered a mean field
approximation in which each oscillator is influenced by a global
mean field ${\bf X}$, with a coupling strength $\sigma k_i$, namely,
\begin{equation}
\dot{\bf x}_i={\bf F}({\bf x}_i)
+ \sigma k_i({\bf X}-{\bf x}_i),  \;\;\; k_i \gg 1.
\label{mean-field}
\end{equation}
 The authors
compared the synchronization of each oscillator to ${\bf X}$ by computing
$\Delta X_i=|{\bf x}_i-{\bf X}|$ and then obtained the  average $\Delta X(k)$
 over all nodes with the same degree $k$. It was shown that out of the complete synchronized state
\begin{equation}
\Delta X (k) \sim k^{-\gamma},
\end{equation}
where the exponent $\gamma\approx 1$, as seen in Fig.~\ref{zk06_fig}.
This result shows that in heterogenous networks, the hubs ($k_i>k_{\mbox{\scriptsize{th}}}$) will synchronize more
closely with the mean field and they will form effective synchronization clusters
($|{\bf x}_i-{\bf x}_j|<\Delta_{\mbox{\scriptsize{th}}}$).
However, there is not a unique threshold to define such effective clusters (see Fig.~\ref{zk06_fig}).
This path to synchronization, i.e.,  the formation of clusters by the hubs,
 was further described later  in \cite{gma07a}, see also Sect. \ref{sect_km_path}.

The formation of such SF or hierarchical  clusters could be
understood from a linear analysis using the MSF formalism.
The  linear variational  equations of (\ref{mean-field})  read
\begin{equation}
\dot{\xi}_i=\left[ D{\bf F}({\bf X}) - \sigma k_i {\bf I} \right]\xi_i,
\;\;\; k_i \gg 1.
\label{zk06_eq_msf}
\end{equation}
They have the same form as (\ref{MSF_mode_i}), except that $\lambda_i$
is replaced by $k_i$ and $D{\bf H}$ by the identity matrix ${\bf I}$.
The MSF  for the coupling function  ${\bf H} ({\bf x})={\bf x}$ is
$\lambda_{\max}(\alpha)=\lambda_1^F -\alpha$, where $\lambda_1^F$ is the
largest Lyapunov exponent of the  isolated oscillator ${\bf F}$.
Thus the  largest Lyapunov exponent
$\lambda_{\max}(k_i)$ of  the linearized Eq. (\ref{zk06_eq_msf})
is a function of $k_i$ and becomes  negative for $\sigma k_i>\lambda_1^F$.
For large $k$ values satisfying  $\sigma k \gg \lambda_1^F$,
we have $\lambda_{\max}(k) \approx - \sigma k$.

Now suppose that the network is not  completely synchronized, but slightly perturbed
from the state of complete synchronization, when the coupling strength $\sigma$  is below the complete synchronization threshold,
or when  there is  noise present in the system. For nodes with large
degree $k$, so that
$\lambda_{\max}(k)\approx - \sigma k$  is large enough in absolute value but negative,
the dynamics of the averaged
synchronization difference
$\Delta X (k)$  over large time scales can be expressed as
\begin{equation}
\frac{d}{dt}\Delta X(k)=\lambda_{\max} (k) \Delta X(k)+c,
\label{analysis}
\end{equation}
where $c>0$ is a constant denoting  the level of perturbation
with respect to the complete synchronization state, and depends either on the
noise level $D$ or on the coupling strength $\sigma$.
From this we get the asymptotic result $\Delta X(k)=c/|\lambda_{\max}(k)|$, yielding
\begin{equation}
\Delta X (k) \sim k^{-1},
\label{scaling}
\end{equation}
which explains qualitatively the numerically observed
scaling (solid lines in Fig.~\ref{zk06_fig}). Interestingly, the same scaling dependence but for the time needed to get back into the fully synchronized state was obtained in \cite{mp04} for a population of Kuramoto oscillators, see Sect. \ref{sect_km_path}.

%

To round off this section, let us mention other works about synchronizability in complex networks that make use of linear criteria
similar to the MSF~\cite{wc02,lc03,c06}. The analysis of the \emph{global} stability of the synchronized state, was first carried out for general graphs~\cite{wc95} and then followed
for specific complex networks~\cite{bbh04a,bbh06,lc06,c07}.
The global stability requires additional constraints on the dynamical properties of the
individual oscillators.
For example, consider the form of coupling as in Eq.~(\ref{MSF_eq_1}),
the requirement imposed by ~\cite{bbh04a,bbh06} is that the following auxiliary system of
the synchronization difference ${\bf X}_{ij}={\bf x}_i-{\bf x}_j$,
\begin{equation}
\dot{{\bf X}}_{ij}=\left[ \int_0^1 D{\bf F}(\beta {\bf x}_j+(1-\beta){\bf x}_i)d\beta -\alpha D{\bf H} \right] {\bf X}_{ij},\nonumber
\end{equation}
be globally stable at the fixed point ${X}_{ij}={\bf 0}$ for  $\alpha>\alpha_c$.
With this requirement, the condition for global synchronization of a network is
\begin{equation}
\sigma > \sigma^* =\max_{l}  (\frac{\alpha_c}{N}b_l),
\end{equation}
where $b_l$ is the sum of the lengths of all chosen paths that pass through a given
link $l$ in the network. Note that $b_l$ is related to both the  betweenness and the path length of the link.
The result, which is for undirected  networks, also holds for directed networks where the input and output degrees are equal for every
node of the network. Finally, in \cite{bbh04a,bbh06} it is also derived the condition needed for  global synchronization in more general cases where each link in  the network may have a different
coupling strength that is allowed to vary in time.

\section{Applications}

The focus of the review up to now has been to revise the main contributions, from theoretical
and computational points of view, to our understanding of synchronization processes in complex networks.
In this section we will overview the applications to specific problems in such different scientific fields as
biology and neuroscience, engineering and computer science, and economy and social sciences.
There are nowadays several problems where the application of the ideas and techniques developed
in relation to synchronization in complex networks are very clear and the results help to understand
the interplay between topology and dynamics in very precise scenarios.
There are other cases, also included here for completeness,
for which most of the applications so far have been developed in simple patterns of interaction, but
extension to complex topologies is necessary because it is its natural description.

\subsection{Biological systems and neuroscience}

In biology, complex networks are found at different spatio-temporal scales:
from the molecular level up to the population level, passing through many intermediate scales of biological systems.
In some of these networks, dynamical interactions between units, which are crucial for our current
understanding of living systems, can be analyzed in the framework of synchronization phenomena developed so far. Here we review some
of these application scenarios where synchronization in networks has been shown to play an essential role.
Thus, at the molecular level we can analyze the evolution of genetic networks and at the population level the dynamics
of populations of species coupled through diffusion along spatial coordinates and through trophic interactions.
Amongst these two extremes we find a clear application in the analysis of circadian rhythms.
On a different context, neuroscience offers applications
also at two different levels, one for the synchronization of individual spiking neurons and the other
for the coupling between cortical areas in the brain.

\subsubsection{Genetic networks}
The finding that a few basic modules are the building blocks of large real
regulatory networks has enabled the design and construction of small synthetic
regulatory circuits to implement particular tasks.
One of the most salient examples of
a synthetic gene network is the "repressilator", that has become one of the
best studied model systems of this kind. The repressilator is a network of
three genes, whose products (proteins) act as repressors of the transcription
of each other in a cyclic way. This synthetic
network was implemented experimentally in the bacterium {\em E. coli}, so that it periodically induces the synthesis of a green fluorescent protein as a
readout of the repressilator state \cite{el00}. It turns out that the temporal fluctuations in
the concentration of each of the three components of the repressilator can be reproduced by a system of six ordinary differential
equations,
\begin{eqnarray}
\frac{d[x_i]}{dt} &=& -[x_i]+\frac{\alpha}{1+[y_{j}]^n},
\\ \frac{d[y_i]}{dt} &=& -\beta ([y_i]-[x_i]),
\end{eqnarray}
where the couples $(i,j)$ assume the values $(1,3)$, $(2,1)$ and $(3,2)$.  The
variable $[x_i]$ is the mRNA concentration encoded by gene $x_i$, and $[y_i]$
is the concentration of its translated protein $y_i$. The parameter $\alpha$ is the promoter rate, the parameter $\beta$
is the ratio of the protein decay rate to the mRNA decay rate,
and time has been rescaled in units of the mRNA lifetime. This system has a
unique steady state which can be stable or unstable depending on the parameter
values and constitutes an illustrative example of the experience gained by
identifying network modules and modeling its dynamical behavior in real
networks.

\begin{figure}[!t]
\begin{center}
\epsfig{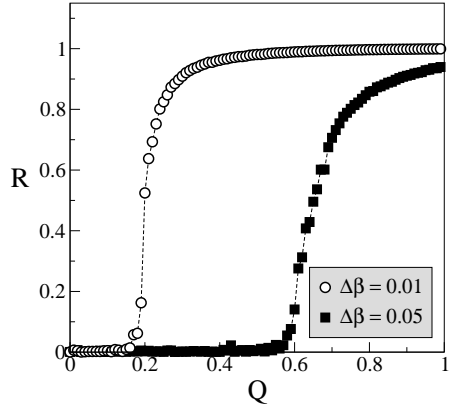}
\end{center}
\caption{Dependency of the order parameter $R$ on the the coupling strength $Q$,
which is linearly proportional to the cell density.
The two curves correspond to different values of the ratio between mRNA and protein lifetimes variance $\Delta\beta$.
From \cite{ges04}.}
\label{repressilator}
\end{figure}

Not surprisingly, the repressilator has called much attention from experts on
biological synchronization, since it offers good prospectives for further
insights into the nature of diverse biological rhythms -whose mechanisms remain to be
understood- which are generated by thousands of cellular oscillators that operate synchronously.
In \cite{ges04} it has been recently proposed a simple modular addition of two proteins to the
repressilator original design that could be used to describe the metabolic oscillations observed in a well mixed suspension of yeast cells. In the new setting, one of the new proteins
can diffuse through the cell membrane, thus providing a coupling mechanism
between cells containing repressilator networks. This inter-cell
communication couples the dynamics of the different cell oscillators (with
different repressilator periods) and thus allows us to study the transition
to synchronization of coupled phase oscillators in a biological system.  In particular, in the limit of infinite cell dilution, the system is made up of a population of uncoupled limit-cycle oscillators. This is not anymore the case when the cell density increases, as now extracellular diffusion provides a mechanism of intercell coupling. As a result, the system shows partial frequency locking of the cells. Finally, if the cell density is large enough, complete locking and synchronized oscillations are observed. The authors proposed an order parameter to measure the degree of synchronization of oscillatory behavior. The dependence of this order parameter $R$ defined as the ratio of the standard deviation of the time series of the average signal to the standard deviation of each individual signal $[x_i]$ averaged over all signals $i$, as a function of the coupling strength, is shown in Fig.\ \ref{repressilator} for different values of the ratio between the mRNA and protein lifetimes width distribution ($\delta\beta$). Note that the phase diagram of the coupled repressilators can be explained by the very same mechanism involved in the transition to synchronization for systems of coupled oscillators (e.g., Kuramoto oscillators) studied in Sec. \ref{sect_KM}. What is relevant here is that the transition from an unsynchronized to a synchronized regime is caused by an increase in cell density and therefore the experimental observation of a synchronizing transition in biological phase oscillators might be achieved.
In fact, it has been  recently designed a simple electronic circuit analogy of a population
of globally coupled repressilators \cite{wbgs06}. They show that coupling is more efficient than externally forcing
for the achievement of synchronization.
In contrast to the existence of a unified rhythm that gives rise to synchronization,
in \cite{kvzk07} it is analyzed the mechanisms of intercell communication that can be responsible for
multirhythmicity in
coupled genetic units.
We foresee that works on this line of research will incorporate more genetical interactions in the near future, being the complex network substrate and the synchronization dynamics key aspects of the whole problem.

\subsubsection{Circadian rhythms}

A circadian rhythm is a roughly 24-hour cycle in the physiological processes of living systems; usually endogenous, or when it is exogenous it is mainly driven by daylight. Understanding circadian rhythms is crucial for some physiological and psychological disorders. A nice description of experiments carried out in human beings, in which their circadian rhythms are altered, can be found in the books by \citeauthor{strogatz03} \cite{strogatz03} and \citeauthor{prk01} \cite{prk01}. Circadian rhythms are known to be dependent on the network of interactions between different subsystems. For example, daylight sensed by eyes and processed by the brain develops a chain of interactions that affects even the behavior of certain groups of cells. On a different scenario, \cite{cj87} reports how non-oscillatory cardiac conducting tissues, when driven rhythmically by repetitive stimuli from their surroundings, produce temporal patterns including phase locking, period-doubling bifurcation and irregular activity.

Synchronization phenomena in complex networks of coupled circadian oscillators has been recently investigated experimentally \cite{fnhmm07} on plant leaves. The vein system is  in this case the complex network substrate of the synchronization process. Plant cells are coupled via the diffusion of materials along two types of connections:
one type that directly connects nearest-neighboring cells and the other type that spreads over the whole plant to transport material among all tissues quickly.
Analyzing the phase of circadian oscillations, the phase-wave propagations and the phase delay caused by the vein network, the authors describe how global synchronization of circadian oscillators in the leaf can be attained. As we have seen throughout this review, the role of the topology of interactions is again fundamental in the development of synchronization.
This work is representative of the new type of applications we can find in the very recent literature
about synchronization in complex networks. This particular case of circadian rhythms in plants might be extended to other living systems, including humans.

\subsubsection{Ecology}

It is a well known observation in nature that fluctuations in animal and plant populations
display complex dynamics.
Mainly irregular, but some of them can show a remarkably cyclical behavior and take place
over vast geographical areas in a synchronized manner \cite{keith63}.
One of the best documented cases of such situation are the population fluctuations
in the Canadian lynx, obtained from the records of the fur trade between 1821 and 1939 in Canada.
Fluctuations in lynx populations show a 10-year periodic behavior from three different regions in Canada
\cite{sctbbkpoyfh99,bhs99,rkl99}.
On the other hand, there are some evidences that the existence of conservation corridors favoring the dispersal of species
and enhancing the synchronization over time
increases the danger of global extinctions \cite{elr00}.

One of the first explanations for such types of behavior was that of synchronous environmental forcing, this is the so-called
Moran effect \cite{moran53}.
There are, however, other explanations for this phenomenon \cite{rr07}, but in any case the problem highlights the importance
of integrating explicitly spatial and trophic couplings into current metacommunity theories \cite{lhmachhsltlg04,mgm07}.
Some efforts along these lines have already been made by considering very simple trophic interaction in spatially extended systems. For example, the authors in \cite{bhs99} analyze a three-level system (vegetation, herbivores, and predators),
where diffusive migration between neighboring patches is taken into account. They find that small amounts of migration are required to induce broad-scale synchronization. Another interesting study is performed in \cite{vandermeer04},
where, again in an extremely simple model, it is found that changing the patterns of interaction between consumers and
resources can lead to either in-phase synchrony or antiphase synchrony.

Nowadays we know, however, about the inherent complexity of food-webs \cite{d06}.
Food webs have been studied as paradigmatic examples of complex networks,
because they show many of their non-trivial topological features.
Furthermore, the existence of conservation corridors affecting the migration between regions
adds another ingredient to the structure of the spatial pattern.
It is precisely this complexity in the trophic interactions
coupled to the spatial dependence that must to be considered in detail in the future to get a deeper understanding of ecological evolution.

\subsubsection{Neuronal networks}

Synchronization has been shown to be of special relevance in neural systems.
The brain is composed of billions of neurons coupled in a hierarchy
of complex network connectivity.
The first issue concerns neural networks at the cellular level.
In the last years, significant progress has been made in the studies about the
detailed interconnections of different types of neurons at the level of
cellular circuits \cite{bdm04,sglmms05,mtwgsw04}.
At this level,  the neuronal networks of mammalian cortex also possess
complex structure, sharing SW and SF features.
Here are two basic neuron types: excitatory principal
cells and inhibitory interneurons. In contrast to the more homogeneous principal
cell population, interneurons are very diverse in terms of morphology and function.
There is an apparently inverse relationship between the number of neurons in various
interneuron classes and the spatial extent of their axon trees--most
of the neurons have only local connections, while a small number
of neurons have long-range axons \cite{bghw04}. These properties
of neuronal networks reflects a compromise between computational needs and wiring
economy \cite{css02,k01}.

On the one hand, the establishment and maintenance of neuronal connections require a
significant metabolic cost that should be reduced, and consequently the wiring length
should be globally minimized. Indeed, the wiring economy is apparent in the distributions
of projection length in neural systems, which show that most neuronal projections are
short \cite{spyb95,bs98}. However, there also exists a significant number of long-distance
projections \cite{sypb94,kh06}.

Large-scale synchronization of oscillatory  neural activity has been believed to play a crucial role in
the information and cognitive  processing \cite{f05}. At the level of cellular circuits,
oscillatory timing can transform unconnected principal cell groups into temporal coalitions,
providing maximal flexibility and economic use of their spikes \cite{kes96}.
Brains have developed  mechanisms for keeping time by
inhibitory interneuron networks \cite{bc95}.
The wiring will be the most economic if the connections were all local. However, in this case
physically
distant neurons are not connected, and synaptic path length and synaptic delays become
exceedingly long for synchronization in large networks.
From previous analysis of synchronization of random networks, we know that
synchronizability (stability of the synchronized state) is optimal in fully random networks with a uniform connectivity
per node, independent of the network size \cite{zmk06}. The same happens if
interneuronal oscillators are coupled \cite{bghw04}.
However, fully random connections irrespective of physical distance are not economic if wiring cost is taken into
account.

In \cite{bghw04}, it was shown with a  model of interneuronal
networks containing local neurons (Gaussian distribution of projection length)
and a fraction of long-range neurons (power law distribution of projection length),
that the  ratio of  synchrony to wiring length is optimized in the SW regime with
a small fraction of long-range neurons. Thus, most wiring is local and neurons
with long-range connectivity and large global impact are rare, as  consistent
with observations.   It was argued that the complex wiring of diverse
interneuron classes could represent an economic solution for supporting global synchrony
and  oscillations at multiple time scales with minimum axon length \cite{bghw04}. While such
mathematical consideration can predict the scaling relationship among the interneuron
classes in brain structures of varying  sizes, understanding the role of complex
neuronal connectivity, most likely mediated by synchronization, is still one of the main
challenges in neuroscience.  The theory reviewed in this article will surely contribute
to their understanding when more systematic information of neuronal connectivity becomes
available in the future.

\subsubsection{Cortical networks}

The application of graph theoretical approaches, and the characterization of the functional activity by neural synchronization, have significantly contributed to
the understanding of complex networks in the brain \cite{bb06,rpbs07}.

On a larger neurophysiological scale, the activity observed experimentally
by electroencephalographs or functional magnetic resonance imaging,
is characterized
by oscillations occurring over a broad spectrum and by synchronization phenomena over a wide range of spatial and
temporal scales.
Reliable databases are available now for large-scale systems level connectivity
formed by long-range projections among cortical areas in the brains of
several animals~\cite{sbhoy99,sckh04}.
Large-scale brain networks are found to be densely connected, with very complex and heterogeneous
connectivity patterns~\cite{sz04,hbosy00,hk04}.
In parallel, the investigation of brain activity has also put significant emphasis on large-scale
functional interactions, characterized by coherence and synchronization between the activity of cortical
regions~\cite{stam04,eccba05,sscpmb05,bmadb06}.
Both the structural and functional connectivity of the brain display SW and SF
features. The relationship between structural and functional connectivity
remains an important open problem in neuroscience.

\begin{figure}
\begin{center}
\epsfig{figure=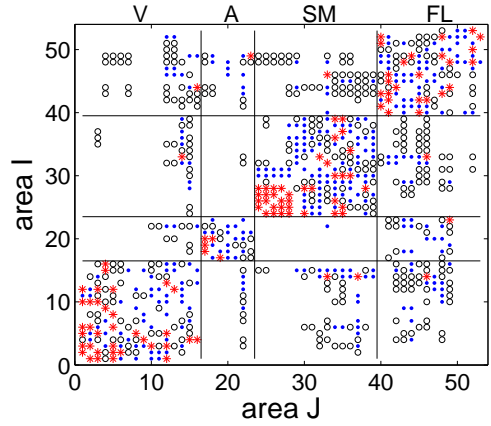,width=0.4\textwidth,clip=,}
\caption{ (Color online) Connection matrix $M^A$ of the  cortical network of the cat brain.
                The different symbols represent different connection
weights: $1$ ($\circ$ sparse), $2$ ($\bullet$ intermediate) and $3$ ($\ast$ dense).
The organization of the system into four topological communities
(functional sub-systems, V, A, SM, FL) is indicated by the solid lines.
From \cite{zzzhk06}).
}
                \label{fig:Network}
\end{center}
\end{figure}

\begin{figure}
\begin{center}
\epsfig{figure=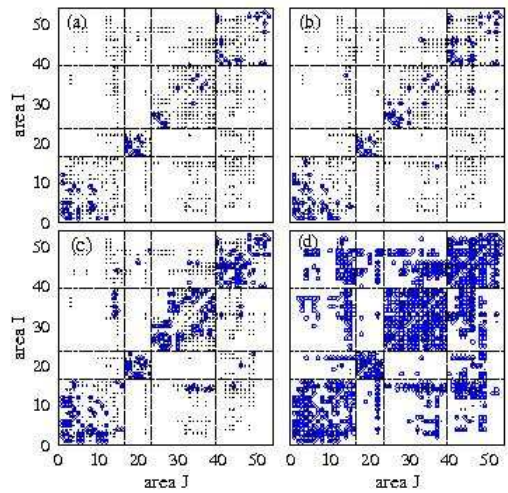,width=0.43\textwidth,clip=,}
   \caption{ (Color online)
                     The functional networks ($\circ)$ at various  thresholds: $R_{th}=0.070$ (a),
                     $R_{th}=0.065$ (b), $R_{th}=0.055$ (c) and $R_{th}=0.019$ (d).
                     The small dots indicate the anatomical connections.
From \cite{zzzhk06}.
           } \label{fig:function_net}
\end{center}
\end{figure}

\begin{figure}
\begin{center}
\epsfig{figure=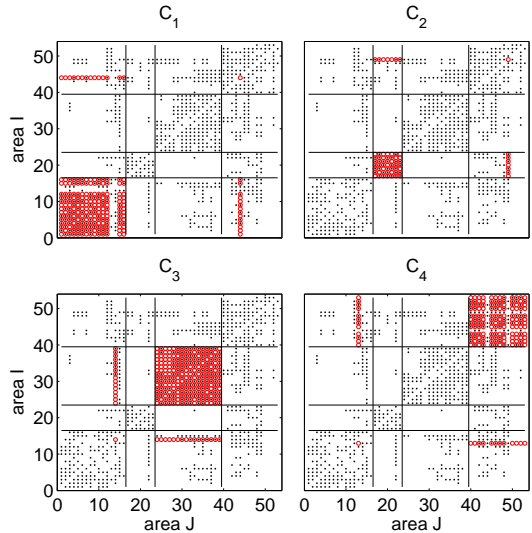, width=0.43\textwidth, clip=,}
             \caption{ (Color online)
             Four major dynamical clusters ($\circ$) in the weak synchronization regime,
             compared to the underlying anatomical connections ($\cdot$).
From \cite{zzk06}.
}
             \label{fig:cluster1}
        \end{center}
\end{figure}

Recent simulations of  synchronization dynamics of brain networks have shed
light on this challenging problem. In a series of papers \cite{zzzhk06,zzk06,zzzhk07}
the dynamics of a realistic cortico-cortical
projection network of the cat has been modeled at the level of functional areas~\cite{sbhoy99}.
At this level,  the network (see Fig.~\ref{fig:Network})
displays a hierarchical cluster organization~\cite{hk04}. There are four prominent clusters that agree
broadly with the four functional subsystems: visual (V), auditory (A), somatomotor (SM), and frontolimbic (FL).
They simulated the network dynamics by a 2-level model: each node (cortical area) is represented
by a SW subnetwork of  neurons (network of networks).
It was  shown that the model possesses  two distinguished regimes, weak
and strong synchronization. In the  weak synchronization regime, the  model displays biologically
plausible dynamical clusters.  The  functional connectivity, obtained by passing the correlation matrix
through various  thresholds, exhibits various levels of organization. The clusters with the  highest levels
of synchronization are from respective functional subsystems (Fig.~\ref{fig:function_net} (a,b))
and are related to specialized functions of the subsystems. The specialized clusters are integrated into
larger clusters through brain areas having many inter-community connections
(Fig.~\ref{fig:function_net} (c,d)). As a whole,
the functional connectivity reveals the hierarchical organization
of the structural connectivity~\cite{zzzhk06}. The dynamics forms
four major clusters (Fig.~\ref{fig:cluster1}), in excellent agreement with the
four functional subsystems~\cite{zzk06}. Furthermore,
brain areas that bridge different dynamical clusters are found to be the areas involved in
multisensory associations. In a comparative study~\cite{zzzhk07}, it was shown
that representing the brain
areas with a periodic, low-dimensional neuronal mass oscillator describing alpha waves~\cite{lhsz74}
cannot resolve these four clusters. The detailed network topology becomes rather irrelevant to
the dynamical patterns which is not very much changed when the network is randomized.
This is the same for the strong synchronization regime in the 2-level model which resembles
epileptic-like activity~\cite{zzk06}.).  This can be understood recalling previous analysis based on random networks where is shown that synchronization is mainly determined by the node intensity ~\cite{zmk06,zk06b} (see Section IV  6). Furthermore, the transition from the weak to the strong synchronization regime shares a similar picture with the Kuramoto model in complex networks~\cite{gma07a, gma07b} (see Section III 4).

Shortly after these works,
a very  similar structure-function relationship was  observed~\cite{hkbs07}.
Each area of the macaque neocortex was represented by a neural mass model in the regime
of spontaneous activity with complicated temporal patterns. It was shown that the
functional connectivity, measured over a very long time, is closely shaped by the underlying
structural connectivity as described  in \cite{zzzhk06,zzk06,zzzhk07}. On short time-scales,
the functional connectivity changes, forming two
anticorrelated clusters, similar to functional networks obtained
from brain  imaging data~\cite{fsvcer06}.

These findings support the idea that the brain is an active network, and it can generate
activity by itself in the absence of external signals.
Classical theories in cognitive neuroscience  viewed the brain as a passive, stimulus-driven
device and  the spontaneous on-going activity of the brain had been regarded as background noise~\cite{efs01}.
It is still customary in data analysis to take the average signals  over many
trials of electroencephalographs as  the event-related activation
and  associate them  with cognitive processes.
In the view of active dynamical brain networks, it has been shown that
the spontaneous on-going  activity imposes significant impact on the selective responses
to stimuli~\cite{efs01,f05}. The intricate relationship between large-scale
structural and functional networks revealed in these works \cite{zzzhk06,zzk06,zzzhk07,hkbs07})
will contribute to  this reorientation of concepts  in neuroscience. On the other hand, excessive activation and synchronization of neural networks have been found to associate with dysfunctions and disorders of the brain, such as  the epileptic seizure~\cite{stam05}  and the Parkinsonian disease~\cite{trwkpvsf07}.  Understanding synchronization in neuronal networks of various level, especially studying the role played by the complex network topology, is crucial  to elucidate  how  normal  brains can maintain desirable levels of synchronization. It will also contribute significantly to biomedical data analysis of pathological brain activities \cite{lfoh07}, for example,  the challenging task of  detecting precursors that can make prediction much before the clear onset of  seizures , and design suitable methods for treatments of neural diseases \cite{pht06}.

\subsection{Computer science and engineering}

Complex networks and synchronization dynamics are relevant in many computer science and engineering problems.
For example, in computer science, synchronization is desirable  for an efficient performance of
distributed systems.
Sometimes, the goal of the distributed system is to achieve a global common state (consensus).
Nowadays these systems are becoming larger and larger and
their topologies more and more complex.
On the other hand, some engineering problems also face the need of maintaining coordination at the level of large
scale complex networks, for example in problems of distribution of information, energy or materials.


\subsubsection{Parallel/Distributed computation}

The simulation of large systems are, nowadays, mainly implemented as parallel
distributed simulations where parts of the system are allocated and simulated on different
processors. A large class of interacting systems
including financial markets, epidemic spreading, traffic, and dynamics of physical systems in
general,
can be described by a set of local state variables allowing a finite number of possible values.
As the system evolves in time, the values of the local state variables change at discrete instants, either
synchronously or asynchronously, depending on the dynamics of the system.
The instantaneous changes in the local configuration are called discrete events,
forming what has been coined as a parallel discrete-event simulation (PDES) \cite{nf94}. The main difficulty of PDES is the absence of a global pacemaker when dealing with asynchronous updates.
This imposes serious problems because causality and reproducibility of experimental results are desired.
In a conservative scheme, processes modeling physical entities are connected via channels that
represent physical links in the target system. Since PDES events are not synchronized via a global clock, they must synchronize by communication between nodes.

\begin{figure}[!t]
\begin{center}
\epsfig{file=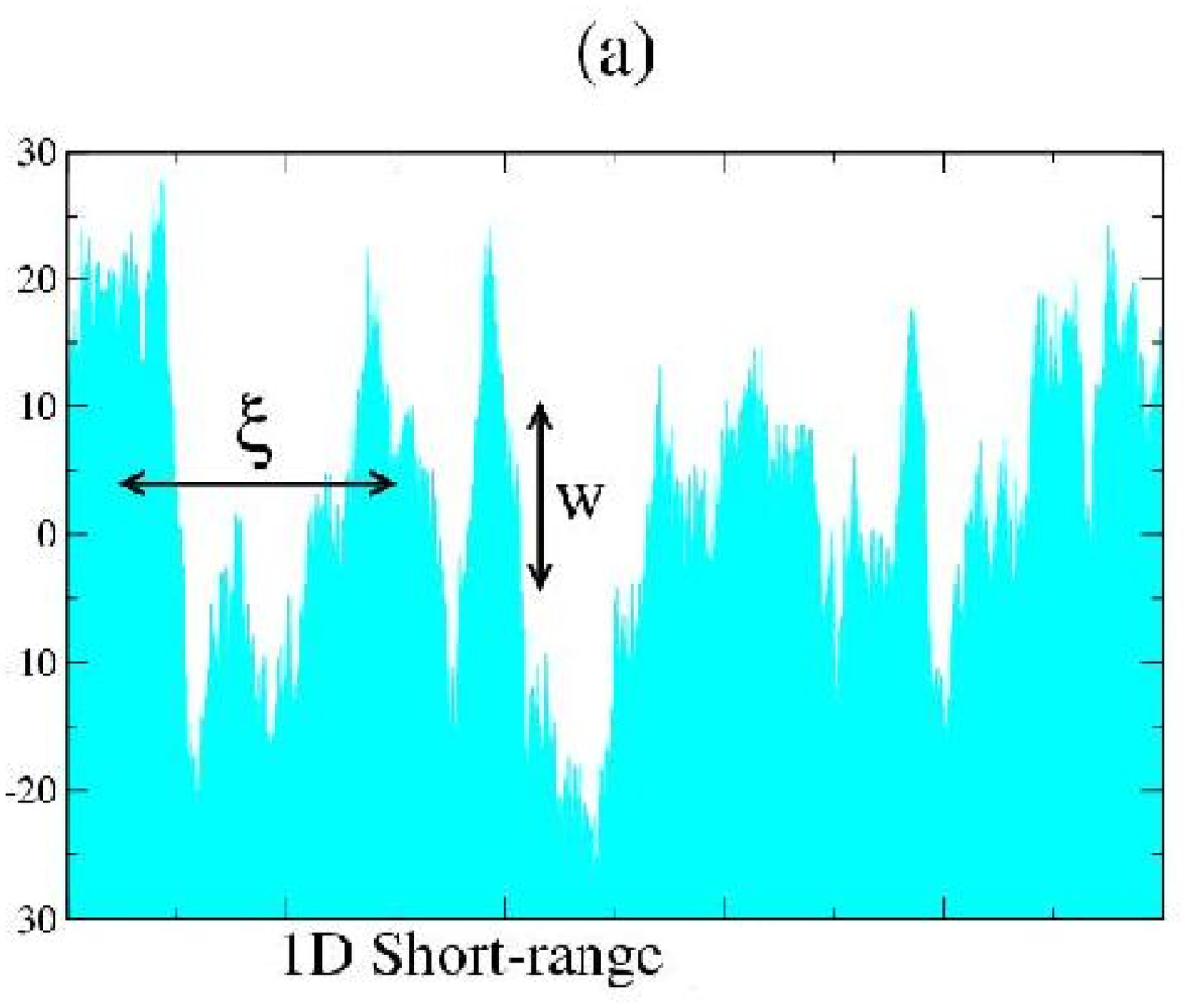,width=3.0in,angle=0,clip=1}
\epsfig{file=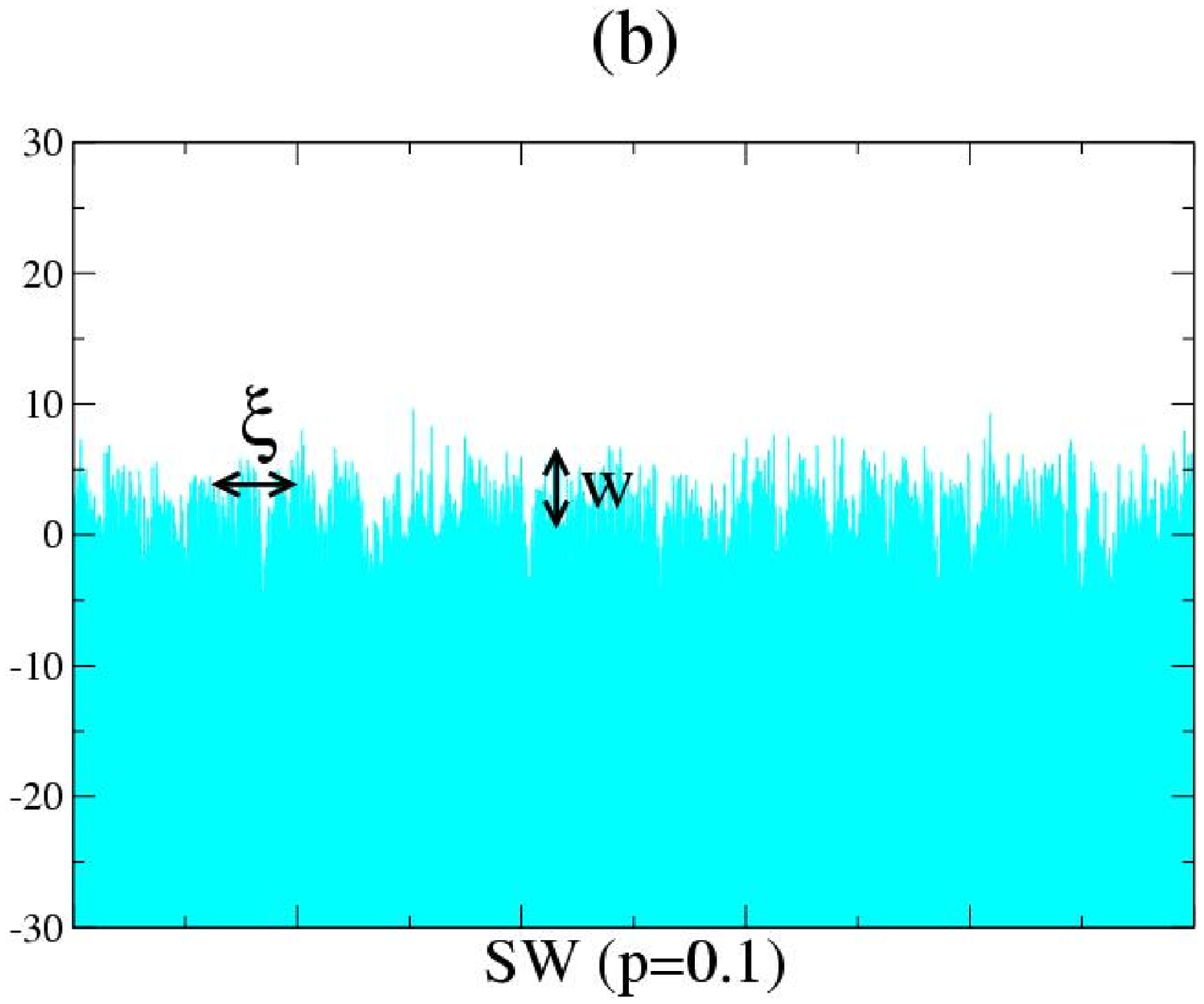,width=3.0in,angle=0,clip=1}
\end{center}
\caption{(Color online) Virtual time horizon snapshots in the steady state for 10000 sites in one dimension.  (a)  For a regular network. The lateral correlation length $\xi$ and width $w$ are shown for illustration in the graph. (b) For the SW network with $p=0.1$,
the heights are effectively decorrelated and both the correlation length and the width are reduced. From \cite{gkntr06}.}
\label{Appli_CS}
\end{figure}

The essentials of a PDES are: a set of local simulated times of the processors and a synchronization schedule. As the
grid-computing networks with millions of processors emerge, fundamental questions of the scalability of the underlying algorithms must be addressed. The properties of a PDES to be scalable are: (i) the virtual time horizon $\{\tau_i(t)\}_{i}^{N_p}$, a set of time simulated instants in $N_p$ processors after $t$ parallel steps, should progress on average with a non-zero rate, and (ii) the width of the time horizon should be bounded when $N_p\rightarrow \infty$.

In \cite{ktnr00}, it was studied the asymptotic scaling properties of a conservative synchronization algorithm in massive PDESs where the discrete events are Poisson arrivals.
They found an interesting analogy between the evolution of the simulated time horizon and the growth of a nonequilibrium surface.\footnote{Note that the first analogy between synchronization processes and the theory of surface growth appeared in \cite{gmsb93}, posteriorly revisited in \cite{mp03} (see \cite{prk01} for a comprehensive review).}
They showed that the steady-state behavior of the macroscopic landscape of the simulated time horizon, in one dimension, is governed by the Edwards-Wilkinson Hamiltonian \cite{ew82}.

The analogy becomes clear by interpreting the virtual times $\tau_i$ as the height of a surface, and defining the width of the simulated times surface as the mean root square of the virtual times with respect to the mean $\tau$. This width
provides a measure for de-synchronization
\begin {equation}
\langle   w^2 \rangle =\langle   \frac{1}{N_p}\sum_{1}^{N_p}[\tau_i(t)-\bar{\tau}(t)]^2 \rangle .
\label{w}
\end{equation}

The problem that now is faced is how the width of the synchronization landscape scales with the number of processors. If the scaling diverges, it means that the synchronization is hardly achievable. In one- and two-dimensional regular lattices,
the width of the synchronization landscape diverges with the number of processors as $w\sim N_{p}^{1/d}$ where $d$ is
the dimension.
This effect can be traced back to the lateral correlation length $\xi$ of the surface that also diverges with the number of processors $\xi \sim N$ \cite{bs95}.
An interesting solution to this problem has been proposed in \cite{kngtr03} and \cite{gkntr06}.
It consists of adding a few random links to the regular lattices resulting in a SW structure. This structure has the effect of de-correlating the lateral length, suppressing large fluctuations in the synchronization surface (roughness), and producing finite average  values of $w$ in the large system-size limit, see Fig. \ref{Appli_CS}. Moreover, the extreme height diverges only logarithmically in this limit. This latter property ensures synchronization in a practical sense in a SW topology of processors.

\subsubsection{Data mining}

The term \emph{data mining} refers to  the process of automatically searching large volumes of data for patterns that provide some useful information for classification. In \cite{mt07} it has been  proposed a new method of data mining based on spontaneous data clustering using a network of locally coupled limit-cycle phase oscillators. The method is closely related to the determination of community structure via synchronization processes devised by several authors \cite{adp06a,bilpr07}. The idea is to encode multivariate data vectors (that are the elements of the database) into vectors of natural frequencies for an oscillators' dynamical model, akin to the KM, expecting that the dynamics of the system will group similar data in clusters of synchronization. More precisely, given $n$ multivariate data points with $m$ degrees of freedom, $\vec{x_i} = (x_i(1), x_i(2), ..., x_i(m))$, $i=1,\ldots, n$, they write the dynamical model:
\begin{equation}
\dot{\theta}_i(l)=x_i(l) + \frac{\sigma}{k_i} \sum_{j=1}^{n} H(d_{i,j}) \sin (\theta_{j}(l)- \theta_{i}(l))
\end{equation}
where $\theta_i(l)$ is the $l$-th component of the phase vector $\vec{\theta_i} = (\theta_{i}(1), \theta_{i}(2), ..., \theta_{i}(m))$, $H(d_{i,j})$ is a function that determines the neighborhood of $\vec{\theta_i} $ based on the distance $d_{i,j}=|\vec{x}_i -\vec{x}_j|$. The determination of the neighborhood provides the network of interactions between oscillators. The proposal by the authors is a neighborhood centered at $\vec{x}_i$ defined by the hyper-sphere of radius $d_0$, being $d_0=\alpha |\vec{x}_i|$ and $\alpha$ a tuning parameter. The function $H(d_{i,j})=1$ if $d_{i,j}\le d_0$ and 0 otherwise. The application of the method in a database of aging status in frail elderly reported in \cite{mt07} shows a good performance of the method, and gives a nice expectative of exploitation of the concepts of synchronization in the area of data mining.

\subsubsection{Consensus problems}

Consensus problems, understood as the ability of an ensemble of dynamic agents to reach a unique and common
value in an asymptotically stable stationary state,
have a long history in the field of computer science, particularly in automata theory and
distributed computation.
In many applications, like for instance cooperative control on unmanned air vehicles, formation control or
distributed sensor networks,
groups of agents need to agree upon certain quantities of interest \cite{om04}.
As a result, it is important to address these problems of agreement within the assumption that
agents form a complex pattern of interactions.
These interactions can be directed or undirected, fixed or mobile, constant or weighted, keeping
then many of the ingredients we have been discussing in this review.
Another interesting fact in this sort of problems is the existence of time delays in the communication
process.

In \cite{om04} the authors define consensus problems on general graphs. Let us consider a dynamic graph in which the
connectivity pattern of the nodes can change in time. At each node, a dynamical agent evolves in time according to the dynamics
\begin{equation}
\dot{x}_i=f(x_i,u_i),
\end{equation}
where $f(x_i,u_i)$ is a function that depends on the state of the unit $x_i$, and on $u_i$ that describes
the influence from the neighbors.
The $\chi$-consensus problem in a dynamical graph is a distributed way to reach an asymptotically stable equilibrium $x^*$ satisfying $x^*_i=\chi(x(0)), \; \forall i$,
where $\chi(x(0))$ is a prescribed function of the initial values (e.g., the average or the minimum values).

They present two protocols that solve consensus problems
in a network of agents:
\begin{enumerate}
\item fixed or switching topology and zero communication time-delay:
\begin{equation}
\dot{x_i}=\sum_{i,j=1}^N a_{ij}(t)(x_j(t)-x_i(t)),
\end{equation}
\item fixed topology and non-zero communication time-delay $\tau_{ij}>0$
\begin{equation}
\dot{x_i}=\sum_{i,j=1}^N a_{ij}(x_j(t-\tau_{ij})-x_i(t-\tau_{ij})).
\end{equation}
\end{enumerate}
We note that the analysis of the asymptotic behavior of such linear system is similar to
the stability analysis performed in the framework of the MSF (Sect. \ref{sect_MSF}).

The authors find very interesting results in terms of network properties.
For instance, a network with fixed topology that is a strongly connected digraph (a subgraph connected
via a path that follows the direction of the edges of the graph).
solves the average consensus problem if and only if all the nodes of the network
have the same indegree as the outdegree,
i.e. $k_i^{\mbox{\scriptsize{in}}}=k_i^{\mbox{\scriptsize{out}}}, \; \forall i$,
as the balanced networks discussed in \cite{bbh06} (see Sect. \ref{subsect_beyond_MSF}).
Furthermore, the performance of the network measured in terms of the speed
in which the global asymptotic equilibrium state is reached, is proportional
to $\lambda_2 (\hat{\mathcal{G}})$ where $\hat{\mathcal{G}}$ is the mirror graph induced by $\mathcal{G}$,
which is defined as an undirected graph with symmetric adjacency matrix $\hat{a}_{ij}=(a_{ij}+a_{ji})/2$.

For a switching topology, they find that if the dynamics of the network is such that any graph along
the time evolution is strongly connected and balanced then the switching system asymptotically converges to
an average consensus.
Concerning time communication-delays, the important result is that if all links have the same time-delay $\tau>0$,
and the network is fixed, undirected and connected, the system solves the average consensus if $\tau \in (0,\pi/2\lambda_N)$.
In this case, in a similar way as discussed in previous applications, there are two tradeoff issues that can be related to problems of network design;
one concerns the robustness of the protocol with respect to time-delays,
and the other to communication cost.

When applying this framework to a certain class of networks, it is found \cite{o05} that the speed of convergence,
as the inverse of $\lambda_2$ (as was also found in synchronization problems \cite{ad07}), can be increased by orders of magnitude by simply rewiring a regular lattice, while this change has a negligible effect on  $\lambda_N$, which measures the robustness to delays of the system.
This can be understood by the eigenvalues of the SW network in Eq. (\ref{bp02_eq5}) as compared to the regular networks in Eq.  (\ref{bp02_eq2}).
Some other variations can be found in the recent literature; e.g., in \cite{wgw07} several network models with physical neighborhood connectivity are analyzed. Depending on the precise rules they discuss on the performance and the robustness of the system.

Due to its importance as an application in computer science, consensus problems are interesting by themselves.
But understanding its \emph{linear} dynamics can be also of great importance in the behavior
of complex populations of units that evolve according to more complex \emph{non-linear} dynamics, as it happens
in many synchronization problems.

\subsubsection{Wireless communication networks}

Another emerging line of research can be found in the field of synchronizing wireless sensor communication networks. Wireless ad-hoc networks are telecommunication infrastructures formed by devices equipped with a short-range wireless technology, such as WiFi or Bluetooth. Unlike wired networks, these networks can be created on the fly to perform a task, such as information routing, environmental sensing, etc. \cite{hekmat06}.
Furthermore, the topology of these networks can be changed dynamically to achieve a desired functionality. From the perspective of fundamental research, these systems provide a clear-cut example of highly dynamic, self-organizing complex systems.

One of the main technological problems in wireless networks is that of synchronize time activity in a decentralized way. Wireless time synchronization is used for many different purposes including location, proximity, energy efficiency, and mobility for example. We will revise in this section two approaches to the problem that have been developed within the scope of the synchronization scenario reviewed so far. Other approaches not clearly connected to synchronization in complex networks to solve this important problem can be found in the specific literature \cite{sy04}. The first approach concerns to the routing and information flow algorithms which require synchronization of the clocks of the nodes of the wireless network to establish a global coordinated time. The topology of these sensor networks is accurately represented by random geometric graphs which are constructed by dropping $n$ points randomly uniformly into the unit square (or more generally according to some arbitrary specified density function on d-dimensional Euclidean space) and adding edges to connect any two points distant at most $r$ from each other. In a very recent work \cite{dgmn08}, synchronization of Kuramoto oscillators in
these networks is studied. They consider a wireless system in which the connections vary at a time scale much shorter
than the time scale associated to the synchronization dynamics, and hence the network is static. Nodes correspond to
devices that have a finite transmission range, and are linked to those nodes that are located within the range.
This procedure gives rise to a two-dimensional random geometric graph, which is characterized by a
high clustering coefficient and  a very large average shortest path length, as compared to ER graphs with the same number of nodes and links.
The remarkable result is that this type of network is very hard to synchronize,
both in terms of the stability of the synchronized state and in terms of the time needed to reach
the completely synchronized state. Although they are very homogeneous, the
smallest non-zero eigenvalue of the Laplacian matrix is very low, providing a clear limitation for synchronizability of Type II. This result points out the limitations concerning synchronizability that raise from this topology, interestingly, just by rewiring a small fraction of the links at random synchronizability is clearly improved, the distances are shortened and at the same time the clustering is decreased, which show up as an increasing of the eigenvalue $\lambda_2$ almost without affecting the largest eigenvalue $\lambda_N$. An extended study about the convergence properties of a similar system
where nodes are represented as discrete-time oscillators, is studied through algebraic graph theory in \cite{ss07}. The authors main finding is that the distributed synchronization problem converges to a unique cluster of synchronized nodes, if and only if the associated weighted directed graph $G$ is strongly connected, i.e. if there is a path from each vertex in the graph to every other vertex.

The second approach concerns the shared resources available in these systems. Communication channels have a finite bandwidth so that the access times of different users should be desynchronized when their number is large and non necessarily constant, this is basically the idea behind any Time Division Multiple Access (TDMA) algorithm to be used over a multi-hop wireless network. In  \cite{drpn07}, the authors proposed a biologically inspired algorithm for desynchronization
in a single-hop network that is in the scope of the review. They consider a set of $N$ nodes (integrate-and-fire oscillators) that generate events with a
common period. The nodes rearrange their phases, just considering the times in which neighboring (in time) units fire,
in such a way that the events become spaced at intervals $T/N$. The final state then corresponds to what is usually called a round-robin schedule.
In this way, the use of the bandwidth without collisions between messages is optimized.
Inspired by this result, in \cite{da08} the authors considered the units to be Kuramoto oscillators with a common frequency.
Introducing some dephase angle in the sinus function and coupling pairs of units along a closed chain,
the authors find new stable configurations different from the completely synchronized state.
Some of these configurations correspond also to the round-robin schedule, which turns out to be very robust under the addition
or deletion of nodes from the system. This finding obtained in the scope of a precise application, is general, and accounts for global synchronization of discrete-time dynamical directed networks.

To end this section we also to point out a different problem, also in the area of wireless mobile sensors, that has been nicely solved by a description in terms of pulse coupled oscillators in complex networks. The problem is that of the decentralized detection of abrupt changes, that in wireless networks can represent failures in communication, attacks to certain sensors or, generally speaking, any change in the sensed activity of purpose. The authors of \cite{hs04} proposed a very simple transmitter with no routing, no multiple access, only a  "pulse position modulation" mechanism. In particular, they assume that the nodes can transmit only through the emission of pulses with constant amplitude. The information of the sensor data and the interaction among them can only be encoded in the timing of the pulse emission. This simplistic but effective configuration leads the problem of decentralized detection of abrupt changes to that of observing the synchronization of pulse coupled oscillators in a network, after a local perturbation of them. The work did not included the specific topology of interactions as a fundamental parameter of the problem, and then do not propose an optimum network scenario for the propagation of the signals, however it clearly points out another technological problem in electrical and computer engineering, that can be solved in the scope of synchronization in complex networks.

\subsubsection{Decentralized logistics}

Logistics is the management of the flow of goods, information and other resources. Sometimes, this management is limited by the capability
of maintaining global knowledge and/or global communication. In this case, the necessity of decentralized coordination mechanisms are mandatory.
In many material flow systems
coordination of tasks in a parallel way is essential for an optimal functioning but difficult to achieve.
Typical examples of this are cross-roads in road traffic \cite{nagatani02} and supply chains in production processes \cite{h03}. Recent work on supply networks has shown
how to treat them as physical transport problems governed by balance equations and
equations for the adaptation of production speeds. Although the nonlinear behavior is different, the linearized
set of coupled differential equations is formally related to those of mechanical or electrical oscillator networks \cite{hlssp04}.
Whereas traditional optimization techniques can be used to setup single nodes, the
inherent topological complexity makes maintenance of coordination at  network-wide level to be practically unsolvable by these methods.
Furthermore, robustness and flexibility, due to continuous changes in demand and failures, are also required for an efficient transportation.

There is an analogy between material transportation in networks and the flow of chemical substances and nutrients in biological organisms,
where synchronization dynamics plays an important role.
In \cite{lkph06} it has been proposed a decentralized control model for material flow networks
with transportation delays and setup-times, based on phase-synchronization
of oscillatory services at the network nodes.

A material transport network is a directed and weighted graph where  the flow of material at nodes
is conserved. Subsets of links are active at different times, and this makes that the activity of the node is periodic
and one can map a continuous phase variable $\theta(t)$ to a discrete service state $M: \theta(t)\rightarrow s(t)$. While the phase angle $\theta$ of the oscillator modeling the intersection varies from 0 to $2\pi$ at a rate $\omega$, all states $s$ are sequentially activated.
To achieve a coordination of the switching states on a network-wide
level, they propose a suitable synchronization mechanism.

\begin{figure}[!t]
    \begin{center}
        \begin{picture}(200,200)(0,0)
        \includegraphics[width=200pt]{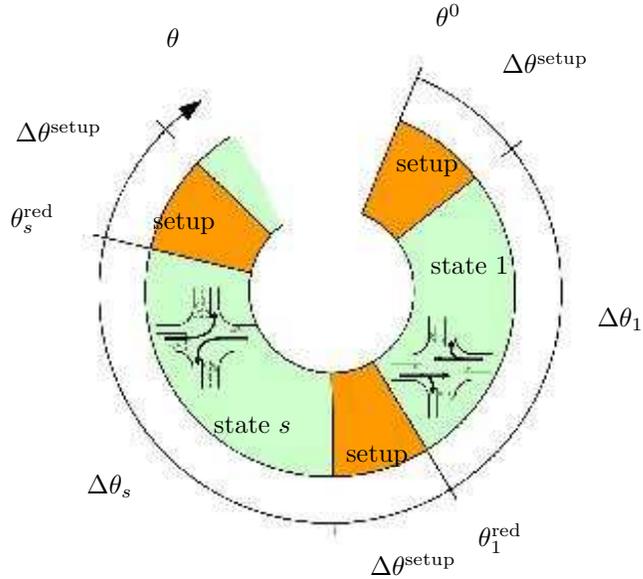}
        \end{picture}
        \begin{picture}(0, 200)(200, 0)
            \put(122,149){setup}
            \put(103,39){setup}
            \put(30,127){setup}
            \put(135,110){state 1}
            \put(53,50){state $s$}
            \put(137,204){$\theta^0$}
            \put(162,186){$\Delta \theta^\mathrm{setup}$}
            \put(198,90){$\Delta \theta_1$}
            \put(153,7){$\theta_1^\mathrm{red}$}
            \put(112,-4){$\Delta \theta^\mathrm{setup}$}
            \put(5,27){$\Delta \theta_s$}
            \put(-24,128){$\theta_s^\mathrm{red}$}
            \put(-22,160){$\Delta \theta^\mathrm{setup}$}
            \put(35,196){$\theta$}
        \end{picture}
    \end{center}
\caption{(Color online) A single intersection adjusts the mapping of the phase-angle $\theta$ to the switching states $s$ locally. Within a complete cycle, each state
$s$ is sequentially activated for a period $\Delta \theta_s$, during which the corresponding non-conflicting traffic lights are set to green. While switching
from one state to another, all traffic lights are set to red for a period of $\Delta \theta^{\hbox{setup}}$. The phase-angle, at which a new cycle starts, is denoted
by $\theta_0$. From \cite{lkph06}.}
\label{map_service_state}
\end{figure}

The authors apply this formalism to the control of traffic lights at intersections of road networks. A single traffic light intersection is modeled by an oscillator where the continuous phase maps to a set of states corresponding to
green lights (see Fig. \ref{map_service_state}). The maximum frequency of the oscillator dynamics is calculated in terms of the load at the different lanes and the setup-time. Global coordination of the network is achieved by synchronizing the local phases and frequencies, requiring to reach a phase-locked state where the phase difference between neighboring sites is fixed. They suggest a coupling
on two different time scales:

\begin{itemize}
\item Adaptation of the phase $\theta$ a la Kuramoto:
\begin{equation}
\dot{\theta_i}=\min \left\{ \omega_i^{{\max}},\omega_i(t)+\frac{1}{T_{\theta}}\sum_{j\in \Gamma_i} \sin
\left( \theta_j(t)-\theta_i(t) \right) \right\}
\end{equation}
where $\omega_i$ is the intrinsic frequency. As long as $\omega_i < \omega_i^{\max}$, $\theta_i$ tries to adjust to the neighboring phases. The constant $T_\theta$ corresponds to the typical time scale for this adaptation.
\item A second decentralized coupling can be used to increase the intrinsic frequencies to approach the possible maximum within a slow time scale:
\begin{equation}
\dot{\omega}_i=\frac{1}{T_{\omega}}
\left( \min_{j \in \Gamma_i} \left\{ \omega_j(t) \right\} + \Delta\omega -\omega_i(t) \right).
\end{equation}
Here the constant parameter $\Delta\omega$ provides a linear drift towards higher frequencies.
\end{itemize}

\begin{figure}[!t]
\begin{center}
\epsfig{file=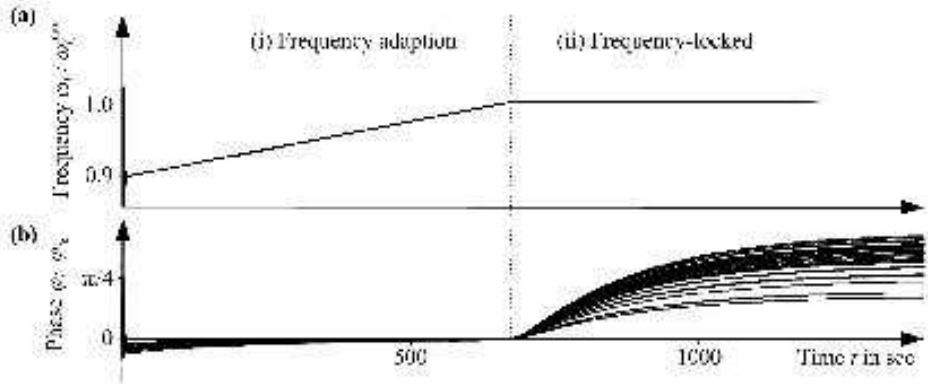,width=0.8\columnwidth,angle=0,clip=1}
\end{center}
\caption{A single intersection of roads is modelled by a phase angle $\theta$.
This continuous phase maps to a discrete set of states $\{ s\}$. Simulation results for a regular lattice road network, where the intersections are defined as oscillators with (a) a frequency $\omega_i$
and (b) a phase$\theta_i$, shows a phase transition towards a synchronized (frequency-locked) state.
From \cite{lkph06}.
}
\label{traffic_synchro}
\end{figure}

Under these assumptions two dynamical regimes are possible (see Fig. \ref{traffic_synchro}).
Starting with a random initial condition (left), the system quickly settles into a state with
growing common frequency and vanishing phase-differences. As soon as the maximum common frequency is found, the system enters the other state with a locked common frequency and phase-differences exponentially
converging towards constant values (right).

The extension of the analysis performed on such a simple setting to more realistic complex transportation networks
is very promising but challenging \cite{hll05}.

\subsubsection{Power-grids}

Power grids are physical networks of electrical power distribution lines of generators and consumers.
In the pioneering paper by \citeauthor{ws98} \cite{ws98} it was
already reported that the power-grid constitutes one of the examples of a self-organized topology that has grown without
a clear central controller. This
topology is indeed very sensitive to attacks and failures. From its topological point of view there are several
analyses on power-grids in different areas of the world \cite{clm04a,cp05} and some models have been proposed
to deal with the cascading process of failures \cite{scl00,clm04b,ste05}.

The principles of electricity generation and distribution are well known.
Synchronization of the system is understood as every station and every piece of equipment  running on the
same clock, which is crucial for its proper operation.
Cascading failures related to de-synchronization can lead to massive
power blackouts \cite{whitepaper}.

Then power production, dissipation, transmission, and consumption represents a dynamical problem and the power grid
can be seen as an example of a system of oscillators \cite{strogatz03}.
Along this line, in \cite{fnp07} it has beeb proposed  a model where each element (generators and machines) is described in terms of a phase that grows linearly in time with
a frequency that is close to the standard frequency of the electric system (50 or 60 Hz).

Consider the power produced at a generator. It can then
be dissipated, accumulated, or transmitted along the electric line. The first two terms (dissipation and accumulation) depend on the
frequency of the generator whereas the last one (transmission) is proportional to the sinus of the phase difference between the generator and the
machine at the other extreme of the line. Then, a simple energy balance equation relates the evolution of the phase (first and second time
derivatives) with sinus of phase differences.

Applying this simplified approach to a networked system of generators and machines, they arrive to a set of Kuramoto-like differential equations
\begin{equation}
\ddot{\theta_i} = -\alpha \dot{\theta_i}+\omega_i + K\sum_{j\in \Gamma (i)}\sin (\theta_j-\theta_i),
\end{equation}
where $\omega_{i}$ is related to the power generated at the element and to the dissipated power,
and $K$, representing the stength of the coupling, is related to the maximum transmitted power.

Within this framework,they analyze, as an application, under which conditions the system is able to restore to a stable operation
after a perturbation in simple networks of machines and generators. To the best of our knowledge this is a first approximation
to the real applicability of the knowledge about synchronization in complex networks to power grids, although the hypothesis along the work
are still very relevant here.

\subsection{Social sciences and economy}

In the last decades, social sciences and economy have become one of the favorite applications for
physicists.
In particular, tools and models from statistical physics can be implemented on what some people has called
\emph{social atoms}, \cite{buchanan07}
i.e. unanimated particles are replaced by agents that take decisions, trade stocks or
play games.
Simple rules lead to interesting collective behaviors and synchronization is one of them, because some of the
activities that individual agents do can become correlated in time due to its interaction pattern, which, in turn,
is clearly another example of the complex topologies considered along the review.

In social systems, however, it is not an easy task to identify the relationship between agents (being humans or collectives in social interactions, stock prices in finances, or countries in the World Trade Web). We alert the reader, that in some cases the quasi-periodic correlated activity is interpreted as synchronization in this scenario. For example, dynamical similarity along economic cycles is understood as synchronization in economic terms. Keeping this in mind, some of the applications presented in this section are weak formulations of synchronization in complex networks, however we think they are interesting because they open the door to stronger formulations in this context.

\subsubsection{Opinion formation}

One of such problems is the study of opinion formation in society.
The underlying idea  is that individuals
(or agents) have opinions that change under the influence
of other individuals giving rise to a sort of collective behavior, grouping together a macroscopic part of the whole population with similar social features \cite{blmch06}. Therefore, the main goal is to figure out whether and when a complete or partial consensus can emerge out of initially different opinions, no matter how long it takes for the consensus to be reached.

In general, the formation of a collective opinion about a certain matter  is not equivalent to a transition to some kind of synchrony, but rather to a transition to an absorbing state. However, a recently proposed model \cite{plr05} makes explicit use of a modified KM (see Sect. III) and thus in this case the formation of groups of opinions can be thought of a synchronization process. Agreement models deal with $N$ individuals characterized by an opinion $x_i$ (either an integer or real number) and a network of contacts that drives the dynamics of opinion formation through deterministic rules \cite{blmch06}. In \cite{plr05} it has been considered the case in which opinions are neither bounded nor periodic, but that two initially different opinions can also diverge when time goes on. Moreover, they also take into account that two quite different individuals tend to interact less by assuming that the coupling between these two individuals is a decreasing function of their opinion differences. Finally, in \cite{plr05}  the main source of heterogeneity is not given by the initial positions, but by different rates of changing individuals' opinions.

Taking all the previous statements into account, it has been proposed the following governing equations for the dynamics of the rate of change of opinions \cite{plr05}
\be
\dot{x}_i=\omega_i+\frac{\sigma}{N} \sum_{j=1}^{N} \alpha\sin(x_j-x_i)\text{e}^{-\alpha|x_j-x_i|}\;,
\label{opinion1}
\ee
where $x_i(t)\in(-\infty,+\infty)$ is the opinion of the $i$th individual at time $t$ and the $\omega_i$'s are the natural or intrinsic opinion changing rates. Note that the interactions are assumed to be all-to-all, though the model can be directly generalized to any other topology. Moreover, the $\omega$'s are drawn from a distribution $g(\omega)$ centered at $\omega_0$ with the same properties as in the KM case (Sect. III). In other words, in \cite{plr05} the authors approach the problem of opinion formation from a radically different point of view in which individuals do not only change their opinions, but also the rate at which these changes take place. It is plausible then to assume a dynamics described by oscillators coupled together. As opinion changing rates depend on the actual interaction between the members of the population, the dynamical model fits quite naively a Kuramoto-like model with the additional constraint that individuals having too far opinions will not likely interact. On the other hand, as opposite to the KM, opinions are not periodic anymore, so that a new order parameter is introduced as
\be
R(t)=1-\sqrt{\frac{1}{N}\sum_{1}^{N}(\dot{x}_j(t)-\dot{X}(t))^2}\;,
\label{opinion2}
\ee
with $\dot{X}(t)$ being the average of $\dot{x}_j(t)$ over all individuals. From Eq.\ (\ref{opinion2}), it follows that $R=1$ if complete synchronization is achieved and $R<1$ when only partial synchronization occurs. Note that synchronization in this context means that the population has reached a unique state of opinion, i.e., it is uniquely polarized and that further changes take places at the same rate. Moreover, when complete synchronization is not achieved, different opinions are present in the system and the population can be regarded as multipolar. The issue is then to investigate under what conditions different emergent behaviors are observed.

Numerical simulations of the model show that there is a phase transition from incoherence to synchrony at a well defined critical coupling $\sigma_c$. It is argued that when $\sigma < \sigma_c$, the society can be thought of as being formed by isolated, non-interacting cultures or groups of opinions, since mixture or agreement is not achievable. On the contrary, when $\sigma \gg \sigma_c$, the system fully synchronizes, giving rise in a social context to a polarized or globalized society where social and cultural differences are constrained into a single way of thinking, notwithstanding the different tendencies to changes of the individuals. Finally, the authors reported that bipolarity is only possible if $\sigma \sim \sigma_c$, although in this case the model shows a rich behavior depending on the way initial opinions are assigned \cite{plr05}. It would be extremely useful to investigate in the future what is the influence of the underlying topology and if the overall picture described above still remains valid. Furthermore, as changes in opinions in a population also implies reshaping of the social structure, the question of how rewiring mechanisms that take into account the actual opinion states of individuals is worth studying in the future.

\subsubsection{Finance}

When reading the economic news, it is not difficult to identify the existence
of economic cycles in which Gross Domestic Products (GDP's), economic sectors, or stock options raise and fall.
Most of the time this does not happen for isolated countries, sectors or options but it occurs
in quite a synchronized way, although some delays are noticeable.

Within the framework of the current review, we are focusing on synchronization in complex networks, and this
is  what we can identify  in many economical sectors: there exists a complicated pattern of
interactions among companies  or countries and the dynamics
of each one is quite complex. But, in contrast to many networks with a physical
background, here we neither know in detail the node dynamics nor its connectivity pattern.
In this situation it is useful to look at the problem from a different angle.
By analyzing some macroscopic outcomes, we get some insight into the
agents' interactions.

In the economic literature, synchronization is measured by a correlation coefficient (see, for instance, \cite{forbes02}), based on the idea that correlated (synchronized) business cycles should generate
correlated returns. The point is to identify what types of interactions lie behind market comovements.
Synchronization is the result from two different effects. On the one hand, there are different types of
common disturbances (world interest rates, oil price, or political uncertainty). On the other hand,
there exist strong interactions between the agents (financial relationships, sector dependencies,
co-participation in director boards, etc.). It is precisely, these interactions that play a crucial role in the synchronized behavior
along economic cycles of tightly connected agents and the analysis of the
correlations can help in shedding light on the strength of the different factors.

The application of networks concepts, mainly that of trees, to economical
systems dates back to the pioneering work by \citeauthor{mantegna99} \cite{mantegna99}, who found a hierarchical arrangement
of stocks through the study of the correlation returns.

Recently, the authors in \cite{ockkk03} have taken a similar approach to analyze the dynamics of markets.
They look at the daily closure prices for a total of $N=477$ stocks traded by the New York Stock Exchange over a period
of 20 years, from Jan 02, 1980 to Dec 31, 1999. The data is smoothed by looking at time windows of given width. As is usually done in the analysis of financial data, the measured quantity is the logarithmic return of the stocks, defined as
\begin{equation}
r_i(t)=\ln P_i(t) - \ln P_i(t-1),
\end{equation}
where $P_i(t)$ is the closure price of stock $i$ at time $t$. To quantify the degree of synchronization of the data, they use the equal time correlation between assets
\begin{equation}
\rho_{ij}(t)=\frac{\langle  r_i(t)r_j(t)\rangle -\langle  r_i(t)\rangle \langle  r_j(t)\rangle }{\sqrt{\left[\langle  r_i^2(t)\rangle -\langle   r_i(t)\rangle ^2\right]\left[\langle  r_j^2(t)\rangle -\langle  r_j(t)\rangle ^2\right]}}
\end{equation}
where $\langle  \ldots \rangle $ stands for a time average over the consecutive trading days.

From these correlations the \emph{asset tree} is constructed. The distance between assets is defined as
\begin{equation}
d_{ij}(t)=\sqrt{2(1-\rho_{ij}(t))}.
\end{equation}
The minimum spanning tree is a simply connected graph with $N-1$ edges, such that the sum of all the distances between connected nodes in the graph $\sum d_{ij}(t)$ is minimum.

This procedure generates a time sequence of asset trees that can be interpreted as a sequence of evolutionary steps
of a single dynamics asset tree. For instance, one can identify the leading asset, that, in most instances,
corresponds to General Electric.

Such a reduction of the  whole set of data retains most of the salient features of the stock market. It is a remarkable fact that during crisis periods the markets are very strongly correlated. In terms of the tree its average length is reduced and the tree is very tightly packed. By reducing the time window, the location of the smallest tree converges to the Black Monday (October 19, 1987).

It is clear that a hidden pattern of interactions between the assets is responsible for such a \emph{synchronized} behavior and that during crashes the interactions are strengthened. The message here is that the degree of synchronization, quantitatively described in terms of the asset correlation, is an indirect measure  of the
existence of strongly connected agents in financial markets.

The goal of measuring correlations in time series of financial data is to identify
synchronized behavior of stocks.
Stocks are synchronized if they are strongly connected by means of some of the interactions we
listed above (sharing directors, capital flow, sector dependency, etc.).
But clustering these stocks according to their correlations is usually a hard task and not very accurate.
In a recent paper \cite{bbdfp05}, it has been proposed a new method based on synchronization of chaotic maps to get
a more precise clustering of the data. They look at correlations between stocks in the usual way ($c_{ij}=<\rho_{ij}(t)>_t$, is the time average of the correlation matrix) but they use these values
to construct, through a nonlinear function, a new matrix $J_{ij}$ that is used as the interaction matrix
between units. The units evolve according to chaotic map dynamics
\begin{equation}
x_i(\tau +1)=\frac{1}{\sum_{i\ne j}J_{ij}}\sum_{i\ne j}J_{ij} f(x_j(\tau)
\end{equation}
and $f(x)=1 -2x^2$ is the logistic map. The next step is to convert, after some equilibration time, the continuous variable $x_i$ into a spin-like one
and compute the mutual information between stocks, $I_{ij}$. Then stocks are clustered by using $I_{ij}$ as
the similarity index, and the most stable partition is that with the highest cluster entropy.
The dendrogram obtained in this way shows, for a particular set of data, a clear partition
between different classes of stocks. However, the analog procedure applied to the original correlations, $c_{ij}$,
shows a "chaining effect" that tends to yield elongated clusters.
In this case, synchronization of chaotic dynamics turns out to be a powerful tool for
the analysis of financial data subjected to similar business cycles.

\subsubsection{World Trade Web}

The World Trade Web (WTW) is another example of economic system that has been widely analyzed
from the network perspective.
Different sources provide data about the trade between countries.
Networks are constructed such that countries are the nodes and trade corresponds to the interaction strengths
between the nodes.
There are several studies that have focused on the \emph{static} complex nature of
the links between countries \cite{sb03} and also the evolution of the statistical properties of the
network \cite{gl05}.
But, from a dynamical point of view, the trade volume between countries is
related  to the internal state
of the nodes, measured, in this case, by the Gross Domestic Product (GDP).

Hence a clear relationship between node dynamics (which can have a cyclical component) and
trade strength appears. In particular, in
\cite{gdacl07} the interplay between the topology of the WTW and the dynamics
of the GDP's was analyzed.

\begin{figure}[!t]
\begin{center}
\epsfig{file=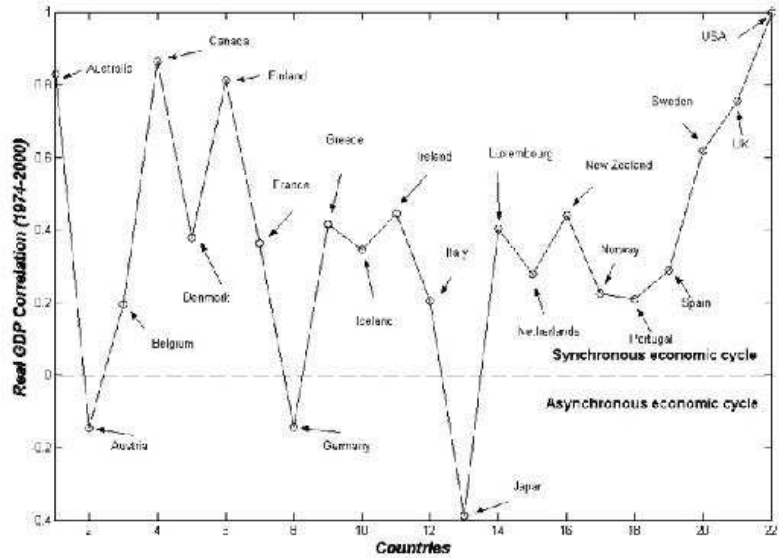,width=0.65\columnwidth,angle=0,clip=1}
\end{center}
\caption{Twenty-two developed countries' economic cycles synchronization phenomena. The positive real GDP correlation means synchronous economic cycles with the US.
From \cite{lyc03}.
}
\label{economic_synchro}
\end{figure}

In none of the previous studies any reference is made to the precise dynamics of the countries
economies. As we have stated before, we find what are usually called \emph{economic cycles}. Economies rise and fall,
although they mainly raise, but without a constant rate,
following rather unpredictable evolutions in time. Due to globalization effects, all economies are
strongly correlated and they will tend to follow a common trend.
In our framework, we can say that following a similar time evolution, economies
are synchronized. This cycle synchronization of economies is a topic of current interest in the
economic literature; in \cite{ccs07} a comparative analysis between
developing and industrial countries is performed, finding that the correlations within the
first group are positive but smaller than within the second group.
Another example is found in \cite{lyc03} where correlations between countries are measured
in a similar way as the financial series reported above. By taking into account that the
US is the largest economy, and also the biggest node in the WTW, they look at the degree
of synchronization between 22 developed countries and the USA (see Fig. \ref{economic_synchro}).

Another clear effect is that particular economies are tightly connected because of
economical agreements or dependence on particular sectors.
In the language of communities this stronger relation can be understood as the existence
of communities in the overall structure of the world economy.

Along these lines we have observed that there is a tight relationship between synchronization of economic
cycles in terms of the GDP and the topological structure of the WTW.
The important question raised here is if the structure of the network that can be constructed from
the empirical data on the cycles correlations, as is done with the finance data, can be mapped into
the WTW also empirically constructed from the world trade transactions.




\section{Perspectives}

As we have seen, the  MSF provides a powerful framework to study the relation between the network architecture
and various types of synchronizability. However, the analysis is mainly limited to the linear stability of the
complete synchronization states. In most realistic systems where synchronization is relevant, complete synchronization
of fully identical oscillators is too ideal, and  very  strong  degrees of synchronization could relate to
pathological activities, such as epileptic seizure in neural systems or social catastrophes. Most likely, various levels
of synchronization are desirable to enable the system to have flexibility and robustness for the emergency of
coordination at different scales.

In this sense, the investigation of synchronization in complex networks is still emerging. We would like to list a few
future research directions which we think are the main challenges in this field that need to be acomplished to
achieve a complete comprehension of the structure-synchronization relations in complex networks.

\begin{itemize}



\item Spectral properties and synchronization processes

In the MSF formulation, we mainly considered the minimal and maximal eigenvalues that are associated to the stability of the synchronization states.
The detailed spectral properties, including the eigenvectors, of the Laplacian/Adjaceny matrix are important when we are interested  in the processes leading to synchronization, or interested in the dynamical patterns emerging after a perturbation occurs. So far studies on detailed spectral properties are still mainly restricted to random networks \cite{clv03,dgms03,rakk05,kk07,bj07,km07} and only a few deal on the relation between synchronization patterns and spectral properties, out of the complete synchronization, see e.g., \cite{roh04,zk06b, adp06b}.

\item Directed networks and synchronization

There is growing interest in characterizing  directed and weighted networks\cite{jtaob00,gl04,bbpv04,zzzsk08,bgm08,roh08}.
The analysis of the spectra of directed and weighed networks\cite{roh06b,ra06}, which in general are complex, and the study
of the impact of the directed characterizations of the networks on synchronization process\cite{nm06a,nm06b,roh05b},  will be key
to  understanding  dynamical organization in  more realistic complex systems.

\item Co-evolution of structure and synchronization

Most of the work has considered the impact of network architectures on the synchronization dynamics.
In many realistic systems, the feedback of dynamics can reshape the
network structures, e.g., in neural systems  through synaptic plasticity \cite{bp01}.
As a result, adaptation, co-evolution and self-organization occur crossing a broad range of scales (see \cite{tb08} for a recent
review on adaptive networks). Adaptation due to synchronization has received increasing interest \cite{ik02,ik03,zk06a,so08},
and this line of research will lead to an integrated understanding of the structure and dynamics of many complex network systems.

\end{itemize}

\section{Conclusions}

Through the current review we have outlined the state of the art towards a {\em theory} of synchronization in complex networks. We emphasize the word theory, because, up to now, physicists have made an effort of characterization that certainly deepened our understanding of the complex connectivity of natural and manmade networks, however, we cannot yet state that we have a {\em theory} of complex networks. The topological characterization may not be useful to make actual predictions which can be contrasted with experiments. To this specific end, the complex network substrate must be enriched and entangled to the functioning of the system, i.e., to the dynamics run on top of it.

The phenomenon of synchronization is one of the paradigmatic observations in different dynamical systems. It is at the heart of some biological processes, and according to the wide variety of applications presented here, it is a plausible abstraction for many other processes in different contexts. The natural approximation to synchronization, from the simplicity of the Kuramoto model, has been explored, and even in this case an intricate set of questions concerning the uniformity of the equations proposed, or the nature of the critical behavior at the onset of synchronization still have to be definitely settled. The main results, however, have helped to understand the nature of the relationship between the topology of interactions and the synchronization of phase oscillators. We foresee the importance of these results in the basis of a theory of neural dynamics of the brain. Although neural dynamics is far more complex than the phase representation reviewed, main features that describe the path towards synchrony in complex networks have been already stated and can thus be considered as a good starting point. However, regarding the subjects revised here, one easily realizes that the work is far from complete and many questions are in the air, e.g. the evolution of synchronization in evolving topologies, the effect on the phenomenon of several kind of disorder, time delays, the presence of noise, etc.

The other main contribution to the problem comes from the MSF formalism. The elegant structure of the formalism, designed for linear systems or nonlinear systems close to the synchronization state, allows us to make theoretical predictions independent of the specificities of the dynamics. This general framework has been deeply studied recently, and provides one of the few mechanisms that allow to make predictions about the evolution of synchronized systems as a function of its specific topology.

We think that the exploration of new mathematical objects, able to merge the information provided by specific dynamics (as for example the Kuramoto model) along to the whole process towards synchronization, together with the fine description of the dynamics near the synchronization manifold, should be the focus of intense research if we aim to provide a general theory of synchronization processes in complex networks. On the other hand, the myriad of applications that can be cast into mathematical models equivalent to those presented along the review, indicates an explosion of activity in different disciplines that will use the conceptual framework of synchronization processes in networked systems as a fundamental playground for understanding its dynamical behavior.

\section*{Acknowledgments}

We thank all those that have actively collaborated with us along the years in the subject. Special thanks go to S. Boccaletti for many suggestions and enlightening discussions on the subject of this review. We are also indebted to J. Garc\'{\i}a-Ojalvo, C.J. P\'{e}rez-Vicente, J. Schmidt,
M. Thiel, for having carefully read the manuscript and for their very helpful comments.
A.A. and A.D.-G. thank funding from Spanish Government
(BFM-2003-08258 and FIS-2006-13321) and Generalitat de Catalunya (2005SGR00889).
JK acknowledges support from the EU Network of Excellence BIOSIM (Contract No. LSHB-CT-2004-005137), the EU Network BRACCIA, the DFG SFB 555, and the BMBF GoFORSYS.
Y.M. is supported by MEC through the Ram\'{o}n y Cajal Program, the Spanish DGICYT Projects FIS2006-12781-C02-01 and
FIS2005-00337, and by the European NEST Pathfinder project GABA (contract 043309). C.S.Z. is supported by  the Faculty Research Grant (FRG) of Hong Kong Baptist University (FRG /07-08/II-08 and FRG/07-08/II-62).

\bibliography{synchro}

\end{document}